\documentclass[12pt]{iopart}
\usepackage{epsfig,amssymb}

\def\spose#1{\hbox to 0pt{#1\hss}}
\def\lesssim{\mathrel{\spose{\lower 3pt\hbox{$\mathchar"218$}}
 \raise 2.0pt\hbox{$\mathchar"13C$}}}
\def\gtrsim{\mathrel{\spose{\lower 3pt\hbox{$\mathchar"218$}}
 \raise 2.0pt\hbox{$\mathchar"13E$}}}
\def\case#1#2{{\textstyle\frac{#1}{#2}}}

\begin{document}

\title{ 
Multicritical behavior in the fully frustrated XY model and related systems
}

\author{Martin Hasenbusch,$^{1}$ Andrea Pelissetto,$^2$ and Ettore
Vicari$\,^1$ } 

\address{$^1$ Dip. Fisica dell'Universit\`a di Pisa and
INFN, Largo Pontecorvo 2, I-56127 Pisa, Italy} 

\address{$^2$
Dip. Fisica dell'Universit\`a di Roma ``La Sapienza" and INFN, \\ P.le
Moro 2, I-00185 Roma, Italy}

\ead{
Martin.Hasenbusch@df.unipi.it, Andrea.Pelissetto@roma1.infn.it,
Ettore.Vicari@df.unipi.it} 

\begin{abstract}
We study the phase diagram and critical behavior of the
two-dimensional square-lattice fully frustrated XY model (FFXY) and of
two related models, a lattice discretization of the
Landau-Ginzburg-Wilson Hamiltonian for the critical modes of the FFXY
model, and a coupled Ising-XY model.  We present 
a finite-size-scaling analysis of the results
of high-precision Monte Carlo
simulations on square lattices $L\times L$, up to $L=O(10^3)$.

In the FFXY model and in the other models, when the transitions are
continuous, there are two very close but separate transitions.  There
is an Ising chiral transition characterized by the onset of chiral
long-range order while spins remain paramagnetic. Then, as temperature
decreases, the systems undergo a Kosterlitz-Thouless spin transition
to a phase with quasi-long-range order.

The FFXY model and the other models in a rather large
parameter region show a crossover behavior at the chiral
and spin transitions that is universal to some extent.  We
conjecture that this universal behavior is due to a 
multicritical point. The numerical data suggest that the relevant
multicritical point is a zero-temperature transition.
A possible candidate is the O(4) point that controls the 
low-temperature behavior of the 4-vector model.

\end{abstract}

%\pacs{74.81.Fa, 64.60.-i, 64.60.Fr, 75.10.Hk}
%74.81.Fa Josephson junction arrays
%64.60.-i general studies of phase transitions
%64.60.Fr Equilibrium properties near critical points
%75.10.Hk classical spin models

%\bigskip
%\noindent
%{\it Keywords\/}:
%Frustrated XY models, Josephson junction arrays,
%Phase transitions and critical phenomena.

\maketitle

% ========================= BODY =========================
%\narrowtext

\section{Introduction}
\label{intro}

In the last few decades there has been considerable interest in the
nature of the phase diagram of the two-dimensional (2-$d$) fully
frustrated XY (FFXY) model, whose Hamiltonian is \cite{Villain-77}
\begin{equation}
{\cal H}_{\rm FFXY} = 
- J \sum_{\langle xy\rangle} {\rm cos} (\theta_x - \theta_y + A_{xy}).
\label{ffxy}
\end{equation}
Here $J>0$, the angle variables $\theta_x$ are defined on the sites of
a regular 2-$d$ lattice, the summation is over nearest-neighbor pairs,
and the quantities $A_{xy}$ are fixed and satisfy the constraint $\sum
A_{xy}=\pi$ around each plaquette.  The FFXY model is
experimentally relevant for Josephson-junction arrays in a magnetic
field.  It should describe the superconducting-to-normal transition at
half a flux quantum per plaquette, see, e.g., 
Refs.~\cite{Ling-etal-96,ML-00,ATPLM-02} and references therein.

As a consequence of frustration, the ground state of the FFXY model
presents an O(2)$\otimes{\mathbb Z}_2$ degeneracy
\cite{Villain-77}. While the O(2) degeneracy is the usual one and is
related to the O(2) global invariance of the Hamiltonian that is
broken in the low-temperature (LT) phase, the additional ${\mathbb
Z}_2$ degeneracy \cite{Villain-77,YD-85}. A variable that distinguishes the
different ground states is the chirality \cite{Villain-77}, which is
defined by
\begin{equation}
\chi_n = {\rm sign} \left[ \sum_{\langle xy \rangle \in \Pi_n} 
{\rm sin} (\theta_x - \theta_y + A_{xy}) \right],
\end{equation}
where $\Pi_n$ is a lattice plaquette and $n$ is the dual-lattice site
at its center.  At zero temperature there are two degenerate ground
states related by a global flip of the chiral variables $\chi_n$.
Therefore, the staggered chirality magnetization defines an order
parameter related to the chiral ${\mathbb Z}_2$ symmetry, which
competes with the spin magnetization in determining the phase diagram
and critical behavior of the FFXY model. Much work has been dedicated
to the study of the phase diagram of the FFXY and related models that
have the same ground-state degeneracy
\cite{YD-85,TJ-83,SS-84,MS-84,CD-85,Halsey-85,CS-85,Minnhagen-85,%
LJNL-86,Korshunov-86,Van-86,BDGL-86,GKP-86,Foda-88,TK-88,%
Scheinine-89,Grest-89,EHKT-89,%
GGPS-89,Henley-89,TK-90,GKLN-91,LKG-91,Nicolaides-91,LGK-91,FPA-91,dN-92,%
LT-92,RJ-92,SGZ-92,Granato-92,GN-93,Lee-94,LL-94,KNKB-94,Olsson-95,LLK-95,%
NGK-95,JR-96,XS-96,GKN-96,CN-96,Olsson-97,BZG-97,BG-97,JPC-97,LLK-97,%
Simon-97,LLKC-97,HH-98,%
Kawamura-98,LSZ-98,LL-98,CVCT-98,BD-98,PP-98,LS-00,FCKF-00,%
CLJ-01,MD-01,CP-01,Nussinov-01,NREK-02,Korshunov-02,ZRR-03,COPS-03,OI-03,JLC-03,GD-05,%
OT-05,PFMT-05}. In spite of that, there is not yet a general consensus
on the critical behavior of these systems. In this paper, we address
again this issue and, by performing simulations on very large lattices
and by means of a careful finite-size scaling (FSS) analysis of the
results, we provide what we believe is the definite answer on this
problem.

Two scenarios have been put forward for the critical behavior
of the FFXY and related 2-$d$ models. In one of them the 
system undergoes two transitions as temperature decreases.
First, chiral modes undergo an
Ising-type transition at $T_{\rm ch}$, characterized by the onset of
chiral long-range order while spins remain paramagnetic. Then, at a
lower temperature $T_{\rm sp}$, spin modes exhibit a transition to
a phase with quasi-long-range order, which coexists with the
long-range order of the chiral modes. This second transition is
expected to belong to the same universality class as the
Kosterlitz-Thouless (KT) transition in the standard 2-$d$ XY model
\cite{KT-73,JKKN-77}. In this scenario spin and chiral modes decouple
and order at different temperatures.  This scenario is supported by
arguments based on a kink-antikink unbinding picture
\cite{Korshunov-02,OT-05}. A second possibility is that spin and
chirality order at the same temperature, $T_c$, where both chiral
long-range order and spin quasi-long-range order set in
simultaneously.  In this case spin and chiral modes may be coupled at
the transition and may give rise to a qualitatively new critical
behavior. Therefore, one may observe values of the chiral critical
exponents that differ from the Ising ones, $\nu=1$ and $\eta=1/4$.
Field-theoretical calculations within the corresponding
Landau-Ginzburg-Wilson (LGW) theory \cite{CP-01,COPS-03} support the
existence of such a new universality class.

Previous results are presented in
Table~\ref{summaryliterature}.  We report results for the fully
frustrated XY model on the square lattice (FFXY$_{\rm sq}$) and on the
triangular lattice (FFXY$_{\rm tr}$), and also for some
generalizations: a model with next-to-nearest--neighbor interactions
(FFXY$_{nn+nnn}$), a nearest-neighbor model with Villain Hamiltonian,
and XY models in which frustration is induced by the competition of
nearest-neighbor and next-to-nearest--neighbor interactions
(FXY$_{J_1,J_2}$) or by a zig-zag pattern of ferromagnetic and
antiferromagnetic couplings (FXY$_{\rm zz}$).  We also include results
for some related models that have the same ground-state degeneracy:
the fractional-charge Coulomb-gas (CG) model that is obtained from the
FFXY model by the usual duality mapping, a model with two coupled XY
fields (2cXY) or with an Ising and XY field (IsXY), the
antiferromagnetic XXZ model with easy-plane exchange anisotropy on the
triangular lattice (FFXXZ$_{\rm tr}$), the frustrated
antiferromagnetic six-state clock model on a triangular lattice
(FA6SC), the 19-vertex model, the quantum ladder of Josephson
junctions (QLJJ), the solid-on-solid model coupled to the Ising model
(SOS-Is), and the corresponding LGW $\phi^4$ theory.  In 
Table~\ref{summaryliterature} we report the method that has been
employed to investigate the phase diagram, the number of transitions
that are observed (whenever there are two transitions, we also give
the difference of the critical temperatures) and the estimates of the
critical exponents at the chiral transition.  Most of the results have
been obtained by Monte Carlo (MC) simulations.  But we should also
mention transfer-matrix (TM) and real-space renormalization-group (RG)
calculations, as well as perturbative field-theoretical (FT) analyses. 
Note that most MC
studies focussed on the finite-size behavior at criticality; only in a
few cases \cite{FPA-91,RJ-92,JR-96,Olsson-97} was the temperature
dependence investigated in the large-volume limit.  In some cases the
critical behavior was investigated by studying the nonequilibrium
relaxation (NER) at criticality.

The results summarized in Table~\ref{summaryliterature} are rather
contradictory.  The most recent MC simulations favor the existence of
two transitions.  If this is correct, the results of the MC
simulations observing only one transition may be reasonably explained
by noting that the two transitions are very close and that high
accuracy is required to disentangle them. If there are two
transitions, chiral and spin modes decouple and thus one expects Ising
behavior at the chiral transition.  But this is not supported by
numerical simulations that have found $\nu\approx 0.8$ in most cases. To
explain this result, one may conclude that there is a single
transition that belongs to a new universality class.  In this case,
the observation of two very close transitions might be explained by
uncontrolled systematic errors in the analysis of the MC data.  Of
course, it is also possible, as discussed in
Refs.~\cite{Olsson-95,Olsson-97,NREK-02}, that the departure of the
chiral exponents from the Ising values is due to a slow crossover towards the
Ising asymptotic behavior, somehow caused by the interaction with the
spin modes, which will eventually give rise to a KT
transition. However, this argument does not explain why the same
estimate $\nu\approx 0.8$ is obtained in several different models.
Finally, there is a third, less conventional possibility: There are
two transitions, but the chiral transition does not belong to the
Ising universality class for some unknown reasons.

\begin{table}
\caption{\label{summaryliterature}
Overview of results for the FFXY and related models.
The shorthands used in the second and third column are
explained in the text.  The fourth column reports the
number of transitions observed, and, in the case of two transitions,
the relative difference between their temperatures, i.e.  $\delta\equiv (T_{\rm
ch}-T_{\rm sp})/T_{\rm ch}$.  The fifth column gives the estimates of
the critical exponents associated with the chiral degrees of freedom.
}
\hspace{-1cm}
\footnotesize
\renewcommand\arraystretch{0.6}
\begin{tabular}{@{}lllll}
\br
Ref.& model & method& transitions& chiral exponents\\
\mr
\protect\cite{TJ-83} (1983) & FFXY$_{\rm sq}$ & MC ($L\leq 32$) &  & consistent with Ising \\

\protect\cite{SS-84} (1984) & FFXY$_{\rm tr}$ & MC ($L\leq 32$) & 2tr & \\ 

\protect\cite{MS-84} (1984) & FFXY$_{\rm sq}$ & MC ($L\leq 45$) & 2tr, $\delta\approx 0.02$ & \\ 

\protect\cite{Halsey-85} (1985) & FFXY$_{\rm sq}$ & & $\delta\geq 0$ & \\

\protect\cite{YD-85} (1985) & FFXY$_{\rm sq}$, CG  & RG & 1tr & \\ 

\protect\cite{CS-85} (1985) & 2cXY & RG & 1tr & \\ 

\protect\cite{LJNL-86} (1986) & FFXY$_{\rm tr}$ & MC ($L\leq 72$) & 1-2tr, $\delta\lesssim 0.01$
& consistent with Ising \\

\protect\cite{Van-86} (1986) & FFXY$_{\rm tr}$ & MC ($L\leq 72$)  & 1tr &   \\

\protect\cite{BDGL-86} (1986) & FFXY$_{\rm sq}$ & MC ($L\leq 100$)  & 1tr &   \\

\protect\cite{GKP-86} (1986) & 2cXY  &  real-space RG  & 1tr &   \\

\protect\cite{TK-88} (1988) & CG   & MC ($L\leq 30$)  & 1tr &   \\

\protect\cite{Grest-89} (1989) & CG  & MC ($L\leq 50$)  & 2tr, $\delta\approx 0.03$ &   \\

\protect\cite{EHKT-89} (1989) & CG  & MC ($L\leq 48$)  & 1tr &   \\

\protect\cite{TK-90} (1990) & FFXY$_{\rm sq}$ & TM MC ($L\leq 12$) & 1tr &  
$\nu\approx 1$, $\eta=0.40(2)$ \\

\protect\cite{GKLN-91,LKG-91} (1991) & FFXY$_{\rm sq}$ & MC ($L\leq 40$) & 1tr &  
$\nu=0.85(3)$, $\eta=0.31(3)$ \\

\protect\cite{GKLN-91,LKG-91} (1991) & FFXY$_{\rm tr}$ & MC ($L\leq 40$) & 1tr &  
$\nu=0.83(4)$, $\eta=0.28(4)$ \\

\protect\cite{Nicolaides-91} (1991) & FFXY$_{\rm sq}$ & MC ($L\leq 128$) & 1-2tr &  
$\nu=1.009(26)$ \\

\protect\cite{LGK-91} (1991) & IsXY ($C=-0.2885$) & MC ($L\leq 32$) & 1-2tr &  
$\nu=0.84(3)$ \\

\protect\cite{FPA-91} (1991) &  FXY$_{J_1,J_2}$ & MC ($L\leq 150$) & 1tr & 
$\nu=0.9(2)$, $\eta=0.4(1)$  \\

\protect\cite{dN-92} (1992) & 19-vertex-Is & TM ($L\leq 7$) & 1tr & 
$\nu=1.0(1)$, $\eta=0.26(1)$ \\

\protect\cite{RJ-92} (1992) & FFXY$_{\rm sq}$ & MC ($L\leq 240$) & 1-2tr, $\delta\gtrsim -0.07$ & 
$\nu=0.875(35)$ \\

\protect\cite{Granato-92,GKN-96} (1992) & QLJJ ($E_x/E_y=1$) & TM QMC &
 & $\nu=0.81(4)$, $\eta=0.47(4)$ \\ 

\protect\cite{Granato-92,GKN-96} (1992) & QLJJ ($E_x/E_y=3$) & TM QMC &
 & $\nu=1.05(6)$, $\eta=0.27(3)$ \\ 

\protect\cite{GN-93} (1993) & FFXY$_{\rm sq}$ & TM MC ($L\leq 14$) & 1tr & 
$\nu=0.80(5)$, $\eta=0.38(2)$ \\

\protect\cite{Lee-94} (1994) & CG  & MC ($L\leq 30$)  & 2tr, $\delta\approx 0.04$ &  
$\nu=0.84(3)$, $\eta=0.26(4)$ \\

\protect\cite{LL-94} (1994) & FFXY$_{\rm sq}$  & MC ($L\leq 48$)  & 2tr, $\delta\approx 0.03$ &  
$\nu=0.813(5)$, $\beta=0.089(8)$ \\

\protect\cite{KNKB-94} (1994) & 19-vertex model & TM  ($L\leq 15$) & 1tr & 
$\nu=0.81(3)$, $\eta=0.28(2)$ \\

\protect\cite{Olsson-95} (1995)  & FFXY$_{\rm sq}$ &
MC ($L\leq 128$) & 2tr, $\delta\approx 0.013$
 & consistent with Ising \\

\protect\cite{NGK-95,GKN-96} (1995) & IsXY ($C=-0.2885$) & TM MC ($L\leq 30$) & 1tr & 
$\nu=0.79$, $\eta=0.40$ \\

\protect\cite{JR-96} (1996) & FFXY$_{\rm sq}$ & MC ($L\leq 128$) &  & $\nu=0.898(3)$  \\

\protect\cite{XS-96} (1996) & FFXY$_{\rm tr}$ & MC ($L\leq 144$) & 2tr, $\delta\approx 0.02$ & 
$\gamma=1.6(3)$, $\beta=0.11(3)$ \\

\protect\cite{Olsson-97} (1997)  & Villain FFXY$_{\rm sq}$ &
MC ($L\leq 256$) & 2tr, $\delta\approx 0.014$ & consistent with Ising \\

\protect\cite{BZG-97} (1997) & Villain FFXY$_{\rm sq}$ &  spin waves, LT phase & 1tr & \\

\protect\cite{BG-97} (1997) & FXY$_{\rm zz}$ ($\rho=0.7$) & 
MC ($L\leq 36$) & 1tr &  $\nu=0.78(2)$, $\eta= 0.32(4)$ \\

\protect\cite{BG-97} (1997) & FXY$_{\rm zz}$ ($\rho=1.5$) & 
MC ($L\leq 36$) & 1tr &  $\nu=0.80(1)$, $\eta= 0.29(2)$ \\

\protect\cite{JPC-97} (1997)  & FFXY, 2cXY, CG & position-space RG & 2tr,
$\delta\approx 0.0005$ & different from Ising \\

\protect\cite{LLK-97} (1997) & SOS-Is & MC ($L\leq 22$) & 2tr & consistent with Ising \\

\protect\cite{Simon-97} (1997) & FXY$_{J_1,J_2}$, CG & RG & 1tr &  \\ 

\protect\cite{LSZ-98} (1998) & FFXY$_{\rm sq}$ & NER$_{\rm std}$ MC ($L\leq 256$) & & 
$\nu=0.81(2)$, $\eta=0.261(5)$ \\ 

\protect\cite{LL-98} (1998) & FFXY$_{\rm tr}$ & MC ($L\leq 60$) & 2tr, $\delta\approx 0.012$ &  
$\nu=0.833(7)$, $\eta=0.25(2)$ \\

\protect\cite{CVCT-98} (1998) & FFXXZ$_{\rm tr}$ & MC ($L\leq 120$) & 2tr, $\delta\lesssim 0.01$ & 
consistent with Ising \\

\protect\cite{BD-98} (1998) & FFXY$_{\rm sq}$ & MC ($L\leq 140$) & 1tr & 
$\nu=0.852(2)$, $\eta= 0.203(6)$ \\

\protect\cite{LS-00} (2000) &  FXY$_{J_1,J_2}$ & MC ($L\leq 150$) & 2tr, $\delta\approx 0.003$ & 
$\nu=0.795(20)$, $\eta=0.25(1)$  \\ 

\protect\cite{FCKF-00} (2000) & FXY$_{nn+nnn}$ &  MC ($L\leq 72$) & 2tr, $\delta\lesssim 0.01$ & 
$\nu=1.0(1)$ \\ 

\protect\cite{CLJ-01} (2001) & FFXY$_{\rm sq}$ &  NER$_{\rm std}$ MC ($L\leq 256$) & & 
$\nu=0.80(2)$, $\eta=0.276(7)$  \\ 

\protect\cite{CLJ-01} (2001) & FXY$_{nn+nnn}$ & NER$_{\rm std}$ MC ($L\leq 256$) & & 
$\nu=0.80(3)$, $\eta=0.282(8)$  \\ 

\protect\cite{NREK-02} (2002) & FA6SC & MC ($L\leq 192$)  
& 2tr, $\delta\approx 0.003$ & consistent with Ising \\

\protect\cite{Korshunov-02} (2002) & FFXY & & 2tr, $\delta> 0$ & \\

\protect\cite{COPS-03,CP-01} (2003) & LGW $\phi^4$ & five-loop FT & 
stable FP & \\

\protect\cite{OI-03} (2003) & FFXY$_{\rm sq}$ & NER MC ($L\leq 2000$) & 2tr, 
$\delta\approx 0.010$ &  $\nu=0.82(2)$, $\eta=0.272(15)$ \\ 

\protect\cite{OI-03} (2003) & FFXY$_{\rm tr}$ & NER MC ($L\leq 2000$) & 2tr, 
$\delta\approx 0.008$ &  $\nu=0.84(2)$, $\eta=0.250(10)$ \\ 

\protect\cite{GD-05} (2005) & FFXY$_{\rm sq}$ & ED MC ($L\leq 180$) & 1tr & 
$\nu=0.9(1)$ \\

\protect\cite{OT-05} (2005) & FFXY$_{\rm sq}$ & MC ($L\leq 128$) & 2tr, $\delta> 0$ &  \\

this work & FFXY$_{\rm sq}$ & MC ($L\leq 1000$) & 2tr, $\delta=0.0159(2)$ & Ising \\

this work & IsXY ($C=0$) & MC ($L\leq 360$) & 2tr, $\delta=0.0167(7)$ & Ising \\

this work & $\phi^4$ ($U=1,D=1/2$) & 
MC ($L\leq 1200$) & 2tr, $\delta= 0.0025(2)$ & Ising \\
\br
\end{tabular}
\end{table}

In this paper we study the phase diagram of the 2-$d$ square-lattice
FFXY and of two related models, a coupled Ising-XY (IsXY) model and a
$\phi^4$ model obtained from a straighforward lattice discretization
of the LGW Hamiltonian for the critical modes of the FFXY model, see,
e.g., Refs.~\cite{LGK-91,YD-85}.  We present MC simulations on square
lattices $L\times L$, up to $L=O(10^3)$.  The phase diagrams and
critical behaviors are obtained by means of a FSS analysis of the MC
results. Short reports already appeared in Ref.~\cite{sr}.  In the LT
phase the ${\mathbb Z}_2$ chiral symmetry is broken and the spin
degrees of freedom show the same quasi-long-range order as in the
2-$d$ XY model. We conclusively show that the square-lattice FFXY
model undergoes two very close but separate transitions: a KT and
then, as temperature increases, an Ising transition with $\delta\equiv
(T_{\rm ch}-T_{\rm sp})/T_{\rm ch}=0.0159(2)$.  The same transition
pattern is observed in the $\phi^4$ and IsXY models when the
transitions are continuous.

Beside confirming the two-transition scenario, we have also 
observed an unexpected crossover behavior that is universal to some 
extent. 
In the FFXY model and in the $\phi^4$
and IsXY models in a large parameter region, the finite-size
behavior at the chiral and spin transitions is model independent, apart from
a length rescaling.  
In particular, the universal approach to the Ising regime at the
chiral transition is nonmonotonic for most observables, and there is a
wide region in which the finite-size behavior is controlled by an
effective exponent $\nu_{\rm eff} \approx 0.8$. This occurs for $L\lesssim
\xi_s^{(c)}$, where $\xi_s^{(c)}$ is the spin correlation length at the
chiral transition, which is usually large in these models; for example,
$\xi_s^{(c)}=118(1)$ in the square-lattice FFXY model.  This explains why 
many previous studies that considered smaller lattices always found 
$\nu\approx 0.8$.  This universal behavior may be explained by the 
presence of a multicritical point, where chiral and spin modes are both
critical. As far as its nature is concerned, our numerical data at the 
chiral transition suggest a zero-temperature multicritical point.
In this case, a possible
candidate is the O(4) multicritical point that is present in the $\phi^4$ lattice
model and controls the low-temperature phase of the 4-vector model.

This paper is organized as follows.  In Sec.~\ref{models} we define
the models investigated in this paper.  
The thermodynamic quantities considered in this study are introduced in 
Sec.~\ref{Defs}. A  brief presentation of our MC simulations
is given in Sec.~\ref{MCsim}.
Sec.~\ref{lowtsec} is dedicated to the study of the LT phase.  In
Sec.~\ref{phasetrffxy} we discuss the critical behavior of the
square-lattice FFXY model. In Secs.~\ref{phasetr} and \ref{isingxy} we
investigate the phase diagram of the $\phi^4$ model and of the IsXY
model respectively. In Sec.~\ref{crossover} we discuss the crossover
behavior at the chiral transition.  Finally, in Sec.~\ref{conclusions}
we summarize the main results of the paper and draw our conclusions.
In \ref{mcalg} we provide some details on the algorithms used in the
MC simulations.  In \ref{elmodeta} we report some results for the LT
phase of the 2-$d$ XY model which are used in the paper.

\section{Models}
\label{models}

The Hamiltonian of the square-lattice FFXY model is 
\begin{equation}
{\cal H}_{\rm FFXY} = - J 
\sum_{\langle xy\rangle} j_{xy} \, \vec{s}_x \cdot \vec{s}_y ,
\label{ffxyim}
\end{equation}
where the two-component spins $\vec{s}_x$ satisfy $\vec{s}_x\cdot \vec{s}_x=1$,
$j_{xy}=1$ along all horizontal lines, while along vertical lines 
ferromagnetic $j_{xy}=1$ and antiferromagnetic $j_{xy}=-1$ couplings
alternate. Here (as in the following models) $J$ plays the role 
of inverse temperature, the Gibbs probability being simply
proportional to $e^{-{\cal H}_{\rm FFXY}}$.

We also consider the $\phi^4$ model on a square lattice. The
Hamiltonian is
\begin{eqnarray}
{\cal H}_{\phi} &=& - 
J \sum_{\langle xy\rangle,i} \vec{\phi}_{i,x}\cdot \vec{\phi}_{i,y} 
+ \sum_{i,x} \left[ \phi_{i,x}^2 + U (\phi_{i,x}^2-1)^2 \right]  
\nonumber \\ &&+ 
2 (U+D) \sum_x \phi_{1,x}^2 \phi_{2,x}^2,
\label{HLi} 
\end{eqnarray}
where $i=1,2$, $\vec{\phi}_{i,x}$ is a real two-component variable,
the first sum goes over
all nearest-neighbor pairs, and $\phi^2_i\equiv \vec{\phi}_i \cdot
\vec{\phi}_i$.  Hamiltonian ${\cal H}_{\phi}$ describes two
identical O(2)-symmetric models coupled by an energy-energy term. 
This model is a straightforward lattice discretization of the LGW
Hamiltonian
\begin{equation}
{\cal H}_{\rm LGW}  = \int d^d x
 \Bigl\{ {1\over2}
      \sum_{a=1,2} \Bigl[ (\partial_\mu \phi_{a})^2 + r \phi_{a}^2 \Bigr]
+ {1\over 4!}u_0 \Bigl( \sum_{a=1,2} \phi_a^2\Bigr)^2
+ {1\over 4} v_0 \phi_1^2 \phi_2^2 \Bigr\}
\label{LGWH}
\end{equation}
($\phi_a$, $a=1,2$, is a two-component vector), which can be obtained by
applying a Hubbard-Stratonovich transformation to the FFXY model,
dropping terms of order higher than four
\cite{LGK-91,CD-85,YD-85}.  Therefore, the lattice $\phi^4$ model
(\ref{HLi})  represents an effective ferromagnetic theory
that is expected to describe the critical modes of the FFXY model.
This approach was already used to investigate the critical behavior of
antiferromagnets on a stacked triangular lattice \cite{CPPV-04}.

It is interesting to note that the symmetry of Hamiltonian (\ref{HLi})
is larger than that of the FFXY model. Indeed, the Hamiltonian is
symmetric under separate rotations of the two fields and under the
${\mathbb Z}_2$ transformation $\phi_{1} \leftrightarrow \phi_{2}$.
Therefore, the overall symmetry is [O(2)$\oplus$O(2)]$\otimes {\mathbb
Z}_2$.  For $D>0$ the ground state corresponds to $\phi_1^2 = 0$ and
$\phi_2^2 \not= 0$ or the opposite, and thus it has the same
degeneracy as the ground state of the FFXY model. Indeed, the
lowest-energy configurations are determined once one fixes which field
does not vanish---thereby breaking the ${\mathbb Z}_2$ interchange
symmetry---and the direction of the nonvanishing one. Thus, the
relation between the FFXY model and the $\phi^4$ Hamiltonian is not
fixed by the symmetry of the original Hamiltonian but rather by the
ground-state degeneracy group, which is the quotient of the symmetry
groups of the model and of the ground state. This is consistent with
the paradigm that relates the universality class to the symmetry
breaking pattern. Note that the extended symmetry of the
LGW Hamiltonian is related to the truncation of the
Hamiltonian to fourth order in the fields and it is lost if
higher-order terms are added. More precisely, the Hubbard-Stratonovich
calculation gives rise to $\phi^6$ and $\phi^8$ terms on the
triangular and square lattice respectively, which break the 
O(2)$\oplus$O(2) symmetry down to O(2) \cite{YD-85}.  These terms are
irrelevant close to four dimensions but it is far from clear that the
same holds in two dimensions. In any case, they do not change the
relevant symmetry breaking pattern.

Due to the ground-state structure, for $J\to \infty$ the
field-interchange symmetry $\phi_{1} \leftrightarrow \phi_{2}$ is
broken and thus this symmetry is the analog of the ${\mathbb Z}_2$
chiral symmetry of the FFXY model.  The corresponding order parameter
is
\begin{equation}
C_x = \phi_{1,x}^2 - \phi_{2,x}^2. 
\label{cxdef}
\end{equation}
For $D=0$ model (\ref{HLi}) is O(4) symmetric. Therefore, it
does not have any transition at finite temperature. Criticality is
observed only for $J\rightarrow \infty$. In this limit the correlation length
$\xi$ increases exponentially, i.e. $\xi\sim e^{c J}$; see, e.g.,
Ref.~\cite{PV-review}.

We also consider the Ising-XY (IsXY) model \cite{GKLN-91}
\begin{equation}
{\cal H}_{\rm IsXY} = 
- \sum_{\langle xy\rangle} 
\left[ \frac{J}{2} (1+\sigma_x \sigma_y) \, \vec{s}_x \cdot \vec{s}_y 
+ C \sigma_x \sigma_y \right] ,
\label{IsXYp}
\end{equation}
where $\sigma_x=\pm 1$ and the two-component spins $\vec{s}_x$ satisfy
$\vec{s}_x\cdot \vec{s}_x=1$.  Here $\vec{s}_x$ and $\sigma_x$
correspond to spin and chiral variables, respectively.  Note that, by
performing the limit $U\to \infty$ and then $D\to \infty$ in the
$\phi^4$ model (\ref{HLi}), one recovers the IsXY model for $C=0$. In
this case, the variables $\sigma_x$ and $\vec{s}_x$ are related with
those of the $\phi^4$ model by
\begin{equation}
\vec{\phi}_{1,x} = \case{1}{2} (1 + \sigma_x) \vec{s}_x, \qquad\qquad
\vec{\phi}_{2,x} = \case{1}{2} (1 - \sigma_x) \vec{s}_x. \qquad\qquad
\label{mapping}
\end{equation}
Models with $C\not=0$ can also be recovered from a $\phi^4$
Hamiltonian. It is enough to add an energy-energy nearest-neighbor
hopping term $\sum_{\langle xy\rangle} (\phi^2_{1,x} \phi^2_{1,y} +
\phi^2_{2,x} \phi^2_{2,y})$.

Apparently, the IsXY model is only invariant under the group
O(2)$\otimes{\mathbb Z}_2$. However, its relation with the $\phi^4$
model indicates that the symmetry is larger. Indeed, for any
value of $C$, the model is invariant under the O(2)$\oplus$O(2)
nonlinear transformations
\begin{eqnarray}
{\vec{s}_x}\!' &=& [\case{1}{2} (1 + \sigma_x) R^{(1)} + 
              \case{1}{2} (1 - \sigma_x) R^{(2)} ] \vec{s}_x,
\nonumber \\
\sigma'_x &=& \sigma_x,
\end{eqnarray}
where $R^{(1)}$ and $R^{(2)}$ are O(2) rotation matrices.

It is possible to argue that in the IsXY model chiral and spin
modes cannot  be critical at the same value of $J$ and $C$,
see, e.g., Ref.~\cite{Korshunov-02}.  Let us assume that, for fixed $C$,
the model
undergoes a continuous phase transition at $J=J_{\rm ch}$, where the
chiral correlation length is infinite, i.e. where $\sigma$
correlations are critical.  Let us consider first the dynamics of
the spins $\vec{s}_x$ for a fixed Ising $\{ \sigma_x \}$
configuration.  Spins $\sigma$ with the same sign form geometrical
clusters and spins $\vec{s}_x$ belonging to
different geometrical (not Fortuin-Kasteleyn) 
clusters do not interact (either directly or indirectly)
because of the prefactor $(1 + \sigma_x \sigma_y)$ in the hopping term
of the spin variables. Thus, the behavior of the spins $\vec{s}_x$ is
completely determined by their behavior on each single cluster.  It is
easy to convince oneself that $J = J_{\rm ch}$ is the percolation
threshold of the clusters.  Indeed, if the clusters percolate at $J^*
< J_{\rm ch}$, for $J = J^*$ Ising spins would already be critical,
since in two dimensions percolation of the geometrical clusters
implies Ising criticality. Now, at the percolation threshold we expect
the clusters to have a fractal dimension that is smaller than 2.  This
is true for the Ising transition \cite{SV-89,JS-05}; we assume here that it
holds in general.  Thus, fields $\vec{s}_x$ interact as on a system
with $d < 2$ and thus cannot be critical. Therefore, at the chiral
transition the spin correlation length must be finite, and, moreover,
$J_{\rm ch}<J_{\rm sp}$.  As we shall see, the MC results that we
shall present provide support to this argument.

\section{Definitions and notations}
\label{Defs}

\subsection{The fully frustrated XY model}
\label{ffxydef}

In the square-lattice FFXY model the ground state is only invariant
under translations of two lattice spacings. Therefore, we divide the
lattice into four sublattices, so that the four sites of each
plaquette belong to different sublattices. Then, we define the spin
correlation function $G_s(x) \equiv \langle \vec{s}_0\cdot \vec{s}_x
\rangle$ only for $x = (2n,2m)$, $n,m$ integers, and its Fourier
transform as $\widetilde{G}_s(q) = \sum_x e^{iq\cdot x} G_s(x)$ where
the sum goes over $x \equiv (2n,2m)$.  The corresponding
susceptibility $\chi_s$ and second-moment correlation length $\xi_s$
are given by
\begin{eqnarray}
&&\chi_s \equiv  \sum_{x = (2n,2m)} G_s(x), \label{chidefffxy}\\
&&\xi_s^2 \equiv  {1\over 4 \sin^2 (q_{\rm min}/2)} 
{\widetilde{G}_s(0) - \widetilde{G}_s(q)\over \widetilde{G}_s(q)},
\label{xidefffxy}
\end{eqnarray}
where $q = (q_{\rm min},0)$, and $q_{\rm min} \equiv 2 \pi/(L/2)$.
We shall consider two RG invariant ratios related to the continuous
spin modes:
\begin{eqnarray}
R_s\equiv {\xi_s\over L},
\qquad\qquad
B_s   \equiv  { \langle ( \mu^2 )^2 \rangle\over
\langle \mu^2 \rangle^2 }, 
\end{eqnarray}
where $\vec{\mu}$ is the magnetization corresponding to 
one of the four sublattices,
i.e. 
\begin{equation}
\vec{\mu}\equiv {4\over V} \sum_{x = (2n,2m)} \vec{s}_{x},
\end{equation}
$V\equiv L^2$ is the volume,
and $\mu^2 \equiv \vec{\mu} \cdot  \vec{\mu}$. 
We also define the helicity modulus $\Upsilon$. 
For this purpose we introduce a twisted term in the Hamiltonian. 
More precisely, we consider the
nearest-neighbor sites $(x, y)$ with $x_1=L$,
$y_1=1$, and $x_2=y_2$, and replace the term $\vec{s}_{x}\cdot
\vec{s}_{y}$ in Hamiltonian (\ref{ffxyim}) with
\begin{equation}
 \vec{s}_{x} \cdot R_{\varphi} \vec{s}_{y} =
 s_{x}^{(1)} \left(s_{y}^{(1)} \cos\varphi  
                         + s_{y}^{(2)} \sin\varphi  \right)+ 
 s_{x}^{(2)} \left(s_{y}^{(2)} \cos\varphi 
                         - s_{x}^{(1)} \sin\varphi \right),
\label{ydef1}
\end{equation}
where $R_{\varphi}$ is a rotation of an angle $\varphi$.  The helicity
modulus is defined as the second derivative of the free energy with
respect to $\varphi$ at $\varphi=0$:
\begin{equation}
\label{ups_infffxy}
\Upsilon \equiv  -  \left .
\frac{\partial^2 \ln Z(\varphi)}{\partial \varphi^2} \right |_{\varphi=0} .
\end{equation}
Chiral modes are related to the ${\mathbb Z}_2$ symmetry that is
broken by the ground state. As discussed in the introduction, a good
order parameter is (we have dropped the sign here and thus $C_n$ is no
longer a ``spin" susceptibility; this is of course irrelevant)
\begin{equation}
C_n \equiv  \sum_{\langle xy \rangle \in \Pi_n} 
j_{xy} {\rm sin} (\theta_x - \theta_y),
\label{chirality}
\end{equation}
where $n$ is the dual lattice site at the center of $\Pi_n$.  
We consider the staggered correlation function
\begin{equation}
G_c(n) \equiv  (-1)^{n_1+n_2} \langle  C_0 \,  C_n \rangle_c,
\label{gcdefffxy}
\end{equation}
the chiral susceptibility $\chi_c$, and the second-moment correlation
length $\xi_c$. They are defined as in Eqs.~(\ref{chidefffxy})
and (\ref{xidefffxy}),
though in this case sums are extended over the whole dual lattice and
$q_{\rm min} \equiv 2 \pi/L$.  Analogously, we define the RG invariant
quantities
\begin{eqnarray}
R_c\equiv {\xi_c\over L}, 
\qquad
B_c \equiv  { \langle \mu_c^4 \rangle\over \langle \mu_c^2 \rangle^2 },
\end{eqnarray}
where the staggered magnetization is defined as
\begin{equation}
\mu_c\equiv {1\over V} \sum_{n=(n_1,n_2)} (-1)^{n_1+n_2} C_n\; .
\label{mcdef}
\end{equation}

\subsection{The $\phi^4$ model}
\label{phi4def}

In the $\phi^4$ model we define the hopping energy density and specific heat as
\begin{equation}
E \equiv {1\over V} \langle {\cal H}_{h}  \rangle,
\qquad 
C \equiv {1\over V} \left( \langle {\cal H}_{h}^2 \rangle -
\langle {\cal H}_h\rangle^2 \right),
\label{eedef}
\end{equation}
where
\begin{equation}
{\cal H}_h = 
\sum_{\langle xy\rangle} \vec{\phi}_{1,x}\cdot \vec{\phi}_{1,y} + 
            \vec{\phi}_{2,x}\cdot \vec{\phi}_{2,y}.
\label{hoppt}
\end{equation}
The spin two-point correlation function $G_s(x)$ is defined as
\begin{equation}
G_s(x) \equiv  \langle \vec{\phi}_{1,0} \cdot \vec{\phi}_{1,x}\;+\;
\vec{\phi}_{2,0} \cdot \vec{\phi}_{2,x} \rangle.
\end{equation}
The corresponding susceptibility $\chi_s$ and second-moment
correlation length $\xi_s$ are given by
\begin{equation}
\chi_s \equiv  \sum_x G_s(x), \qquad \qquad 
\xi_s^2 \equiv  {1\over 4 \sin^2 (q_{\rm min}/2)} 
{\widetilde{G}_s(0) - \widetilde{G}_s(q)\over \widetilde{G}_s(q)},
\label{chixidef}
\end{equation}
where $\widetilde{G}_s(q)$ is the Fourier transform of $G_s(x)$, $q =
(q_{\rm min},0)$, and $q_{\rm min} \equiv 2 \pi/L$.  We shall
consider two RG invariant ratios related to the continuous spin modes:
\begin{eqnarray}
R_s\equiv {\xi_s\over L},
\qquad
B_{s\phi}   \equiv  { \langle ( \sum_i \mu_i^2 )^2 \rangle\over
\langle \sum_i \mu_i^2 \rangle^2 }, 
\label{bsphi}
\end{eqnarray}
where 
\begin{equation}
\vec{\mu}_i \equiv  {1\over V} \sum_x \vec{\phi}_{i,x},
\qquad \mu_i^2 \equiv \vec{\mu}_i \cdot  \vec{\mu}_i. 
\end{equation}
As before, we consider the helicity modulus $\Upsilon_i$ associated
with each variable $\vec{\phi}_{i,x} = (\phi_{i,x}^{(1)},
\phi_{i,x}^{(2)})$, defined analogously to Eqs.~(\ref{ydef1}) and
(\ref{ups_infffxy}).  The total helicity modulus is given by $\Upsilon
\equiv \Upsilon_1 + \Upsilon_2$.

The ``chiral'' modes related to the ${\mathbb Z}_2$ field-interchange
symmetry are associated with the quadratic operator defined in
Eq.~(\ref{cxdef}).  We consider the corresponding connected two-point
correlation function
\begin{equation}
G_c(x) \equiv  \langle C_0 \; C_x \rangle_c,
\label{gcdef}
\end{equation}
the chiral susceptibility $\chi_c$, the second-moment correlation
length $\xi_c$ defined as in Eq.~(\ref{chixidef}), and the ratios
\begin{eqnarray}
R_c\equiv {\xi_c\over L}, 
\qquad
B_c \equiv  { \langle \mu_c^4 \rangle\over \langle \mu_c^2 \rangle^2 },
\qquad \mu_c \equiv  {1\over V} \sum_x C_x.
\end{eqnarray}

\subsection{The Ising-XY model}
\label{isxydef}

In the case of the IsXY model (\ref{IsXYp}) we consider the correlation
functions $G_s(x) \equiv
\langle \vec{s}_0\cdot \vec{s}_x \rangle$ and $G_c(x) \equiv \langle
\sigma_0\cdot \sigma_x \rangle$ and the corresponding observables that
are defined as in the $\phi^4$ model.  For $C = 0$ it is easy to
verify that $G_s(x)$ and $G_c(x)$ exactly correspond to the
correlation functions in the $\phi^4$ model.  Indeed, using mapping
(\ref{mapping}), $C_x = \sigma_x$, so that chiral correlations
correspond to correlations of the Ising variables. On the other hand,
mapping (\ref{mapping}) gives
\begin{equation}
G_s(x) = \case{1}{2} \langle (1 + \sigma_x \sigma_y) 
     \vec{s}_x \cdot \vec{s}_y\rangle.
\end{equation}
However, since Hamiltonian (\ref{IsXYp}) is invariant under the
transformations $\vec{s}_x \to \sigma_x \vec{s}_x$, we have
\begin{equation}
\langle \sigma_x \sigma_y \vec{s}_x \cdot \vec{s}_y\rangle = 
 \langle  \vec{s}_x \cdot \vec{s}_y\rangle.
\end{equation}
Therefore, 
\begin{equation}
G_s(x) = \langle  \vec{s}_x \cdot \vec{s}_y\rangle. 
\end{equation}
Note that the Binder parameter $B_{s\phi}$, defined as in
Eq.~(\ref{bsphi}) using the mapping (\ref{mapping}), is not the
natural one in terms of the spin variables $\vec{s}_x$. Indeed,
mapping (\ref{mapping}) gives
\begin{equation}
B_{s\phi} = {1\over4} {\langle (\mu_s^2)^2 +  2 \mu_s^2 \mu_{\sigma s}^2 
          + (\mu_{\sigma s}^2)^2 \rangle \over 
    \langle \mu_s^2 \rangle^2},
\end{equation}
where 
\begin{equation}
\vec{\mu}_s = {1\over V} \sum_x \vec{s}_x ,\qquad\qquad
\vec{\mu}_{\sigma s} = {1\over V} \sum_x \sigma_x\vec{s}_x .
\end{equation}
The conventional Binder parameter is instead\footnote{
In terms of the original $\phi$ variables, we have 
$B_s = B_{s\phi} + 2 \langle \mu_1^2 \mu_2^2\rangle /
   (\sum_i \langle \mu_i^2\rangle)^2$. 
Note that $B_s > B_{s\phi}$.}
\begin{eqnarray}
{B}_s = \langle (\mu_s^2)^2\rangle/ \langle \mu_s^2 \rangle^2.
\end{eqnarray}
In the HT phase we have $B_{s\phi} = 3/2$ and ${B}_s = 2$
in the thermodynamic limit.

\section{Monte Carlo simulations}
\label{MCsim}

We perfomed MC simulations of the FFXY model (\ref{ffxy}), of the
$\phi^4$ model (\ref{HLi}) for $U=1$ and $D$ in the range $1/20\le D
\le 99$, and of the IsXY model for $-8\le C \le 0.3$.  We considered
$L\times L$ square lattices with periodic boundary conditions, up to
$L=O(10^3)$.  The phase diagram and critical behavior is
investigated by analyzing the FSS behavior of several quantities.  The
total CPU time used in this study was approximately 10 CPU years of a
single 64-bit Opteron 246 (2Ghz) processor.

In the MC simulations of the FFXY model, we used a local algorithm
based on a mixture of Metropolis and overrelaxed (microcanonical)
updates, as suggested in Ref.~\cite{PP-98}.  In the case of the IsXY and
$\phi^4$ model
the updating algorithm was based on mixtures of Metropolis,
overrelaxed (microcanonical), and single-cluster \cite{Wolff-89}
updates.  More details on the MC algorithms are reported in \ref{mcalg}.  
At the chiral transition, single-cluster updates have little influence
on the autocorrelation times, while the overrelaxation updates are only
able to speed up the simulation but do not change the dynamic critical
exponent. This is always approximately equal to 2, as it is expected in the 
case of a purely local dynamics. This means that very long simulations
are needed to obtain  reliable statistics. For instance, in order 
to obtain approximately 2500 independent configurations for 
the FFXY model on a lattice with $L=1000$, we used approximately 
800 days of a single 64-bit Opteron 246 processor. Analogously,
the runs on the largest lattices for the IsXY model 
($C=0$, $L=360$, 4000 independent configurations) and for the 
$\phi^4$ model ($D=1/2$, $L=1200$, 150 independent configurations)
took approximately 100 days in both cases.
A substantial reduction of the critical slowing down is
achieved at the spin transition and in the LT phase, 
because the overrelaxed and
the cluster algorithm are very effective in dealing with the 
spin modes.

\section{The low-temperature phase}
\label{lowtsec}

In the high-temperature (HT) phase, in which symmetry is not broken, the
spin and chiral correlation functions $G_s(x)$ and $G_c(x)$ decay
exponentially at large distances. In the LT phase
($J$ large) instead,
the spin correlation function $G_s(x)$ is expected to
decay with a power law, giving rise to quasi-long-range order.
Indeed, according to the Mermin-Wagner theorem \cite{MW-66}, any
magnetization breaking a continuous symmetry is forbidden in two
dimensions.  On the other hand, the discrete ${\mathbb Z}_2$ symmetry
may be broken with a nonvanishing chiral magnetization.  In this
section we study the main features of the LT phase.  We shall see that
${\mathbb Z}_2$ symmetry is broken and that O(2) quasi-long-range
order is realized in the whole LT phase. The critical behavior of the
spin variables is controlled by a line of Gaussian fixed points that are
exactly those that control the LT phase of the standard XY model.

\begin{table}
\caption{\label{lowtl} 
Results for $B_c$, $R_s$, $\Upsilon$, $\chi_s$, and $\eta_{\rm eff}$ in the
LT phase. In the case of the $\phi^4$ model we report $\xi_{\rm la}/L$,
$\Upsilon_{\rm la}$, and $\chi_{\rm la}$ that converge to $R_s$,
$\Upsilon$, and $\chi_s$ as $L\to\infty$.  The exponent $\eta_{\rm
eff}$ corresponds to $\eta(L,2L)$ defined in
Eq.~(\protect\ref{etal1l2}).  
}
\footnotesize
\begin{tabular}{@{}llrlllll}
\hline
\multicolumn{1}{c}{$$}&
\multicolumn{1}{c}{$J$}&
\multicolumn{1}{c}{$L$}&
\multicolumn{1}{c}{$B_c$}&
\multicolumn{1}{c}{$R_s$}&
\multicolumn{1}{c}{$\Upsilon$}&
\multicolumn{1}{c}{$\chi_s$}&
\multicolumn{1}{c}{$\eta_{\rm eff}$}\\
\hline
FFXY & 2.4 &  64 & 1.00232(2)  & 0.9886(12) & 1.0800(10) & 508.9(2) &  0.1495(6) \\
     &     & 128 & 1.000598(3) & 0.9992(12) & 1.0749(9)  & 1835.2(5)&  0.1480(5) \\
     &     & 256 & 1.0001503(8)& 1.0020(13) & 1.0771(8)  & 6625(2)  &   \\
FFXY & 2.3 & 64  & 1.00712(9)  & 0.8945(12)  & 0.9123(13) & 445.8(2) &  0.1817(8) \\
     &     & 128 & 1.001882(12)& 0.9048(9)   & 0.9066(14) & 1572.0(5) &  0.1775(7) \\
     &     & 256 & 1.000478(3) & 0.9104(12)  & 0.9029(12) & 5562(2) &  0.1760(7)  \\
     &     & 512 & 1.0001207(6)& 0.9133(12)  & 0.9059(13) & 19686(6)&  0.1746(7) \\
     &     & 1024& 1.0000302(3)& 0.9149(14)  & 0.9052(13) & 69766(27)&   \\
FFXY & 2.26 &  128 & 1.00447(5) & 0.8374(10) & 0.7985(16) &1399.4(7)& 0.2053(9) \\
     &      &  256 & 1.001165(8) & 0.8431(9)  & 0.7926(16) &4855(2) & 0.2038(7) \\
     &      &  512 & 1.000291(2) & 0.8447(10) &0.7904(16) &16862(6) &0.2031(9) \\
     &      & 1024 & 1.0000739(3)& 0.8461(15) &0.7888(14) &58594(30)& \\

$\phi^4$, $D=1/2$ & 1.50 & 32 & 1.0435(5) &  1.0236(13) & 1.2079(14) & 730.2(6) & 0.1431(14)\\
& & 64  &  1.01275(14)  & 1.0420(14) & 1.1923(12) & 2645.0(1.4) & 0.1389(10) \\
& & 128 &  1.00342(3)   & 1.0473(17) & 1.1867(10) & 9609(4)     & 0.1353(8) \\
& & 256 &  1.000873(8)  & 1.0528(17) & 1.1878(10) & 34996(12)   &  0.1339(7) \\
& & 512 &  1.000216(2)  & 1.0521(16) & 1.1850(10) & 127579(39)  & 0.1341(6) \\
& & 1024&  1.0000548(4) & 1.0536(16) & 1.1855(10) & 465026(150) & \\

$\phi^4$, $D=1/2$ & 1.48 & 32 & 1.0922(13) & 0.9194(15) & 1.0232(19) &    634.1(9) & 0.195(3)\\
& &   64 & 1.0382(9)    & 0.9207(15) & 0.9810(19) &   2215(3) & 0.180(2) \\ 
& &  128 & 1.0123(3)    & 0.9275(14) & 0.9635(16) &   7824(7) & 0.1718(15) \\
& &  256 & 1.0034(2)    & 0.9337(15) & 0.9570(13) &  27782(17) & 0.1682(11) \\
& &  512 & 1.000832(10) & 0.9341(15) & 0.9539(12) &  98897(46) & 0.1669(11) \\
& & 1024 & 1.000216(5)  & 0.9386(26) & 0.954(2) & 352383(220)& 0.1675(16) \\
& & 2048 & 1.0000530(9) & 0.9377(28) & 0.953(2) &1254983(1100)& \\
\hline
\end{tabular}
\end{table}

\begin{figure}[tb]
\centerline{\psfig{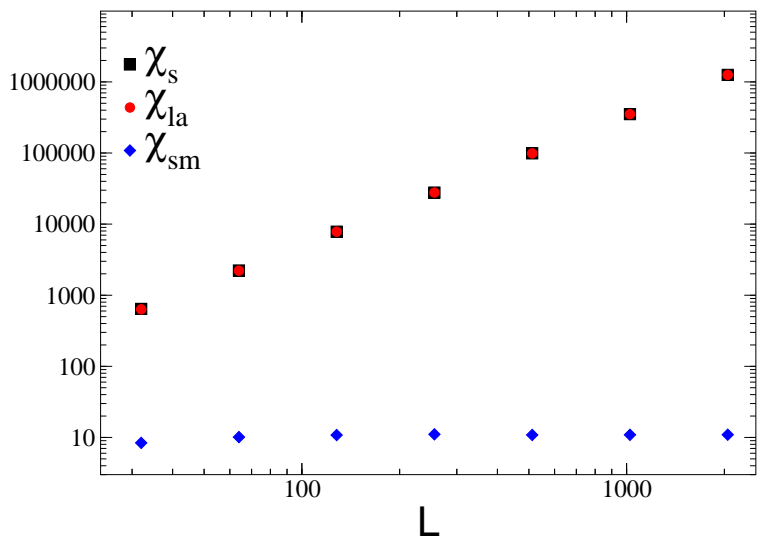}}
%\vspace{2mm}
\caption{ 
Magnetic susceptibilities $\chi_s$, $\chi_{\rm la}$, and
$\chi_{\rm sm}$ for $U=1$, $D=1/2$, and $J=1.48$.  The data for $\chi_s$
and $\chi_{\rm la}$ can hardly be distinguished on the scale of the figure
since differences are very small.  }
\label{chilt}
\end{figure}

Direct evidence for the breaking of the ${\mathbb Z}_2$ chiral
symmetry in the LT phase is provided by the finite-size behavior of
the chiral Binder parameter $B_c$.  In the LT phase we should have
$B_c\rightarrow 1+O(1/V)$ when $L\rightarrow \infty$.  MC simulations
(see Table~\ref{lowtl}) show this behavior in all models that we have
considered.

Let us discuss the $\phi^4$ model in more detail.  We distinguish the
fields $\phi_{i,x}$ from the size of the corresponding quantity $Q_i
\equiv \sum_x \phi_{i,x}^2$. For each configuration, we define the
large (small) field $\phi_{{\rm la},x}$ ($\phi_{{\rm sm},x}$) to be
the field $\phi_{i,x}$ that has the maximum (minimum) value of $Q_i$.
Then, we define the corresponding two-point functions $G_{\rm la}(x) =
\langle \vec{\phi}_{{\rm la},0} \cdot \vec{\phi}_{{\rm la},x} \rangle$
and $G_{\rm sm}(x) = \langle \vec{\phi}_{{\rm sm},0} \cdot
\vec{\phi}_{{\rm sm},x} \rangle$, the magnetic susceptibilities
$\chi_{\rm la}$ and $\chi_{\rm sm}$, and the second-moment correlation
lengths $\xi_{\rm la}$ and $\xi_{\rm sm}$, defined according to
Eq.~(\ref{chixidef}).  Our numerical results show that, as $L$
increases, small and large fields effectively decouple. Indeed,
$\chi_{\rm sm}$ converges to a constant, while $\chi_{\rm la}$
diverges as $L^{2-\eta}$.  Therefore, in the large-$L$ limit, since
$\chi_s = \chi_{\rm la} + \chi_{\rm sm}$, we have $\chi_{\rm
la}/\chi_s=1+O(L^{-2+\eta})$.  Fig.~\ref{chilt} shows estimates of
$\chi_s$, $\chi_{\rm sm}$, and $\chi_{\rm la}$ for $U=1$, $D=1/2$, and
$J=1.48$ in the LT phase [as we shall see, for $D=1/2$ there are two
very close transitions, an Ising transition at $J_{\rm ch}= 1.4668(1)$
and a KT transition at $J_{\rm sp}= 1.4704(2)$].  It is evident that
the large field is critical for any $J$, unlike the small one. We
also checked the behavior of the correlation lengths: while $\xi_{\rm
la}$ diverges as $L$, the small component $\xi_{\rm sm}$ remains
finite. As a consequence $\xi_{\rm la}/\xi_s = 1 + O(L^{-2+\eta})$, as
it was the case for the magnetic susceptibility.  Finally, we computed
the helicity moduli $\Upsilon_{\rm la}$ and $\Upsilon_{\rm sm}$.  The
helicity modulus of the small field vanishes within error bars
starting from relatively small lattice sizes (for example, for $L\ge 64$
for $J=1.48$ and $D=1/2$).  Therefore, also for this quantity we have
$\Upsilon\equiv \Upsilon_1+\Upsilon_2\approx \Upsilon_{\rm la}$.
These results show convincingly that the large-distance behavior in
the LT phase of the $\phi^4$ model is effectively determined by only
one of the two fields $\phi_i$, with a complete decoupling of the
other one.  This reflects the fact that the ${\mathbb Z}_2$ chiral
symmetry is broken with a spontaneous magnetization $M_C \not= 0$.

\begin{table}
\caption{ \label{lowt}
Estimates of $R_s\equiv \xi_s/L$, $\Upsilon$, and $\eta$ in
the LT phase. 
The column $L_{\rm max}$ reports the size of the largest lattice used.  
The estimate of $\eta$ that appears in the fifth
column is obtained by using the finite-size behavior of 
$\chi_s\sim L^{2-\eta}$.  
Those given in the last two columns are obtained by using the estimates of
$R_s$ and $\Upsilon$ and the relations among these variables that are
obtained in the XY model.  }
\footnotesize
\begin{tabular}{@{}llrlllll}
\hline
\multicolumn{1}{c}{}&
\multicolumn{1}{c}{$J$}&
\multicolumn{1}{c}{$L_{\rm max}$}&
\multicolumn{1}{c}{$R_s$}&
\multicolumn{1}{c}{$\Upsilon$}&
\multicolumn{1}{c}{$\eta$ [$\chi$]} &
\multicolumn{1}{c}{$\eta$ [$R_s$]}&
\multicolumn{1}{c}{$\eta$ [$\Upsilon$]}\\
\hline
FFXY & 2.4  & 256 & 1.0020(13)   & 1.0771(8)  & 0.1480(5)   & 0.1478(4)& 0.14776(11) \\
     & 2.3  &1024 & 0.9141(12)   & 0.9052(13) & 0.1750(5)   & 0.1752(4)& 0.1758(3) \\
     & 2.26 &1024 & 0.847(2)     & 0.7872(13) & 0.2023(11)  & 0.2015(7)& 0.2021(5) \\

IsXY $C=0$ & 
1.52  & 512 & 0.8951(11) & 0.8750(6) & 0.1817(7)   & 0.1821(4)& 0.18190(13) \\

$\phi^4$, $D=4/3$ & 1.49 & 500 & 0.936(4)   &  & 0.1689(13)  & 0.1677(13)  &  \\  
$\phi^4$, $D=1/2$ & 3 & 300 & 3.050(10)  &  & 0.01717(10) & 0.01698(11)  &  \\
    & 2 & 400 & 1.903(5)   &  & 0.04301(9)  & 0.0431(2)  &  \\
    & 1.8 & 400 & 1.642(4)   &  & 0.0571(4)   & 0.0575(3)  &  \\
    & 1.6 & 400 & 1.320(3)   &  & 0.0877(4)   & 0.0877(4)  &  \\
    & 1.5 & 1024  & 1.0536(16) & 1.1855(10) & 0.1341(6)  & 0.1346(4)  & 0.13425(12) \\
    & 1.48 & 2048 & 0.938(3)   & 0.953(2)   & 0.1675(16) & 0.1672(10) & 0.1670(4) \\
$\phi^4$, $D=4$   & 1.51    & 200 & 1.04(10)  &  & 0.139(2) & 0.138(3) & \\
    & 1.505   & 300 & 1.002(10) &  & 0.148(2) & 0.148(3)  &  \\
    & 1.50    & 200 & 0.945(6)   &  & 0.166(4) & 0.165(2) &  \\
    & 1.495   & 400 & 0.875(15)  &  & 0.194(6) & 0.190(6) & \\
$\phi^4$, $D=49$  & 1.578  & 512  & 1.020(2)    & 1.221(3)  & 0.146(2)   & 0.1433(5)& 0.1420(4) \\
    & 1.577  & 1024 & 0.867(7)    & 0.849(14) & 0.183(14)   & 0.193(4)& 0.188(3) \\
$\phi^4$, $D=99$  & 1.602   & 256 & 1.170(4)   & 1.434(5) & 0.1108(14) & 0.1105(7) & 0.1110(4) \\
    & 1.601   & 512 & 1.142(5)   & 1.372(4) & 0.1184(18) & 0.1156(9) & 0.1160(5) \\
    & 1.6005  & 512 & 1.113(3)   & 1.324(6) & 0.120(3)   & 0.1214(6) & 0.1202(5) \\
    & 1.6003  & 512 & 1.100(5)   & 1.288(5) & 0.131(4)   & 0.1241(10)& 0.1235(5) \\
\hline
\end{tabular}
\end{table}

\begin{figure}[tb]
\centerline{\psfig{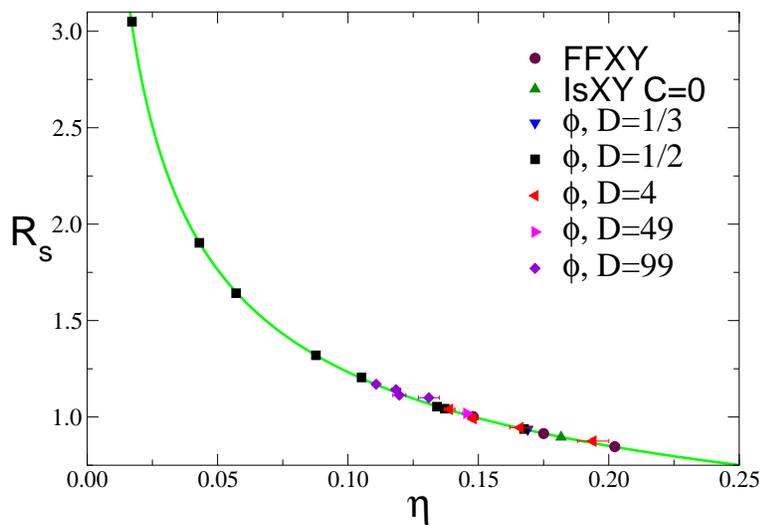}}
\caption{ 
Estimates of $R_s$ vs $\eta$ in the LT phase. 
The continuous line represents the prediction obtained by assuming that the 
LT phase is described by the same Gaussian line of fixed points
as the XY model; see \ref{elmodeta}.}
\label{xiloeta}
\end{figure}

We now prove that the long-distance behavior of the spin correlation
function is analogous to the one of the standard XY model, i.e. it is
controlled by the same line of Gaussian fixed points with varying critical
exponent $\eta$.  For this purpose, we show that, in the large-$L$
limit, the exponent $\eta$ (computed from $\chi_s\sim L^{2-\eta}$),
$R_s\equiv \xi_s/L$, and $\Upsilon$ satisfy the same universal FSS
relations as the corresponding quantities in the XY model.  These
relations, that are reported in \ref{elmodeta}, are universal
once the shape of the lattice and the boundary conditions are fixed:
in our case we consider a $L\times L$ square lattice with periodic
boundary conditions.

In order to verify these relations, for each $L$ we compute
$R_s$, $\Upsilon$, and an effective estimate of $\eta$.
Since the susceptibility behaves as $\chi_s\sim L^{2-\eta}$, we 
define 
\begin{equation}
\eta(L_1,L_2)\equiv  2 - {\rm ln}[\chi(L_1)/\chi(L_2)]/{\rm ln} (L_1/L_2).  
\label{etal1l2}
\end{equation}
In the $\phi^4$ and IsXY models we use the results for $\chi_{\rm la}$,
$\xi_{\rm la}/L$, and $\Upsilon_{\rm la}$ because they have smaller
scaling corrections with respect to the analogous quantities $\chi$,
$\xi_s/L$, and $\Upsilon$.  Some results are reported in
Table~\ref{lowtl}. In most of the cases, as $L$ increases, 
all quantities approach a
constant within error bars, indicating that we have reached the
infinite-volume limit within errors. When convergence
is not clearly observed within errors, we extrapolate the data
assuming the scaling corrections that are expected in the LT phase of
the XY model: we perform an extrapolation of the form $a + b
L^{-\zeta}$, with $\zeta = 1/\eta-4$ for $\Upsilon$ and $\zeta= {\rm
Min}(2-\eta,1/\eta-4)$ for $\chi_s$ and $R_s$, see \ref{elmodeta}.

Infinite-volume estimates of $\eta$ are reported in Table~\ref{lowt}.
Beside the values obtained from the magnetic susceptibility, we also
report the estimates obtained from $R_s$ and $\Upsilon$ 
using the XY relations.  In all cases the estimates are in 
good agreement.  In Fig.~\ref{xiloeta} we plot $\xi_s/L$ versus $\eta$
as obtained from the susceptibility,
for several models in the LT phase.
The results agree quite precisely with the curve obtained in the
standard XY model, see \ref{elmodeta}.  Thus, we conclude that the LT
phase of the FFXY, $\phi^4$, and IsXY models is controlled by the same line
of Gaussian fixed points that are relevant for the XY model.

In the XY model the LT phase becomes unstable against dissociation of
vortices when $\eta=1/4$, and \cite{Has-05} $\xi_s/L=R_{\rm
KT}=0.750691...$, $\Upsilon=\Upsilon_{\rm KT}=0.636508...$. At these
values, a KT transition occurs and the system becomes disordered.  We
expect that a similar mechanism is realized in the models considered
here, as long as the chiral ${\mathbb Z}_2$ symmetry is broken. With
decreasing $J$, $\eta$ increases. If $\eta = 1/4$ is reached, vortices
disorder the system and thus we expect a phase transition. If the
transition is continuous and the chiral ${\mathbb Z}_2$ magnetization
does not vanish at the transition, then the phase transition is
expected to be of KT type. An Ising or first-order transition should
then occur at a smaller value of $J$, where also chiral variables
disorder. Otherwise, if the ${\mathbb Z}_2$ magnetization vanishes, 
the transition should belong to a new universality class. Note that
this possibility is excluded in the IsXY model by the argument
reported in Sec.~\ref{models}.  If a transition occurs for $\eta<1/4$,
it cannot be a KT one.  Thus, it must be of first order or a
continuous transition belonging to a new universality class.  In the
latter case, however, we expect a single transition, since a new type
of transition can only be due to the interaction of spin and chiral
modes that become critical at the same value of $J$.

\section{Phase transitions in the fully frustrated XY model}
\label{phasetrffxy}

In this section we investigate the phase diagram of the square-lattice 
FFXY model.  We show that there
are two very close but separate transitions, a KT transition at $J =
J_{\rm sp}$ involving the spin degrees of freedom and an Ising
transition at $J_{\rm ch}< J_{\rm sp}$ associated with the breaking of
the chiral ${\mathbb Z}_2$ symmetry.

\subsection{Finite-size scaling at the chiral transition}
\label{fssffxy}

We perform a FSS analysis using the method proposed in
Ref.~\cite{Hasenbusch-99} and also discussed in Ref.~\cite{CHPRV-01}.
Instead of computing the various quantities at fixed Hamiltonian
parameters, we compute them at a fixed value of a chiral---this
guarantees that we are observing the chiral transition---RG invariant
quantity; in our specific case we choose $R_c\equiv \xi_c/L$. This
means that, given a MC sample generated at $J = J_{\rm run}$, we
determine, using standard reweighting techniques, the value $J_{R_c}$
such that $R_c(J = J_{R_c}) = R_{c,\rm fix}$.  All interesting
observables are then measured at $J = J_{R_c}$; their errors at fixed
$R_c$ are determined by a standard jackknife analysis. The
pseudocritical temperature $J_{R_c}$ converges to $J_{\rm ch}$ as
$L\to \infty$.  This method has the advantage that it does not require
a precise knowledge of the critical value $J_{\rm ch}$ (an estimate
is only needed to fix $J_{\rm run}$ that should be close to $J_{\rm
ch}$).  Moreover, for some observables the statistical errors at fixed
$R_c$ are smaller than those at fixed $J$ (close
to $J_{\rm ch}$).  This is due to
cross correlations and to a reduction of the effective autocorrelation
times.  The ratio $E/E_{{\rm fix}R_c}$ of the errors at $J_{R_c}$
computed at fixed $J$ and at fixed $R_c$ turns out to be roughly
independent of $L$: $E/E_{{\rm fix}R_c}\approx 1.6, 4.3, 1.9, 1.0$
respectively for $\chi_c$, $B_c$, $d B_c/dJ$ and $\Upsilon$.

\begin{table}
\caption{ 
\label{resffxy}
Results at fixed $R_c=R_{\rm Is}$ for the FFXY model. Estimates of 
the pseudocritical coupling $J_{R_c}$,
the chiral magnetic susceptibility $\chi_c$, the chiral Binder
parameter $B_c$, the derivatives of $R_c$ and $B_c$ with respect to
$J$, the spin correlation length $\xi_s$, the derivative of $R_s$ with respect
to $J$, and the helicity modulus $\Upsilon$. 
}
\hspace{-5mm}
\footnotesize
\begin{tabular}{@{}rllllllll}
\hline
\multicolumn{1}{c}{$L$}&
\multicolumn{1}{c}{$J_{R_c}$}&
\multicolumn{1}{c}{$\chi_c$}&
\multicolumn{1}{c}{$B_c$}&
\multicolumn{1}{c}{$d R_c/dJ$}&
\multicolumn{1}{c}{$-d B_c/dJ$}&
\multicolumn{1}{c}{$\xi_s$}&
\multicolumn{1}{c}{$dR_s/dJ$}&
\multicolumn{1}{c}{$\Upsilon$}\\
\hline 
8 & 2.1642(16)  & 28.38(3) & 1.1913(7) & 2.099(10) &
0.903(8) & 5.430(8) & 0.805(5) & 0.8628(14) \\ 

12 & 2.1735(11) & 56.99(7) & 1.1820(6) & 3.14(2) & 
1.323(12) & 8.051(12) & 1.179(8) & 0.7833(15) \\
  
16 & 2.1797(8)  & 93.09(11) & 1.1839(7) & 4.34(2) &
1.885(2) & 10.533(14) & 1.653(12) & 0.730(2) \\
  
24 & 2.1849(6)  & 185.7(3)  & 1.1854(8) & 6.98(4) &
3.06(3) & 15.26(2) & 2.53(2) & 0.649(2)  \\
  
32 & 2.1887(3) & 303.4(3) & 1.1872(6) & 9.79(4) &
4.39(3) & 19.72(2) & 3.45(2) & 0.591(2) \\
  
48 & 2.1931(4) & 607.1(9) & 1.1934(8) &  16.08(10) &
7.43(8) & 28.00(4) & 5.43(4) & 0.514(2) \\
  
64 & 2.1968(3) & 1004(2) & 1.1952(10) & 23.1(2) & 
10.77(14) & 35.75(7) & 7.48(7) & 0.457(3) \\
  
96 & 2.2004(2) & 2038(4) & 1.2009(10) & 38.7(3) &
18.4(2) & 49.75(9) & 11.61(8) & 0.384(3) \\
  
128 & 2.20242(8) & 3385(4) & 1.2007(6) & 54.8(3) & 
25.6(2) & 61.92(8) & 15.67(8) & 0.325(2) \\
  
192 & 2.20451(8)  & 6932(14) & 1.2002(9) & 88.6(6) & 
40.2(5) & 81.9(2) & 22.8(2) & 0.235(4) \\
  
256 & 2.20529(4) & 11469(20) & 1.1962(8) & 120.5(9) & 
53.0(6) & 96.0(3) & 27.8(3) & 0.173(4) \\
  
384 & 2.20610(5) & 23414(61) & 1.1853(11) & 178(2) &
72.8(1.3) & 112.6(5) & 31.8(5) & 0.083(5) \\
  
512 & 2.20623(5) & 38450(152) & 1.1793(13) & 232(3) &
94(2) & 117.8(7) & 31.1(6) & 0.031(6) \\

800 & 2.20630(6) & 83390(573) & 1.172(2) & 346(8) & 
133(5) & 118.0(1.2) & 22.7(1.1) & 0.003(8) \\

1000 & 2.20630(4) & 123208(659) & 1.171(2) & 445(8) &
171(5) & 118.3(9) & 18.0(1.3) & $-$0.005(6) \\
\hline
\end{tabular}
\end{table}

The value $R_{c,\rm fix}$ can be specified at will as long as it 
positive,\footnote{The method can be used with any RG invariant
quantity. For instance, one could also use the Binder parameter.
The fixed value can be taken arbitrarily between the corresponding
HT and LT values.}
though $R_{c,\rm fix} = \bar{R}_c$ ($\bar{R}_c$ is the
critical-point value) improves the convergence for $L\to\infty$
\cite{Hasenbusch-99,CHPRV-01}. Since in the two-transition scenario
the chiral transition is predicted to be in the Ising universality
class, we choose $R_{c,\rm fix}=R_{\rm Is}$ where $R_{\rm
Is}=0.9050488292(4)$ is the universal value of $\xi/L$ at the critical
point in the 2-$d$ Ising universality class \cite{SS-00}.  
This choice does not bias our analysis in favor of the Ising
nature of the chiral transition.  
For any chosen value
(as long as it is positive) and whatever the
universality class of the chiral transition is (it may also coincide
with the spin transition), $J_{R_c}(L)$ converges to
$J_{\rm ch}$ for $L\to\infty$. Indeed, in the FSS limit
$R_c = f_R[L^{1/\nu} (J - J_{\rm ch})]$, where $\nu$ and 
$f_R(x)$ depend on the universality class of the transition. 
Therefore, fixing $R_c$ is equivalent to fixing 
$X \equiv L^{1/\nu} [J_{R_c}(L) - J_{\rm ch}]$. 
Critical exponents determined by using FSS do not depend
on the chosen value $R_{c,\rm fix}$, while other finite-size
quantities do. For instance, 
the Binder cumulant satisfies the analogous relation
$B_c = f_B[L^{1/\nu} (J - J_{\rm ch})]$, so that 
$B_c$ converges to $f_B(X)$ in the limit $L\to \infty$ at fixed $R_c$.
By fixing $R_c$ to the critical Ising value,
we will be able to perform an additional consistency check.
If the chiral transition belongs to the Ising universality
class, then $X = 0$ (apart from scaling corrections) and we should find that
any RG-invariant quantity
converges to its critical-point value in the Ising model.

If the transition is
unique also spin variables should have a finite nontrivial limit: for
instance, $R_s$ at fixed $R_c$ should converge to a nonvanishing
constant. On the other hand, if the chiral transition occurs in a
region in which the spin variables are still disordered, the spin
correlation length should be finite and thus $R_s$ at fixed $R_c$
should converge to zero as $1/L$. 

\begin{figure}[tb]
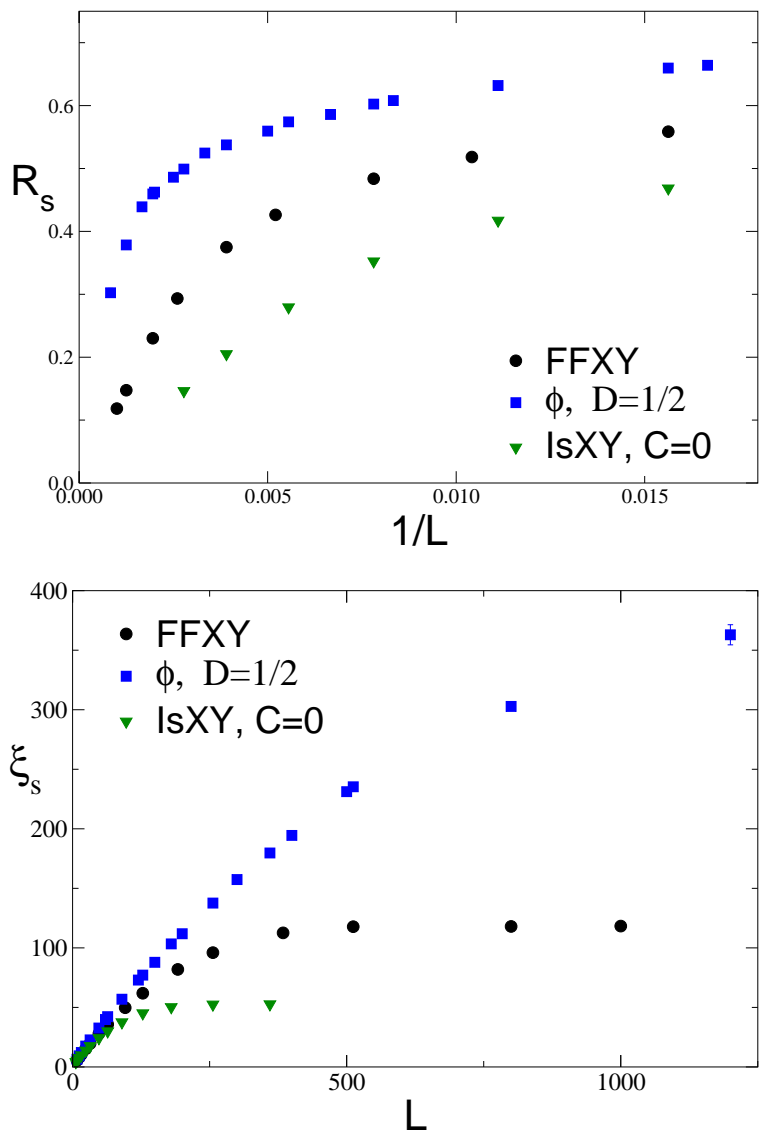

\centerline{\psfig{width=10truecm,angle=0,file=rs.eps}}
\vspace{4mm}
\centerline{\psfig{width=10truecm,angle=0,file=xis.eps}}
%\vspace{2mm}
\caption{ 
$R_s\equiv \xi_s/L$ (above) and $\xi_s$ (below) at fixed $R_c$ versus
$1/L$ and $L$ respectively, for the FFXY model, the $\phi^4$ model at
$D=1/2$, and the IsXY model at $C=0$.  }
\label{rs}
\end{figure}

We performed simulations on quite large lattices, up to $L=1000$.  
Results are reported in Table~\ref{resffxy}. In
Fig.~\ref{rs} we show the spin correlation length $\xi_s$ as a function of $L$.
It clearly converges to a constant as $L\to \infty$. We can thus 
estimate the infinite-volume spin correlation length
$\xi_s^{(c)}$ at the chiral transition:
$\xi_s^{(c)}=118(1)$.  Moreover, the helicity modulus $\Upsilon$ at
the chiral transition appears to vanish in the large-$L$ limit, see
Table~\ref{resffxy}, consistently with the fact that spin variables
are disordered.  These results definitely show that spin modes are not
critical at the chiral transition, and therefore that, for $J > J_{\rm
ch}$, there is a phase in which the chiral ${\mathbb Z}_2$ symmetry is
broken while spins remain paramagnetic.

\begin{figure}[tb]
\centerline{\psfig{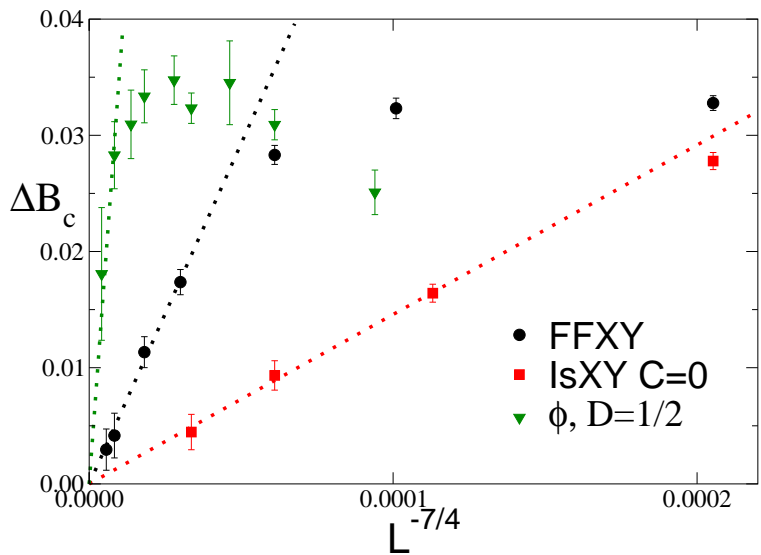}}
\caption{
$\Delta B_c\equiv B_c - B_{\rm Is}$ at fixed $R_c=R_{\rm Is}$ vs $L^{-7/4}$,
for the FFXY model, the $\phi^4$ model at $D=1/2$, and the IsXY model
at $C=0$. $B_{\rm Is} = 1.167923(5)$ is the value of the 
Binder parameter at the critical point in the Ising model
\cite{SS-00}. The dotted lines correspond to a linear fit of the data for
the largest available lattices.
}
\label{bcffxy}
\end{figure}

If spin and chiral modes decouple, the chiral transition should belong
to the Ising universality class.  The best evidence for that is
provided by the finite-size behavior of the chiral Binder cumulant $B_c$.  In
the case of an Ising transition, the asymptotic behavior of $B_c$ is
expected to be
\cite{SS-00,CCCPV-00,CHPV-02}
\begin{equation}
B_c = B_{\rm Is} + O(L^{-\omega}) + O(L^{-2+\eta}),
\label{bcsc}
\end{equation}
where \cite{SS-00} $B_{\rm Is}=1.167923(5)$. The first correction is
due to the leading irrelevant operator, the second one to the analytic
background terms.  Since for unitary Ising models
\cite{CCCPV-00,CHPV-02} $\omega=2$ and $2-\eta=7/4$, the
dominant scaling corrections are those related to the analytic
background.  The estimates of $B_c$ reported in Table~\ref{resffxy}
show a peculiar intermediate nonmonotonic crossover: for 
$L\lesssim \xi_s^{(c)}\approx 118$, 
$B_c$ is approximately constant and equal to 1.19-1.20, while the
asymptotic regime given by Eq.~(\ref{bcsc}) is observed only when
$L\gg\xi_s^{(c)}$. In Fig.~\ref{bcffxy} we plot $\Delta B_c\equiv
B_c-B_{\rm Is}$ versus $L^{-7/4}$.  The data 
clearly approach the Ising value as $L$ increases.  A linear
fit to $a + b L^{-7/4}$ of the data 
with $L\ge 384$ gives $B_c=1.1678(16)$.  We also mention that a
linear fit of $\Delta B_c$ to $b L^{-7/4}$ for $L\ge 384$ gives
$b=585(32)$ with $\chi^2/{\rm d.o.f.}\approx 0.2$, where 
d.o.f. is the number of degrees of freedom of the fit.  These results
provide a rather strong evidence of the Ising nature of the critical
behavior.

\begin{figure}[tb]
\centerline{\psfig{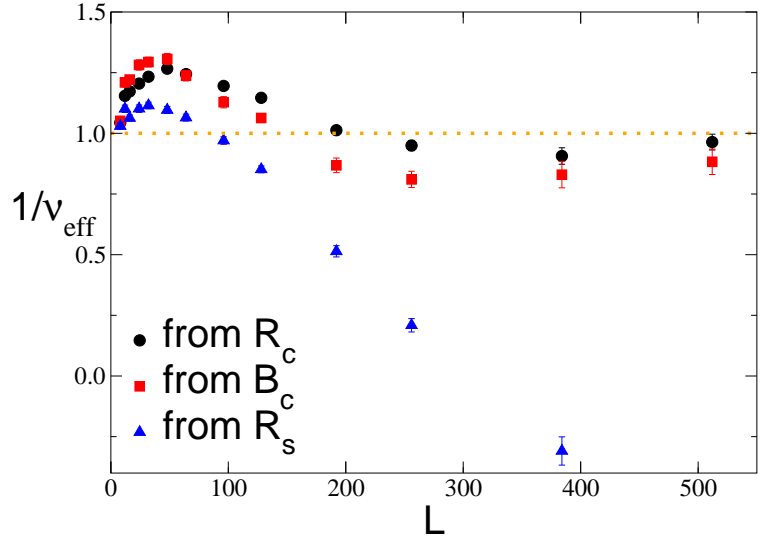}}
\caption{Effective exponent $1/\nu_{\rm eff}$ as obtained from
the derivatives of $R_c$, $B_c$, and $R_s$, for the
FFXY model.  
The dotted line corresponds to the Ising value $\nu=1$. 
}
\label{nueffxy}
\end{figure}

\begin{figure}[tb]
\centerline{\psfig{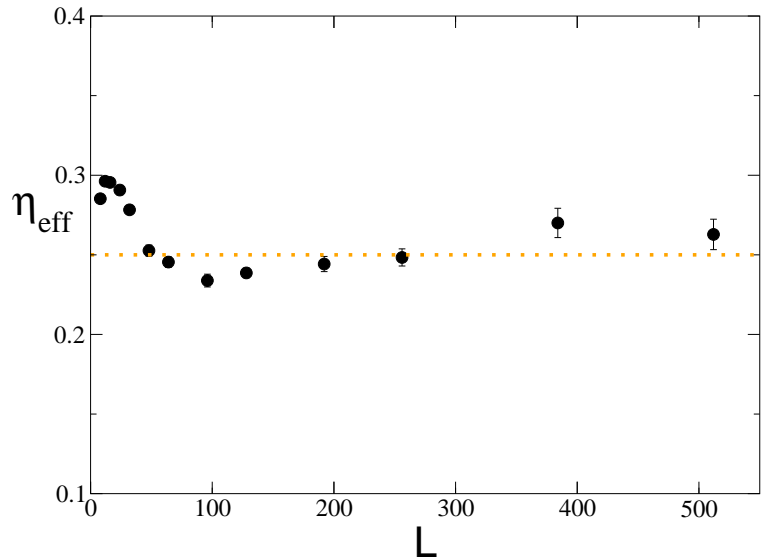}}
\caption{Effective exponent $\eta_{\rm eff}$ as obtained from
$\chi_c$, for the FFXY model.  
The dotted line corresponds to the Ising value $\eta=1/4$. }
\label{etaeffxy}
\end{figure}

In order to determine the critical exponents, we consider the
derivatives of $R_c$ and $B_c$ with respect to $J$ at fixed $R_c$. In
the large-$L$ limit, they behave as
\begin{equation}
{dR_c\over dJ}\sim {dB_c\over dJ}\sim L^{1/\nu},
\label{scaldr}
\end{equation}
where $\nu$ is the chiral exponent associated with the correlation
length.  Then, given an interval $L_1\le L \le L_2$, with $L_2/L_1
\approx 2$, we compute an effective chiral exponent $1/\nu_{\rm
eff}(L_1)$ by fitting the available data for ${dR_c/dJ}$ and
${dB_c/dJ}$ in the interval. In some cases, only two results are
available and we use
\begin{equation}
1/\nu_{\rm eff}(L_1) \equiv  {\ln d S/dJ|_{L_2} - \ln d S/dJ|_{L_1}\over 
   \ln (L_2/L_1)},
\label{nueffdef}
\end{equation}
where $S$ is either $R_c$ or $B_c$.  The results are shown in
Fig.~\ref{nueffxy}.  The exponents $1/\nu_{\rm eff}(L)$ obtained from
$R_c$ and $B_c$ show a similar behavior. They are nonmonotonic and,
for $20\lesssim L\lesssim 70$, they are approximately constant and
equal to 1.3, which corresponds to $\nu_{\rm eff}\approx 0.8$.  This
is consistent with the result $\nu\approx 0.8$ obtained in previous
studies that considered lattice sizes in this range
\cite{TK-90,GKLN-91,LKG-91,GN-93,LL-94}. For $L\gtrsim
\xi_s^{(c)}=118$, the effective exponent $1/\nu_{\rm eff}(L)$
decreases rapidly, undershoots the Ising value $\nu = 1$, which is
then approached from below.  In Fig.~\ref{nueffxy} we also report
$1/\nu_{\rm eff}(L)$ computed by using $R_s$. For small values of $L$,
we find $\nu_{\rm eff}(L)\approx 0.9$---this explains why spin modes
have sometimes been thought to be critical at the Ising
transition---while for $L\gtrsim \xi_s^{(c)}$ it starts to decrease
towards the asymptotic value $-1$, corresponding to a finite
$\xi_s^{(c)}$ at the chiral transition.  In Fig.~\ref{etaeffxy} we
show the effective chiral exponent $\eta_{\rm eff}$ defined
analogously to $\nu_{\rm eff}$ in terms of $\chi_c$, which should
behave as $L^{2-\eta}$.  The exponent is first larger than the Ising
value, then becomes smaller, and only for $L\gtrsim 500$ appears to
approach the Ising value $\eta = 1/4$.

\begin{figure}[tb]
\centerline{\psfig{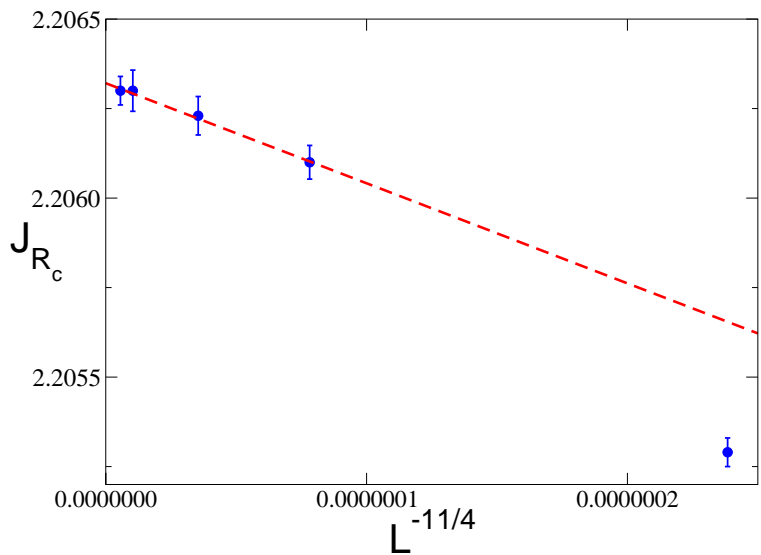}}
\caption{
$J_{R_c}$ versus $L^{-11/4}$ for the FFXY model.
The dashed line corresponds to a linear fit of the data for
the largest available lattices.
}
\label{jcffxy}
\end{figure}

The critical value $J_{\rm ch}$ of the hopping coupling at the chiral
transition can be determined from the finite-size behavior of
$J_{R_c}$.  Assuming that the transition is of Ising type, the
large-volume behavior of $J_{R_c}$ is given by
\begin{equation}
J_{R_c} = J_{\rm ch} + O(L^{-1/\nu-\omega}) + O(L^{-1/\nu-2+\eta}),
\label{jrcsc}
\end{equation}
where the first correction is due to the leading irrelevant operators,
while the second correction is due to the analytic
background~\cite{CCCPV-00,CHPV-02}.  Note that in Eq.~(\ref{jrcsc}) we
use the fact that the fixed value $R_{c,\rm fix}$ is the critical
Ising value $R_{\rm Is}$. If $R_{c,\rm fix}\neq R_{\rm Is}$, the leading
corrections are of order $L^{-1/\nu}$ \cite{Hasenbusch-99,CHPRV-01}.
According to Eq.~(\ref{jrcsc}), since $\omega = 2$ and $\eta = 1/4$,
the leading corrections to the infinite-volume limit are of order
$L^{-11/4}$.  In Fig.~\ref{jcffxy} we plot $J_{R_c}$ versus
$L^{-11/4}$.  In analogy with what is observed for the other chiral
quantities, the asymptotic behavior sets in only for $L\gg
\xi_s^{(c)}$. A linear fit of the data with $L \ge 384 $ gives $J_{\rm
ch}=2.20632(5)$. This estimate is significantly more precise than
earlier results: $J_{\rm ch}=2.2051(24)$ (Ref.~\cite{OI-03}), $J_{\rm
ch}=2.2002(10)$ (Ref.~\cite{CLJ-01}), $J_{\rm ch}=2.212(5)$
(Ref.~\cite{Olsson-95}), and $J_{\rm ch}=2.203(10)$
(Ref.~\cite{LL-94}).

In conclusion, a careful FSS analysis on lattice sizes up to $L=1000$
is consistent with an asymptotic critical behavior
belonging to the Ising universality class, as expected within the
two-transition scenario. However, before observing the asymptotic
Ising behavior, we find a peculiar nonmonotonic crossover
behavior. This is related to the fact that the spin correlation length
$\xi_s^{(c)}$ is large at the chiral transition.  For $L\lesssim
\xi_s^{(c)}$ spin modes appear as being critical, as also indicated by
the effective exponent $\nu_{\rm eff}$ that is obtained from
$dR_s/dJ$, which, for $L
\lesssim \xi_s^{(c)}$, behaves as the exponent obtained from chiral
observables, see Fig.~\ref{nueffxy}.  We will further discuss these
crossover behaviors in Sec.~\ref{crossover}.

\subsection{The spin transition}
\label{spintrffxy}

Since spins are paramagnetic at the chiral transition, the
quasi-long-range order that characterizes the LT phase requires a spin
transition at a larger value of $J$. Since the LT phase is controlled
by the same line of fixed points that are relevant for the XY model,
and chiral modes do not play any role at the transition, it is natural
to conjecture that it belongs to the KT universality class.

In order to determine the critical hopping parameter $J_{\rm sp}$ of
the spin transition, we exploit the predicted asymptotic behavior of
the helicity modulus $\Upsilon$ and of the ratio $R\equiv \xi/L$ at
the KT transition.  On a square lattice $L\times L$ at the critical
temperature we have for $L\to\infty$ \cite{Has-05}
\begin{eqnarray}
&&\Upsilon = 0.63650817819... + \frac{0.318899454...}
{\ln L + C_\Upsilon} + ...,
\label{Yxy} \\
&&R = 0.7506912... + \frac{0.212430...}{\ln L + C_R}+ \;... ,
\label{rxy} 
\end{eqnarray}
where $C_R$ and $C_\Upsilon$ are nonuniversal constants and the
neglected corrections are of order $\ln\ln L/\ln^2 L$.  

\begin{table}
\caption{\label{betaKTffxy}
Results for $J_{\rm sp}$ in the FFXY model.}
\begin{indented}
\item[]
\begin{tabular}{@{}rrll}
\hline
\multicolumn{1}{c}{$L_1$}&
\multicolumn{1}{c}{$L_2$}&
\multicolumn{1}{c}{from $R_s$}&
\multicolumn{1}{c}{from $\Upsilon$}\\
\hline 
  128 & 256  & 2.236(5)   &  2.2405(12)  \\
  256 & 512  & 2.2400(10) &  2.2415(5)  \\
  512 & 1024 & 2.2409(3)  &  2.2415(5) \\
\hline
\end{tabular}
\end{indented}
\end{table}

The critical point $J_{\rm sp}$ is obtained by determining the value of $J$ 
where the behavior
of $\Upsilon$ and $R_s$ is given by Eqs.~(\ref{Yxy}) and (\ref{rxy}).
In practice, we first obtain an approximate estimate of $J_{\rm sp}$
by looking for the value of $J$ at which $\Upsilon$ and $R_s$ are both
close to the theoretical large-$L$ value. Then, we perform simulations at two
values of $J$, $J_1$ and $J_2$, that are close to the estimated
$J_{\rm sp}$, and determine $\Upsilon$ and $R_s$ for $J_1 < J < J_2$ by
linear interpolation.  Finally, for each pair of sizes $L_1$ and
$L_2$, we require $\Upsilon$ and $R_s$ at $J = J_{\rm sp}$ to be
exactly given by the previous expressions.  In this way we obtain two
nonlinear equations for two free parameters: the critical value
$J_{\rm sp}$ and the constant $C$.  Table~\ref{betaKTffxy} shows the
results for several pairs of lattices with $L_2/L_1 = 2$.  In spite of
the expected slow convergence, the results obtained from $R_s$
and $\Upsilon$ are consistent.  This provides a nontrivial
check of the KT nature of the transition.
The helicity modulus $\Upsilon$ shows
the least $L$ dependence and provides our final estimate: $J_{\rm
sp}=2.2415(5)$. Previous estimates of $J_{\rm sp}$ are: $J_{\rm
sp}=2.227(5)$ (Ref.~\cite{OI-03}), $J_{\rm sp}=2.2422(5)$ 
(Ref.~\cite{Olsson-95}), and $J_{\rm sp}=2.273(10)$ 
(Ref.~\cite{LL-94}).  We also estimate the infinite-volume chiral second-moment
correlation length $\xi_c^{(s)}$ at the spin transition:
$\xi_c^{(s)} =  8.0(5)$.  The Ising and
KT transitions are very close: we have
\begin{equation}
\delta \equiv  {J_{\rm sp}-J_{\rm ch}\over J_{\rm ch}} = 0.0159(2).
\end{equation}

\section{Phase transitions in the $\phi^4$ model}
\label{phasetr}

\begin{figure}[tb]
\centerline{\psfig{width=12truecm,angle=0,file=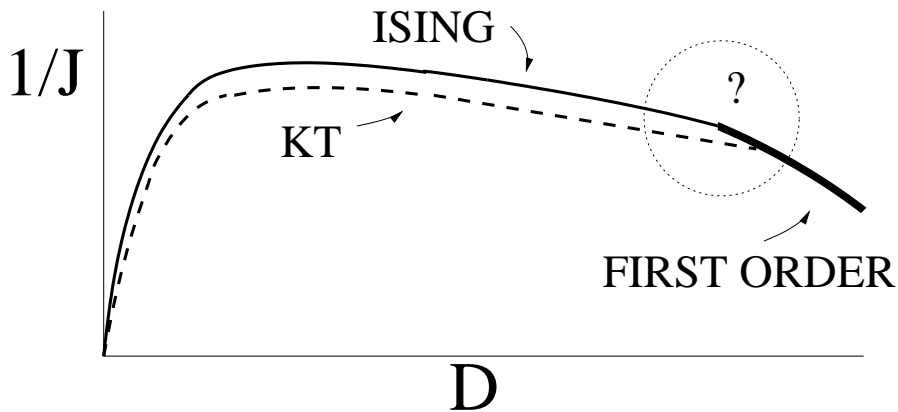}}
\vspace{2mm}
\caption{Sketch of the phase diagram of the $\phi^4$ model (\ref{HLi}) 
for $U=1$ and $D > 0$ in the plane $J^{-1}$-$D$. 
The continuous, dashed, and thick continuous 
lines represent  Ising, KT, and first-order transition lines.
The distance between the Ising and KT lines is amplified;
otherwise, the two transitions could not be distinguished on the scale of the 
figure.
The phase diagram within the circled region is unknown. 
In the figure we show one possibility, in which the 
two continuous transitions connect to the 
unique first-order line.
}
\label{phased}
\end{figure}

In this section we study the phase diagram of the $\phi^4$ model for
$U=1$. We only consider the case $D > 0$, since only in this case is the
ground-state degeneracy the same as in the FFXY model.  We show
that there is a unique first-order transition for 
$D > D^*$, $34 < D^* \lesssim 49$. 
On the other hand, for sufficiently small $D$, we
provide evidence of two very close transitions, a KT transition at $J
= J_{\rm sp}$ associated with the breaking of the continuous degrees
of freedom and an Ising transition at $J_{\rm ch}< J_{\rm sp}$
associated with the breaking of the interchange (chiral)
${\mathbb Z}_2$ symmetry.
In Fig.~\ref{phased} we show the phase diagram of the $\phi^4$ model
for $U=1$ and $D> 0$, which emerges from the MC results we shall
present in this section.
We do not know
how the  two continuous transition lines connect to the unique
first-order one.  There are several possibilities.  Under the
hypothesis that no unique continuous transition occurs, as expected in
the IsXY model (see Sec.~\ref{models}), we may have: (i) the Ising
line becomes of first order at $D=D_{\rm ch} < D^*$, and then the KT
line meets the first-order line at $D^*$ (this possibility is
presented in Fig.~\ref{phased}); (ii) as before, but interchanging
Ising and KT lines; (iii) both the Ising and KT lines become of first
order at $D=D_{\rm ch}$ and $D=D_{\rm sp}$ respectively, with $D_{\rm
ch},D_{\rm sp}<D^*$, and then they meet at $D=D^*$.  Our data suggest
that the intermediate region between the one characterized by two
continuous transitions and the one with a unique first-order
transition should be  relatively small, i.e.  $D_{\rm ch}$, $D_{\rm sp}$,
$D^*$ are close to 49 and larger than 34.  However, we cannot
exclude that the system undergoes a unique continuous transition at
some value of $D$ or in some interval of $D$ in this intermediate
region.  In this case two further possibilities may be realized: the
Ising and KT line meet and then, a single transition line starts,
which can be either of first order or of second order (turning
eventually to first order).

\begin{figure}[tb]
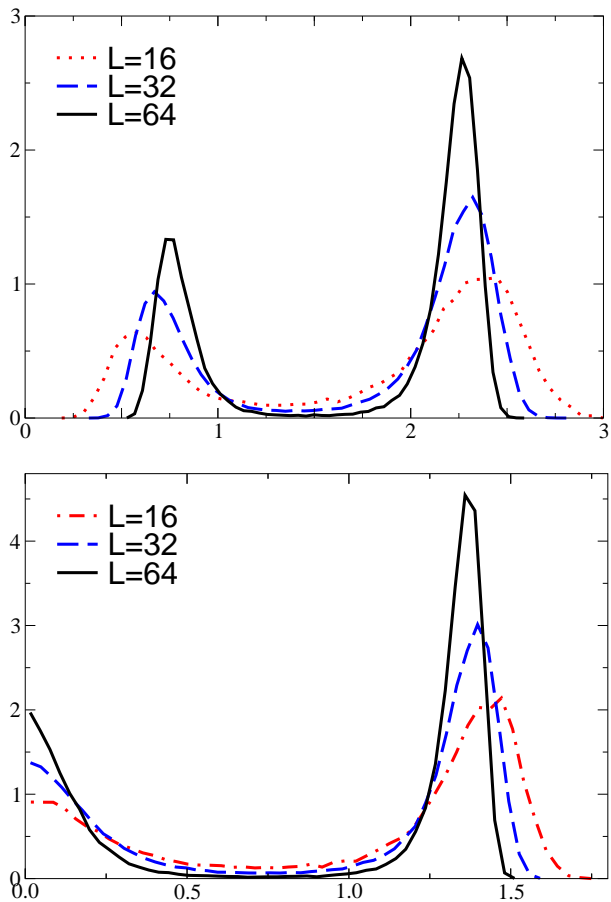

\centerline{\psfig{width=8truecm,angle=0,file=hy100.eps}}
\vspace{4mm}
\centerline{\psfig{width=8truecm,angle=0,file=hyma100.eps}}
\caption{ Distributions of the hopping-energy density (above) and of the
chiral magnetization (below) for $D=99$ 
($\phi^4$ model) at the pseudocritical transition point 
$J_{\rm pc}(L)$.  }
\label{hy100}
\end{figure}

\subsection{First-order transition for large $D$}
\label{firstorder}

\begin{table}
\caption{ \label{betac1o}
Results for $D=99$ and 49.
Here $J_{\rm pc}(L)$ is the pseudocritical $J$ obtained by using the 
equal-weight method, $P(E_{{\rm min}})$ is the minimum of the 
normalized energy distribution between the two peaks, 
$J_{\rm max}(L)$ is the value of $J$ corresponding to the maximum of the 
specific heat, and $C_{{\rm max}}$ the maximum of the
specific heat.
}
\footnotesize
\begin{indented}
\item[]
\begin{tabular}{@{}rrllll}
\hline
\multicolumn{1}{c}{}&
\multicolumn{1}{c}{$L$}&
\multicolumn{1}{c}{$J_{\rm pc}(L)$}&
\multicolumn{1}{c}{$P(E_{{\rm min}})$}&
\multicolumn{1}{c}{$J_{\rm max}(L)$}&
\multicolumn{1}{c}{$C_{{\rm max}}/V$}\\
\hline 
$D=99$ &  16  &  1.5975(3) & 0.120(10)  & 1.5956(4)& 0.663(5) \\
   &  24  &  1.59840(14) & 0.069(23)  & 1.59766(13)& 0.600(4) \\
   &  32  &  1.59942(7)  & 0.053(5)   & 1.59898(7) & 0.569(2) \\ 
   &  48  &  1.59994(4)  & 0.032(5)   & 1.59973(4) & 0.539(2) \\ 
   &  64  &  1.60018(4)  & 0.018(3)   & 1.60006(4) & 0.522(2) \\ 

$D=49$ &   48  &1.57636(6) & 0.152(13) & 1.57608(6) & 0.283(2) \\
   & 64  &1.57661(2) & 0.181(8)  & 1.57643(2) & 0.2523(13) \\
   & 96  &1.576771(15) & 0.174(14) & 1.576692(15) & 0.2145(13) \\
  & 128  &1.576853(11) & 0.151(23) & 1.576801(11) & 0.196(2) \\
\hline
\end{tabular}
\end{indented}
\end{table}

For sufficiently large values of $D$ there is a unique first-order
transition from the disordered phase to the LT phase.  Evidence for a
first-order transition is provided by the presence of two peaks in the
distribution $P(E)$ of the hopping-energy density $E$ defined in
Eq.~(\ref{eedef}).  We determine the position of the transition by
using the equal-weight method for the two phases applied to the energy
density.  In practice, for each lattice size we compute $E_{{\rm
min}}$, the energy corresponding to the minimum between the peaks of
the histogram. Then, a given configuration is considered as a LT
configuration if $E> E_{{\rm min}}$, a HT
configuration otherwise.  Finally, for each lattice size we determine
a pseudo-critical value $J_{\rm pc}(L)$ by requiring that for
$J=J_{\rm pc}(L)$ the ratio of the probabilities of LT and
HT configurations is some fixed number $w$. In our case
the value $w=2$ seems the most appropriate one, since the LT phase is
expected to be twofold degenerate.  Note that $J_{\rm pc}(L)$
converges to the infinite-volume transition value $J_{c}$ for any
chosen value $w$, with corrections that are generically of order
$1/V$. The position of the transition point can also be identified by
considering the value $J_{\rm max}(L)$ where the specific heat $C$
takes its maximum value.  Also $J_{\rm max}(L)$ is expected to
converge to $J_c$ with corrections of order $1/V$.

\begin{figure}[tb]
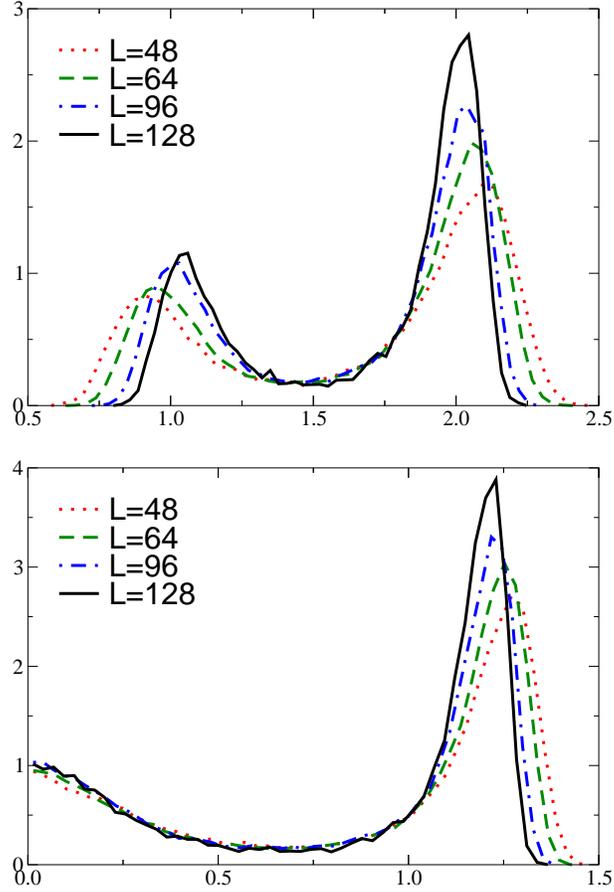

\centerline{\psfig{width=8truecm,angle=0,file=hy50.eps}}
\vspace{4mm}
\centerline{\psfig{width=8truecm,angle=0,file=hyma50.eps}}
\caption{ Distributions of the hopping energy density (above) and of the
chiral magnetization (below) for $D=49$ and several
lattice sizes at the corresponding $J_{\rm pc}$.  }
\label{hy50}
\end{figure}

To identify the first-order transition region, we have performed
simulations for $D = 34$, 49, and 99.  The distributions of the
hopping energy and of the chiral magnetization for $D = 99$ shown in
Fig.~\ref{hy100} clearly indicate that the transition is of first
order. They have a deep minimum between the two peaks which decreases
exponentially as $L\to\infty$, consistently with a first-order
transition. For instance, the minimum of the energy distribution,
$P(E_{{\rm min}})$, is expected to behave as $e^{-2\sigma L}$, where
$\sigma$ is the interface tension.  Fitting the estimates of
$P(E_{{\rm min}})$ reported in Table~\ref{betac1o} to $c e^{-2\sigma
L}$, where $\sigma$ is the interface tension, we obtain the estimate
$\sigma=0.017(2)$, which is rather small. The critical value $J_{c}$
can be estimated by fitting the results for $J_{{\rm pc}}$ and $J_{\rm
max}$ reported in Table~\ref{betac1o} to $J_{c} + a/V$.  We obtain
$J_{c}=1.60042(5)$ from $J_{\rm pc}$ and $J_{c}=1.60040(6)$
from $J_{\rm max}$. Finally, from the maximum value of $C$ one can
also estimate the latent heat $\Delta\equiv E_{+}-E_{-}$, where
$E_{\pm}$ are the values of the hopping energy $E_h$ at the peaks of
the energy distribution. Using $\Delta^2=4 C_{{\rm max}}/V +
O(1/V)$, we obtain $\Delta \approx 1.4$.

\begin{figure}[tb]
\centerline{\psfig{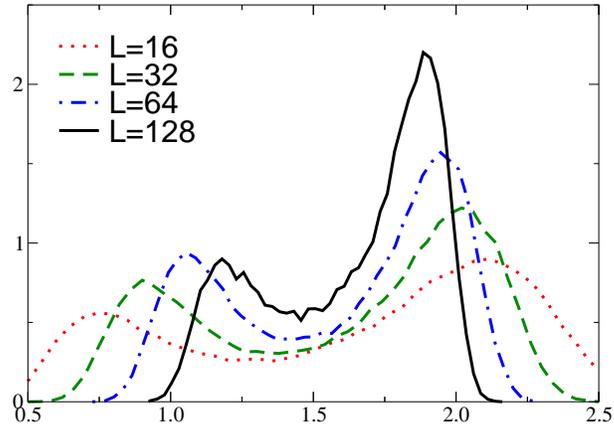}}
\caption{ 
Hopping-energy distribution for $D=34$ and several lattice sizes at
the corresponding $J_{\rm pc}$.
}
\label{hy35}
\end{figure}

The analysis of the chiral and spin degrees of freedom does not
provide evidence of other transitions.  On the LT side of the
first-order transition, i.e. for $J\to J_c\approx 1.6004$ from above,
the data for $D=99$ reported in Table~\ref{lowt} give $\eta\approx
0.12$, which is much below the value
$\eta = 1/4$ where vortex dissociation destabilizes the LT XY phase.
In Table~\ref{lowt} we also report a result for $J=1.6003$. This value
of $J$ is on the HT side.  On the other hand, the MC run
started from a LT configuration and no drastic change of the
thermodynamic variables was observed during the run. Thus, the result
corresponds to an average taken in the metastable
phase.  Finally, we mention that also the finite-size behavior of the
derivative of the chiral Binder parameter with respect to $J$ at
$J_{\rm pc}$ is consistent with the behavior expected at a first-order
transition \cite{VRSB-93}.  Indeed, we find $dB_c(L)/dJ \sim
L^{1/\nu_{\rm eff}}$ with $1/\nu_{\rm eff}=2.0(1)$, in agreement with
$1/\nu_{\rm eff}=d=2$, which is 
valid in the FSS limit at a first-order transition.

With decreasing $D$, the first-order transition becomes weaker and
weaker and eventually disappears.  The results for $D=49$ reported in
Table~\ref{betac1o} and the distributions shown in Fig.~\ref{hy50}
suggest that this occurs for $D\approx 49$. Although two peaks are
still present, $P(E_{{\rm min}})$ is roughly constant with $L$ and does
not show the exponential decrease expected at a first-order
transition.  This is the typical behavior of energy distributions
close to a tricritical point where the first-order transition line
ends; see, e.g., Ref.~\cite{WN-95}. An analogous behavior is observed
in the distributions of the chiral magnetization shown in
Fig.~\ref{hy50}.  It is interesting to observe that two peaks in the
energy and chiral-magnetization distributions are also observed for
smaller values of $D$ on lattice sizes of order $L\sim 10^2$. However,
in this case the minimum value of the energy distribution increases.
This is clearly observed for $D=34$, see Fig.~\ref{hy35}.  For $D=49$
the analysis of the finite-size behavior of the derivative of the
chiral Binder parameter gives $1/\nu=1.84(16)$. This estimate suggests
a value smaller than 2, the value expected at a first-order transition,
and it is consistent with the corresponding value at the
tricritical Ising point, which is $1/\nu=9/5$ \cite{LS-84}.  We do not
find evidence of other transitions for $D=49$.

\subsection{Continuous transitions}
\label{conttra}

\subsubsection{Standard finite-size scaling analysis.}
\label{sfssa}

\begin{figure}[tb]
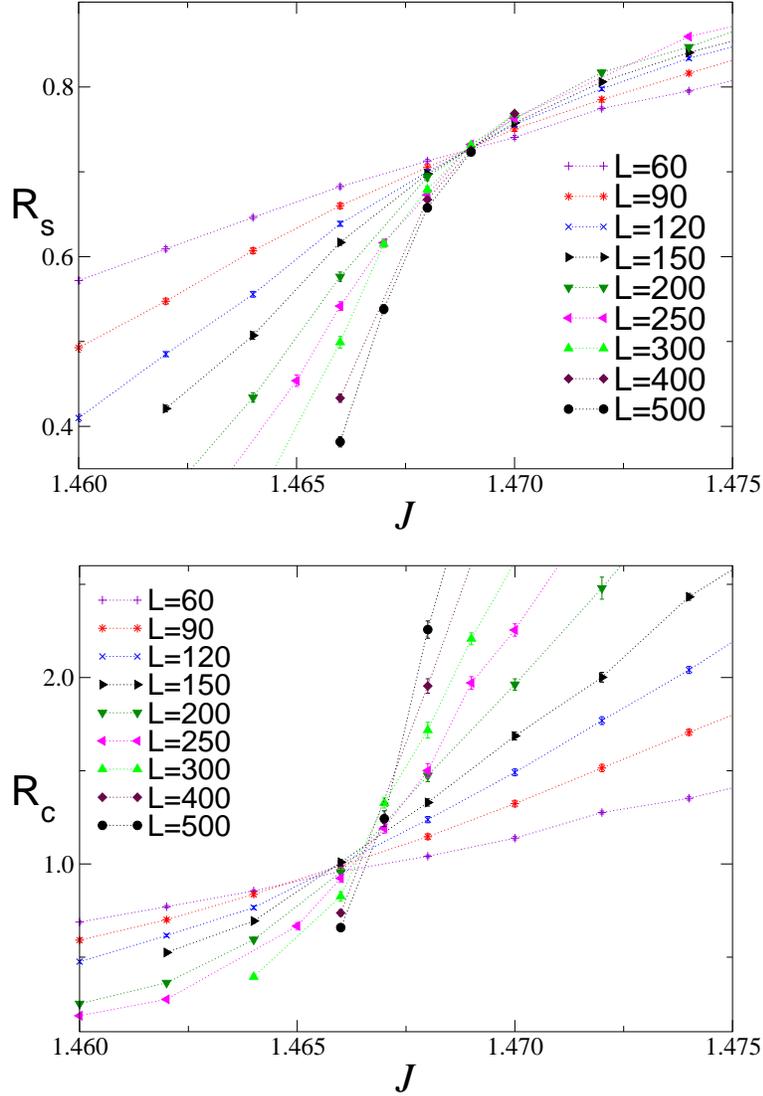

\centerline{\psfig{width=10truecm,angle=0,file=xil3o2.eps}}
\vspace{4mm}
\centerline{\psfig{width=10truecm,angle=0,file=xilch3o2.eps}}
\caption{ 
$R_s\equiv \xi_s/L$ (above) and $R_c\equiv \xi_c/L$ (below) 
versus $J$ for $D=1/2$.
The dotted lines connecting different sets of data are drawn 
to guide the eye.}
\label{xil3o2}
\end{figure}

Let us first discuss the results for $D=1/2$.  In Fig.~\ref{xil3o2} we
plot $R_s\equiv \xi_s/L$ and $R_c\equiv \xi_c/L$ versus $J$ for
$L\le 500$ in the critical region.  The crossing points
correspond to different, though very close values of $J$: $J_{\rm
cross}\approx 1.4688$ for $R_s$ and $J_{\rm cross}\approx 1.4668$ for
$R_c$.  Apparently, chiral and spin variables have different crossing
points, as also confirmed by the analysis of the Binder parameters:
$J_{\rm cross}\approx 1.4680$ for $B_{s\phi}$, and $J_{\rm cross}\approx
1.4668$ for $B_c$.  This difference may be due to scaling corrections
or to the presence of two different transitions. In order to
investigate this issue we perform a scaling analysis. We fit the data
for ${\cal R } = R_s$, $B_{s\phi}$, $R_c$, and $B_c$ with
\begin{equation}
{\cal R }(L, J) = f[(J - J_{c}) L^{1/\nu}],
\end{equation}
where $f(x)$ is approximated by a polynomial. Fits of the chiral
variables $R_c$ and $B_c$ for $L\ge 200$ give $\nu=0.78(2)$ and
$\nu=0.80(3)$ with a reasonable, though not good, $\chi^2$, i.e.
$\chi^2/{\rm d.o.f.} \approx 2$.  The analysis of the spin variables is
much less stable.  We obtain $\nu=0.9$-1.2 from $B_{s\phi}$ and $\nu\approx
1.5$ for $R_s$. These results suggest that spin and chiral variables
undergo two different, though close transitions. Indeed, if the
transition is unique, all observables have a FSS behavior with the
{\rm same} exponent $\nu$. Moreover, the numerical results apparently
exclude that the chiral transition is in the Ising universality class.
Indeed, the exponent $\nu$ appears to be different from the Ising one
$\nu_{\rm Is} = 1$ and, at the crossing point, $R_c$ and $B_c$ take
the values $\bar{R}_c= 1.15(3)$ and $\bar{B}_c=1.116(5)$, that differ
from the Ising values \cite{SS-00} $R_{\rm Is}=0.9050488292(4)$,
$B_{\rm Is}=1.167923(5)$.

Similar results are also obtained for other values of $D$.  For
example, the same FSS analysis of the data at $D=1/3$ up to $L=600$
gives $J_{\rm cross}\approx 1.4768$ (chiral observables), $\nu=0.79(1)$
(from $R_c$), $\nu=0.80(1)$ (from $B_c$), and $\bar{R}_c= 1.11(2)$ and
$\bar{B}_c=1.116(5)$; $R_s$ and $B_{s\phi}$ have crossing points at $J\approx
1.4791$ and $J\approx 1.4785$, with exponent $\nu=0.9$-1.2.  For $D=1$,
$R_c$ and $B_c$ have a crossing point at $J\approx 1.4610$, with
exponents $\nu=0.78(2)$ and $\nu=0.80(2)$ respectively, and
$\bar{R}_c= 1.17(3)$ and $\bar{B}_c=1.132(8)$; $R_s$ and $B_{s\phi}$ have
crossing points at $J\approx 1.4632$ and $J\approx 1.4625$, with
exponents $\nu=1.2$-1.4.

In conclusion, the analysis presented here apparently favors the
presence of two transitions. The first one is associated with the
chiral degrees of freedom with an exponent $\nu\approx 0.8$. This is
similar to what has been observed in the past in similar systems, see
Table~\ref{summaryliterature}.  In the next Section, we shall show
that the result $\nu\approx 0.8$ is a crossover effect: The
chiral transition belongs to the Ising universality class.

\begin{table}
\caption{\label{resd1o2}
Estimates at fixed $R_c=R_{\rm Is}$ of the pseudocritical coupling
$J_{R_c}$, the chiral magnetic susceptibility $\chi_c$, the chiral
Binder parameter $B_c$, the derivatives of $R_c$ and $B_c$ with
respect to $J$, the ratio $R_s$ and its derivative with respect to
$J$, and the total helicity modulus $\Upsilon$. They refer to the
$\phi^4$ model with $D=1/2$.  }
\hspace{-8mm}
\footnotesize
\begin{tabular}{@{}rllllllll}
\hline
\multicolumn{1}{c}{$L$}&
\multicolumn{1}{c}{$J_{R_c}$}&
\multicolumn{1}{c}{$\chi_c$}&
\multicolumn{1}{c}{$B_c$}&
\multicolumn{1}{c}{$d R_c/dJ$}&
\multicolumn{1}{c}{$-d B_c/dJ$}&
\multicolumn{1}{c}{$R_s$}&
\multicolumn{1}{c}{$dR_s/dJ$}&
\multicolumn{1}{c}{$\Upsilon$}\\
\hline 
8    & 1.5320(9)  & 63.76(15)   & 1.185(2)    & 4.10(7) & 0.185(8) 
     & 0.8124(9)  & 1.93(3)  & 1.470(2)   \\

12   & 1.4976(4)  & 107.7(2)    & 1.1687(9)   & 6.45(6) & 0.340(8)
     & 0.7798(7)  & 2.89(2)  & 1.269(2)   \\

16   & 1.4824(3)  & 158.8(2)    & 1.1659(11)  & 9.05(9)  & 0.572(10)
     & 0.7578(7)  & 3.94(4)   & 1.119(2)  \\

24   & 1.4717(2)  & 283.6(5)    & 1.1633(8)   & 14.48(13) &  1.12(2) 
     & 0.7303(7)  & 6.04(5) & 0.968(2) \\

32   & 1.46766(15)& 437.5(6)    & 1.1658(8)   & 20.7(2) &  1.90(3)
     & 0.7107(7)  & 8.28(8)  & 0.878(2) \\

48   & 1.46539(8) & 827.1(1.1)  & 1.1694(7)   & 33.8(3) &  3.95(7)
     & 0.6810(8)  & 13.07(11)  & 0.769(2)   \\

60   & 1.46499(11)& 1188(3)     &  1.1726(8)  & 44.8(4) & 5.98(10) 
     & 0.6641(7)  & 16.87(15)    & \\
  
64   & 1.46498(9) & 1320(3)     &  1.1743(9)  & 49.1(5)  & 6.87(12) 
     & 0.6597(8)  & 18.3(2)   & 0.696(2)  \\

90   & 1.46490(10)& 2323(8)     &  1.1811(11) & 76.2(9) &  13.2(3)  
     & 0.6319(12) & 27.5(3)  &    \\

120  & 1.46523(7) & 3790(11)    &  1.1829(11) & 107.4(1.0) &  21.6(4) 
     & 0.6080(10) & 37.1(4) &  \\
  
128 & 1.46528(5)  & 4223(9)     &  1.1855(9)  & 117.6(1.0)  & 24.7(4) 
    & 0.6024(9)   & 40.3(4) & 0.555(3) \\

150 & 1.46546(7)  & 5539(19)    & 1.1906(14)  & 147(2)  & 33.4(7)
    & 0.5859(14)  & 49.7(6)  &    \\
  
180 & 1.46568(5)  & 7622(23)    & 1.1910(15)  & 185(2)  & 45.0(1.0)
    & 0.5741(12)  & 60.2(8)  & 0.497(3)   \\

200 & 1.46579(6)  & 9132(40)    & 1.193(2)  & 208(3) & 55.5(1.4)
    & 0.559(2)    & 67.9(1.2)  &    \\
  
256 & 1.46601(3)  & 14046(37)   & 1.1988(13)  & 289(3)  & 87.3(1.4) 
    & 0.5376(11)  & 90.5(9)   & 0.421(3)   \\
  
300 & 1.46624(8)  & 18742(148)  & 1.202(4)    & 364(9)   & 119(5) 
    & 0.525(3)    & 109(3)   &    \\

360 & 1.46625(3)  & 25534(100)  & 1.2002(13)  & 438(5) &  159(3)
    & 0.499(2)  & 129.9(1.2)  & 0.352(4)    \\
  
400 & 1.46636(3)  & 30907(139)  & 1.203(2)   & 514(8) &  192(5)
    & 0.486(2)  &  145(2)   &     \\

500 & 1.46646(3)  & 45794(283)  & 1.208(3)    & 682(13)     &  279(9)
    & 0.462(3)    & 183(4)  &   \\ 
  
512 & 1.46651(3)  & 48282(338)  & 1.201(2)  & 696(13)  &  283(5)
    & 0.459(3)    & 186(3)  & 0.280(6)   \\

600 & 1.46664(6)  & 64317(882)  & 1.199(3)  & 834(30)  &  379(18)
    & 0.439(6)    & 209(4)  & 0.252(8)   \\
  
800 & 1.46666(2)  & 104255(603) & 1.196(3)    & 1095(25)     &  564(15) 
    & 0.379(2)    & 252(3) & 0.176(5)  \\
  
1200& 1.46675(4)  & 214664(3746)& 1.186(6)    & 1630(91) &  967(57)
    & 0.302(7)    & 317(17)  & 0.097(7)   \\
\hline
\end{tabular}
\end{table}

\subsubsection{Finite-size scaling  at fixed $R_c$.}
\label{fssrc}

In order to check whether the appearence of two very close transitions
is just due to uncontrolled systematic errors, or the value $\nu\approx
0.8$ of the chiral exponent is an effective exponent associated with a
preasymptotic regime, we performed additional simulations on larger
lattices.  We follow here the same strategy used in Sec.~\ref{fssffxy}
for the FFXY model, i.e. we perform a FSS analysis at fixed
$R_c=R_{\rm Is}$.

\begin{figure}[tb]
\centerline{\psfig{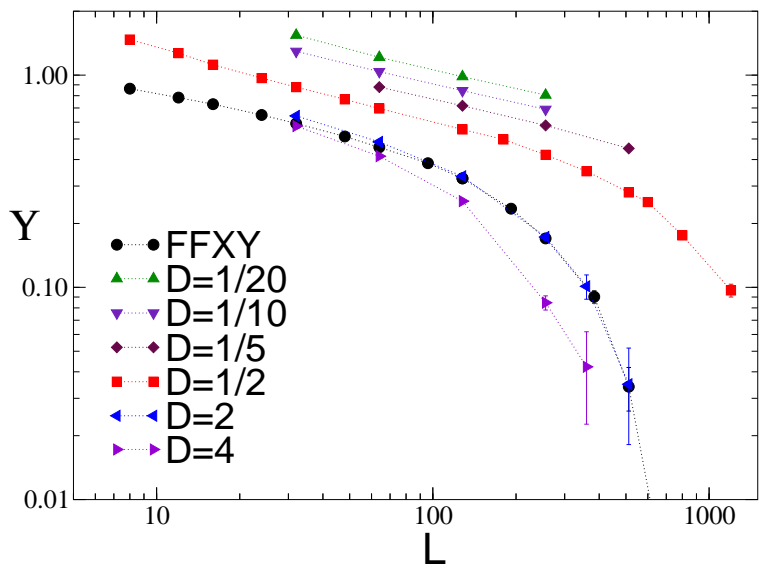}}
\caption{ 
Helicity modulus $\Upsilon$ at fixed $R_c$ for several values of $D$ 
in the $\phi^4$
model. For comparison, we also show the data for the FFXY model.
The dotted lines connecting different sets of data are drawn 
to guide the eye. Note that the data for $D=2$ overlap with those 
for the FFXY model.
}
\label{yupsilonphi4plain}
\end{figure}

In Table~\ref{resd1o2} we report the data for $D=1/2$ up to $L=1200$.
The results for $R_s$ and $\Upsilon$ at fixed $R_c$, which are also
plotted in Fig.~\ref{rs} and \ref{yupsilonphi4plain} respectively,
strongly favor the two-transition picture: $R_s$ and $\Upsilon$ are
clearly decreasing and do not show any evidence of convergence to a
nonzero value.  Note that the spin correlation length $\xi_s^{(c)}$ at
the chiral transition is rather large.  The results for the largest
lattices suggest $\xi_s^{(c)} \gtrsim 360$, explaining the strong
crossover effects.  The large correlation length $\xi_s^{(c)}$ also
suggests that the spin transition should occur at values of $J$ that
are very close to $J_{\rm ch}$. This will be confirmed in 
Sec.~\ref{spintrphi4}.

Similar results are obtained for other values of $D$,
$D=1/20,1/10,1/5,1/3,1,2,4$, performing the same FSS analysis 
for lattices with 
$L\le 600$. In Fig.~\ref{yupsilonphi4plain} we show the helicity
modulus $\Upsilon$ for several values of $D$. In all cases 
$\Upsilon$ appears to decrease, consistently with the fact that it
should eventually vanish in the large-$L$ limit.

\subsubsection{Ising critical behavior at the chiral transition.}
\label{icb}

\begin{figure}[tb]
\centerline{\psfig{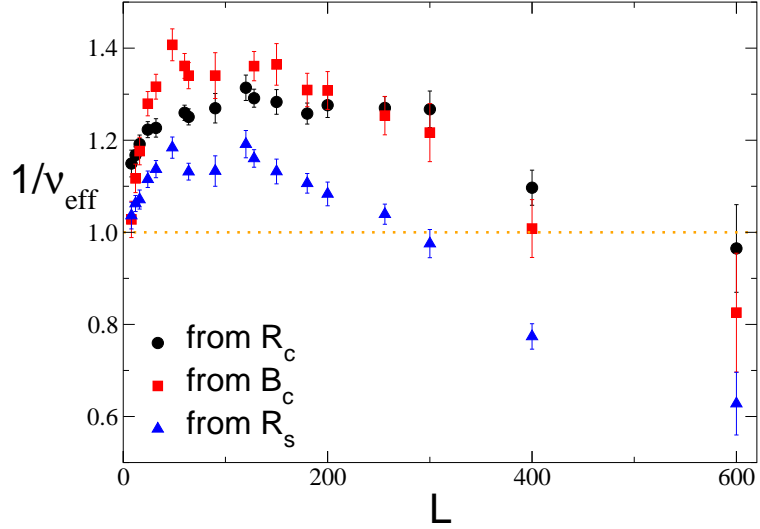}}
\caption{Effective exponents $1/\nu_{\rm eff}$ as obtained
from the derivatives of $R_c$, $B_c$, and $R_s$, for
$D=1/2$.  The dotted line corresponds to the Ising value $\nu=1$.
}
\label{nueff1o2}
\end{figure}

\begin{figure}[tb]
\centerline{\psfig{width=10truecm,angle=0,file=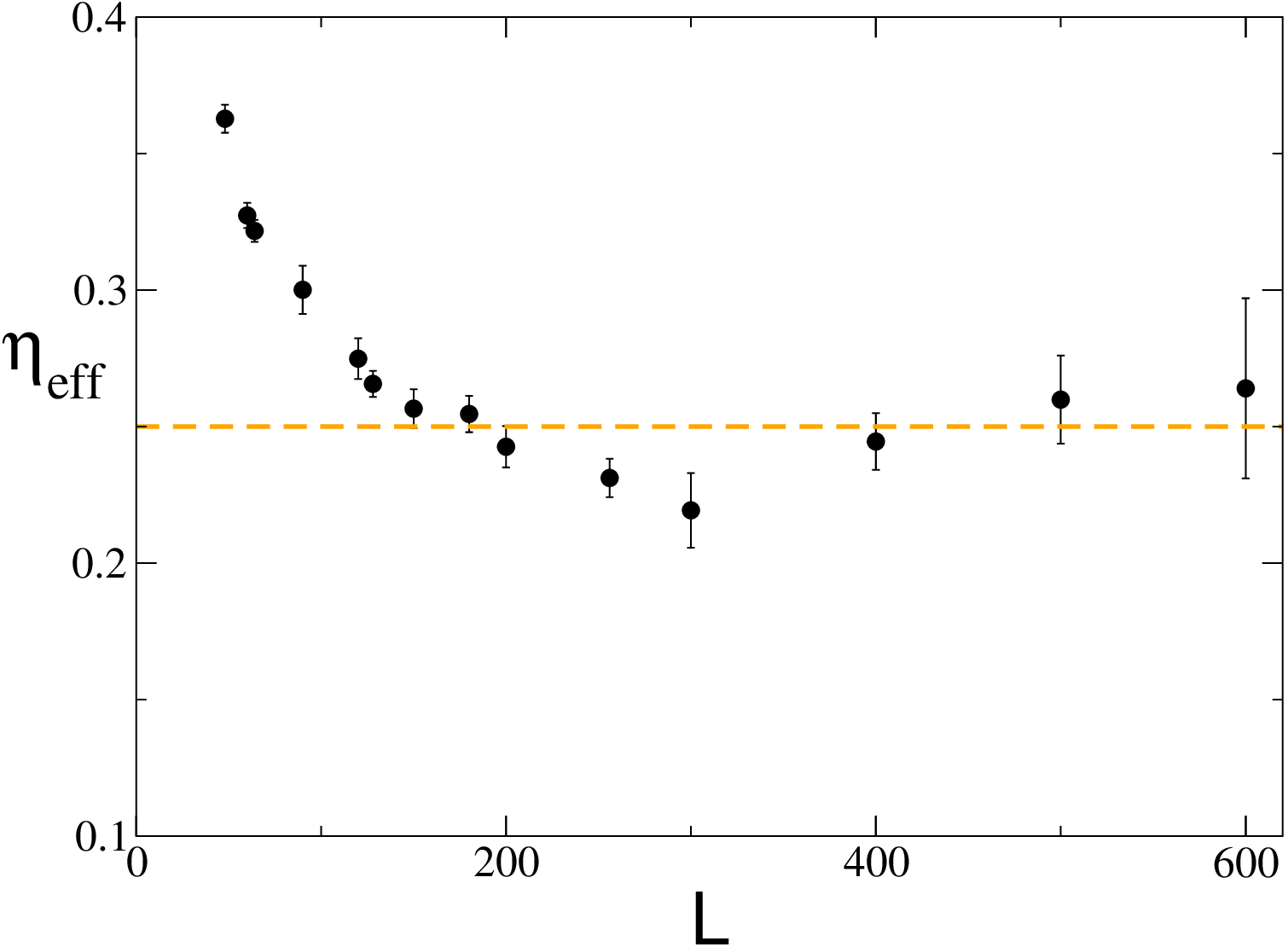}}
\caption{Effective exponent $\eta_{\rm eff}$ as obtained
from $\chi_c$, for $D=1/2$.  
The dotted line corresponds to the Ising value $\eta=1/4$.
}
\label{etaeff1o2}
\end{figure}

We now analyze the critical behavior at the chiral transition. In
particular, we wish to understand whether the results of the FSS
analysis of Sec.~\ref{sfssa}, indicating a critical behavior different
from the Ising one, are confirmed by a FSS analysis that includes
larger lattices. As in Sec.~\ref{fssffxy}, we consider the chiral
effective exponents $1/\nu_{\rm eff}$ obtained from $R_c$ and
$B_c$. The results for $D=1/2$ are shown in Fig.~\ref{nueff1o2}.  The
curves for $1/\nu_{\rm eff}(L)$ show first an extended region,
corresponding approximately to $50\lesssim L \lesssim 300$, in which
$\nu_{\rm eff}\approx 0.8$. Then, the FSS curves decrease rapidly
towards values which are substantially consistent with the Ising
exponent $\nu=1$.  In Fig.~\ref{etaeff1o2} we show the results for the
effective chiral exponent $\eta_{\rm eff}$ defined by using
$\chi_c\sim L^{2-\eta}$, in analogy with the definition of $\nu_{\rm
eff}$.  The exponent approaches
the Ising value $\eta=1/4$ as $L$ increases.  Analogous crossovers are
observed in other quantities, in the Binder cumulant $B_c$, see
Fig.~\ref{bcffxy}, and in the pseudocritical $J_{R_c}$.  The Ising
regime, in which $B_c \approx B_{\rm Is} + c L^{-7/4}$ and $J_{R_c} =
J_{\rm ch} + a L^{-11/4}$, cf. Eqs.~(\ref{bcsc}) and (\ref{jrcsc}), is
observed only when $L\gtrsim 600$.  From linear fits of the data for
the largest available lattice sizes, we obtain the estimate $J_{\rm
ch}=1.4668(1)$.

\begin{figure}[tb]
\centerline{\psfig{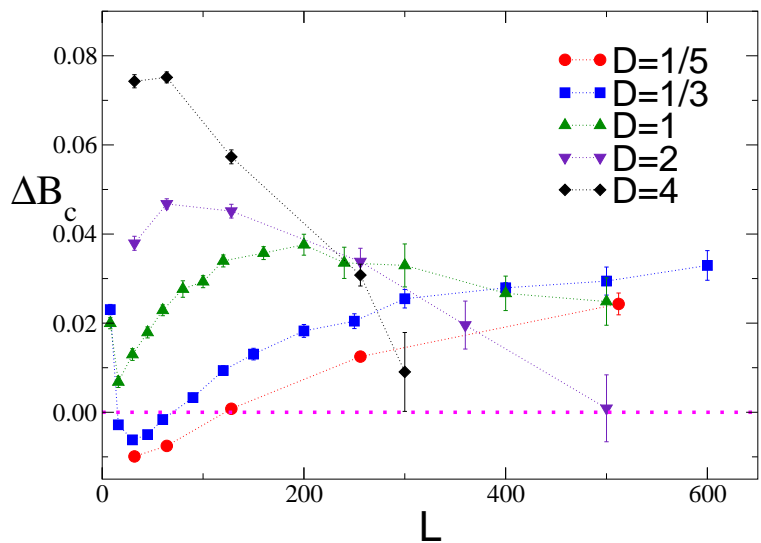}}
\caption{ 
$\Delta B_c\equiv B_c - B_{\rm Is}$ at fixed $R_c=R_{\rm Is}$ for 
several values of $D$ in  the $\phi^4$ model.
$B_{\rm Is} = 1.167923(5)$ is the value of the
Binder parameter at the critical point in the Ising model
\cite{SS-00}.
}
\label{bicphi4}
\end{figure}

\begin{figure}[tb]
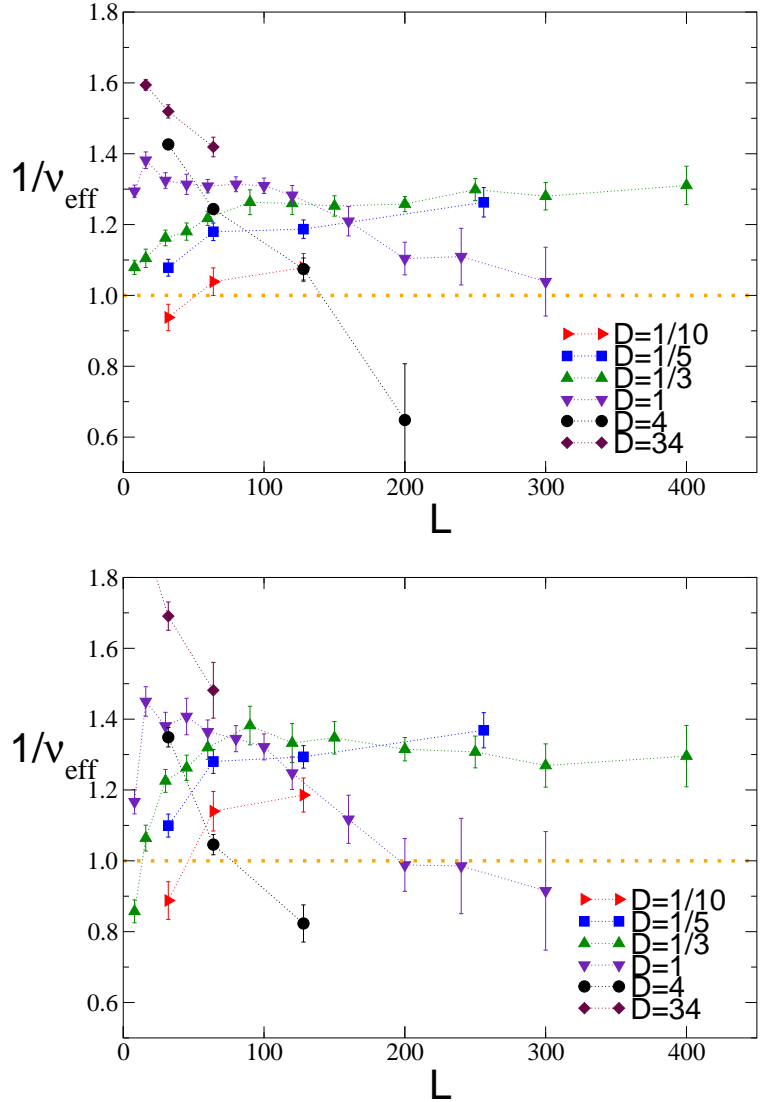

\centerline{\psfig{width=10truecm,angle=0,file=nueffphi4plain.eps}}
\vspace{4mm}
\centerline{\psfig{width=10truecm,angle=0,file=nueffphi4plainbi.eps}}
\caption{ 
$1/\nu_{\rm eff}$ for several values of $D$ in  the $\phi^4$ model,
from $R_c$ (above) and $B_c$ (below). 
}
\label{nueffphi4plain}
\end{figure}

\begin{figure}[tb]
\centerline{\psfig{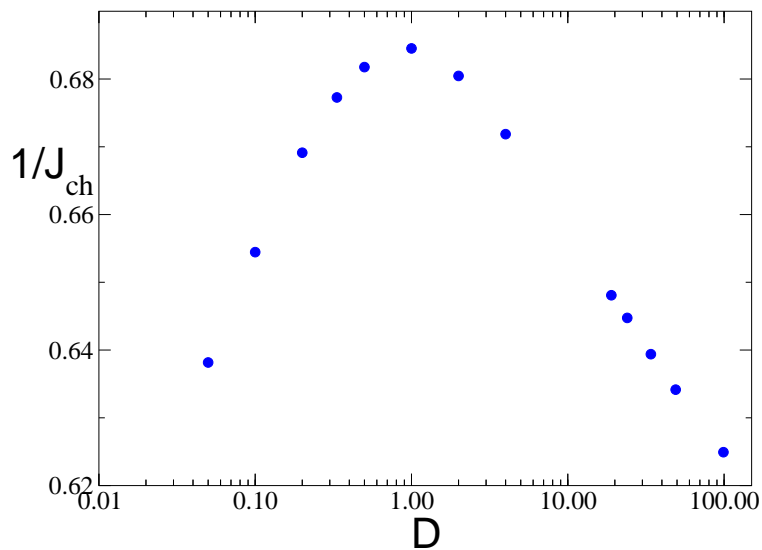}}
\caption{ 
Critical values of the hopping parameter at the
chiral transition for several values of $D$.
}
\label{bcdil}
\end{figure}

In Figs.~\ref{bicphi4} and \ref{nueffphi4plain} we show the chiral
Binder cumulant and the effective exponents $1/\nu_{\rm eff}$ derived
from $R_c$ and $B_c$ for other values of $D$.  For $0<D
\lesssim 2$ the behavior is similar to that observed for $D=1/2$ and
in the FFXY model.  In particular, also here $B_c$ and the exponent 
$\nu_{\rm eff}$ are respectively approximately equal to 1.19-1.20 and 0.8 in
a quite large range of values of $L$. For these values of $D$ we do 
not observe the approach to the Ising value. This is not 
surprising since, as we shall show in Sec.~\ref{crossover}, 
$\xi_s^{(c)}$ increases as $D\to 0$ and thus our data do not satisfy
the condition $L\gg \xi_s^{(c)}$.
For $D\gtrsim 2$ the
approach appears qualitatively different and, in particular, no
plateau at $\nu\approx {\rm 0.8}$ appears.  In Table~\ref{betaiskt},
see also Fig.~\ref{bcdil}, we give the values of $J_{\rm ch}$ for
several values of $D$.

\begin{table}
\caption{ 
\label{betaiskt}
Critical values of $J$ at the chiral and KT transitions,
as explained in the text.
}
\begin{indented}
\item[]
\begin{tabular}{@{}lllll}
\hline
\multicolumn{1}{c}{}&
\multicolumn{1}{c}{$J_{\rm ch}$}&
\multicolumn{1}{c}{$J_{{\rm sp}}$} &
\multicolumn{1}{c}{$J_{\rm sp,low}$ from  $R_s$}&
\multicolumn{1}{c}{$\delta$}\\
\hline 
$\phi^4$, $D=1/3$ &  1.4765(1) &           &  1.4792(1)  [$L=600$] & $\gtrsim  0.0018$ \\
$\phi^4$, $D=1/2$ &  1.4668(1) & 1.4704(2) &  1.46971(2) [$L=1024$]& 0.0025(2) \\
$\phi^4$, $D=1$   &  1.4609(1) &           &  1.4642(1)  [$L=400$] & $\gtrsim  0.0023$ \\
$\phi^4$, $D=2$   &  1.4696(1) &           &  1.4736(1)  [$L=400$] & $\gtrsim  0.0027$ \\ 
$\phi^4$, $D=4$   &  1.4884(1) &           &  1.4918(1)  [$L=300$] & $\gtrsim  0.0023$ \\
IsXY, $C=0$   &  1.4684(1) & 1.493(1) & 1.4888(2)  [$L=512$] & 0.0167(7)  \\
FFXY   &  2.20632(5) & 2.2415(5)  & 2.2348(4) [$L=1024$] &  0.0159(2) \\
\hline
\end{tabular}
\end{indented}
\end{table}

In conclusion, a careful FSS analysis on lattice sizes up to $L=1200$
shows that the behavior close to the chiral transition in the $\phi^4$
model with $D=1/2$ is similar to what has been observed in the FFXY
model. Also in this case the chiral transition should belong to the
Ising universality class, as expected within the two-transition
scenario. We expect that, as in the FFXY model, Ising behavior can be
clearly observed only when $L\gg\xi_s^{(c)}$.  Since,
$\xi_s^{(c)}\gtrsim 360$ in this case, much larger lattices would be
needed to obtain a result as reliable as that of the FFXY model. 
Similar results are obtained for other values of $D$. In particular, for
$0< D \lesssim 2$ we observe a rather extended region in which 
$\nu_{\rm eff} \approx 0.8$, as it was already observed in the FFXY model.
We will discuss again this issue in Sec.~\ref{crossover}.

\subsubsection{The spin transition.}
\label{spintrphi4}

In order to determine the critical hopping parameter $J_{\rm sp}$ of
the spin transition for $D=1/2$, we use the same method applied to the
FFXY model in Sec.~\ref{spintrffxy}. We assume that the transition
belongs to the KT universality class and exploit the predicted
asymptotic behavior of $\Upsilon$ and $R_s\equiv \xi_s/L$ at the KT
transition, cf. Eqs.~(\ref{Yxy}) and (\ref{rxy}). Effective estimates
are reported in Table~\ref{betaKT1o2}. They allow us to obtain the
final estimate $J_{\rm sp}=1.4704(2)$.  We also mention that the
chiral second-moment correlation length $\xi_c^{(s)}$ [which can be
obtained from the chiral connected correlation function (\ref{gcdef})]
is $\xi_c^{(s)} =  22(3)$ at the KT transition.  As expected, the
Ising and KT transitions are very close, indeed $\delta \equiv J_{\rm
sp}/J_{\rm ch} - 1 =0.0025(2)$.

For other values of $D$ we do not have enough data to perform the
previous analysis. Nonetheless, we are still able to obtain a lower
bound for $J_{\rm sp}$.  We assume only that, in the limit
$L\rightarrow \infty$, $R_s=0.7506912...$ and
$\Upsilon=0.63650817819...$ for $J=J_{\rm sp}$, and that $R$ and
$\Upsilon$ decrease with $L$, as suggested by Eqs.~(\ref{rxy}) and
(\ref{Yxy}) and confirmed by our data. Then, for each $L$, we
determine, using the results for $\xi_s/L$ and $\Upsilon$, the
value $J$ for which $R_s$ and $\Upsilon$ are equal to the theoretical
infinite-volume predictions. These results provide lower bounds on the
correct critical $J_{\rm sp}$. If we apply this procedure to the data
with $D=1/2$ and $L=1024$ we obtain $J_{\rm sp,low}= 1.46971(2)$ from
$\xi_s/L$, and $J_{\rm sp,low}=1.46957(2)$ from $\Upsilon$, that are
not far from the result obtained above, $J_{\rm sp} = 1.4704(2)$.  In
Table~\ref{betaiskt} we report lower bounds for other values of $D$
and estimates of the relative difference $\delta \equiv J_{\rm
sp}/J_{\rm ch} - 1$. In all cases $\delta$ is of the 
order of $10^{-3}$.

\begin{table}
\caption{ \label{betaKT1o2}
Critical values of $J$ at the KT transition for the 
$\phi^4$ model with $D=1/2$.}
\begin{indented}
\item[]
\begin{tabular}{@{}rrll}
\hline
\multicolumn{1}{c}{$L_1$}&
\multicolumn{1}{c}{$L_2$}&
\multicolumn{1}{c}{from $R_s$}&
\multicolumn{1}{c}{from $\Upsilon$}\\
\hline 
  128  &  256  & 1.4696(1)  &  1.4709(3)  \\
  256  &  512  & 1.4700(1)  &  1.4704(1)  \\
  512  & 1024  & 1.4705(2)  &  1.4704(1) \\
\hline
\end{tabular}
\end{indented}
\end{table}

\subsection{The phase diagram for  small values of $D$}
\label{o4mc}

Our numerical simulations provide evidence of two transitions for
$D=1/3,1/2,1,2,4$. In addition, relatively short MC simulations at
$D=1/20$, $1/10$, $1/5$, and $34$ appear consistent with the two-transition
scenario, although the evidence is less robust since our simulations
used smaller lattice sizes.  A natural conjecture is that the
two-transition scenario extends from $D\lesssim D^*$, $34 < D^*
\lesssim 49$, down to $D\to 0$.  For $D=0$ Hamiltonian ${\cal H}_\phi$
is O(4) symmetric.  According to the Mermin-Wagner theorem
\cite{MW-66}, it does not have any transition at finite temperature,
and criticality is observed only for $J\rightarrow \infty$. Therefore,
the Ising and KT transition lines should meet at $D=0$, $1/J=0$, which
is a multicritical point.

At a generic multicritical point with two even relevant parameters, 
such as the reduced temperature $t\equiv
T/T_{\rm mc}-1$ and a generic scaling field $g$ (where $T=T_{\rm mc}$
and $g=0$ is the position of the multicritical point), the singular
part of the free energy is given by \cite{KNF-76,LS-84}
\begin{equation}
{\cal F}_{\rm sing} \sim t^{2-\alpha} B(X), \qquad X=gt^{-\phi},
\label{scalmc}
\end{equation}
where $\alpha$ and $\phi$ are the specific-heat and crossover
exponents. This expression is not very useful in our case, since
$T_{\rm mc} \equiv 1/J_{\rm mc} = 0$ and $\xi\sim e^{cJ}$.
However, it is easy to rewrite 
Eq.~(\ref{scalmc}) in a different form that can be easily generalized.
If $\xi(t,g=0)$ is the correlation length for $g = 0$, and $\nu$ is 
the corresponding critical exponent so that $\xi(t,g=0)\sim t^{-\nu}$,
we can write 
\begin{equation}
{\cal F}_{\rm sing} \sim  \xi(t,g=0)^d A(Y), \qquad Y=g\xi(t,g=0)^{\rho}, 
\label{scalmc2}
\end{equation}
where $\rho\equiv\phi/\nu$. In this expression there is no explicit
$t$ dependence and thus we can use it for $D\to 0$.

In the theory described by ${\cal H}_\phi$, $g$ is essentially the
parameter $D$.  The crossover exponent $\rho$ is related to the RG
dimension of the perturbation $\sum_x \phi_{1,x}^2 \phi_{2,x}^2$ at the
O(4) fixed point.  In two dimensions, $\rho$ is equal to the engineering
dimension of the operator, so that $\rho = 2$, 
with logarithmic corrections.  The 
perturbative analysis is analogous to that presented in Ref.~\cite{CMP-01},
in which the effect of a spin-$n$ perturbation in the 3-vector model
is discussed. The correct scaling variable is $Y \equiv D \xi^2 (\ln \xi)^{-6}$,
or, using the fact that $\xi \sim J^{-1/2} e^{cJ}$, 
\begin{equation}
\hat{Y} \equiv D (c J)^{-7} \exp (2 c J).
\end{equation}
The power appearing in the logarithm, which is universal, has been determined
by using the perturbative results of Ref.~\cite{CP-94}, which gives 
the anomalous dimension of any dimension-zero spin-$n$ perturbation of 
the $N$-vector model. In the limit $U\to \infty$, in which the $\phi^4$
model reduces to the 4-vector model with a spin-4 perturbation, we also
have $c = \pi$.

A standard scaling assumption in multicritical theories~\cite{KNF-76,LS-84} is
that, along the critical lines meeting at the multicritical point, the
scaling variables $Y$ and $\hat{Y}$ are asymptotically constant. This
implies that the Ising and KT critical lines, corresponding to 
$J = J_{\rm ch}(D)$ and $J = J_{\rm sp}(D)$,
approach the O(4) multicritical point according to
\begin{equation}
D (c J_{\rm ch,sp})^{-7} \exp (2 c J_{\rm ch,sp}) = \hat{Y}_{\rm ch,sp},
\label{scalcrli}
\end{equation}
where $\hat{Y}_{\rm ch}$ and $\hat{Y}_{\rm sp}$ are two different constants,
with $\hat{Y}_{\rm sp}>\hat{Y}_{\rm ch}$. It follows
\begin{equation}
c J_{\rm ch,sp}\approx {1\over 2} \ln (\hat{Y}_{\rm ch,sp}/D) + 
            {7\over2} \ln \left[{1\over 2} \ln (\hat{Y}_{\rm ch,sp}/D)\right], 
\label{Eq43}
\end{equation}
and 
\begin{equation}
\delta \equiv  {J_{\rm sp}\over J_{\rm ch}}-1
\sim  {\ln (\hat{Y}_{\rm sp}/\hat{Y}_{\rm ch})\over 
       \ln \hat{Y}_{\rm ch}/D}\; . 
\end{equation}
We finally mention that for $D<0$ we expect
only one KT transition line starting from $1/J=0, D=0$, where both
fields $\phi_{i,x}$ become critical simultaneously.
The same RG arguments used above tell us that
$J_{{\rm sp},D<0} \sim \ln (1/|D|)$.

Unfortunately, we are not able to check numerically these predictions.
Indeed, this requires to work in a regime in which $J_{\rm ch}$ and
$J_{\rm sp}$ are large and this happens only for very tiny values of
$D$. But for very small values of $D$ very
large lattices are needed in order to observe the asymptotic behavior.  For
$D\lesssim 1$, the values of $J_{\rm ch}$ increase with decreasing
$D$, see Fig.~\ref{bcdil}, but $D\approx 0.1$ seems still far from the
asymptotic regime in which Eq.~(\ref{Eq43}) holds. 
Note that the presence of the O(4) point influences the values 
of $\nu_{\rm eff}$ for the lowest values of $D$.
Indeed, see Fig.~\ref{nueffphi4plain},
$\nu_{\rm eff}$ is larger than one for small values of $L$. 
This is due to the influence of the O(4)
multicritical point, for which $1/\nu=0$.

Finally, we would like to point out that an O(4) zero-temperature
multicritical behavior is also expected in the 
antiferromagnetic XXZ model on a triangular lattice.
Its Hamiltonian is 
\begin{equation}
{\cal H}_{\rm XXZ} = J \sum_{\langle xy\rangle} s^{1}_x s^{1}_y  + s^{2}_x s^{2}_y 
+ A s^{3}_x s^{3}_y ,
\label{anisH}
\end{equation}
where $\vec{s}_x$ is a three-component vector satisfying $\vec{s}_x
\cdot \vec{s}_x=1$.  In the easy-axis regime $A<1$, the ground-state
structure is analogous to that of the FFXY model \cite{SH-92,CVCT-98},
and therefore two close Ising and KT transitions are expected.  For
$A=1$ one recovers the isotropic Heisenberg antiferromagnetic model,
which is not expected to present any finite-temperature transition,
but it should become critical only for $J\rightarrow \infty$, see,
e.g., Refs.~\cite{WEA-95,CADM-01}.  Due to the effective enlargement of the
symmetry from O(3)$\otimes$O(2) to O(4) \cite{ADM-92}, the asymptotic
behavior for $J\rightarrow\infty$ is expected to be the same as that
of the O(4) vector model.

\section{Phase transitions in the Ising-XY model}
\label{isingxy}

\begin{figure}[tb]
\centerline{\psfig{width=12truecm,angle=0,file=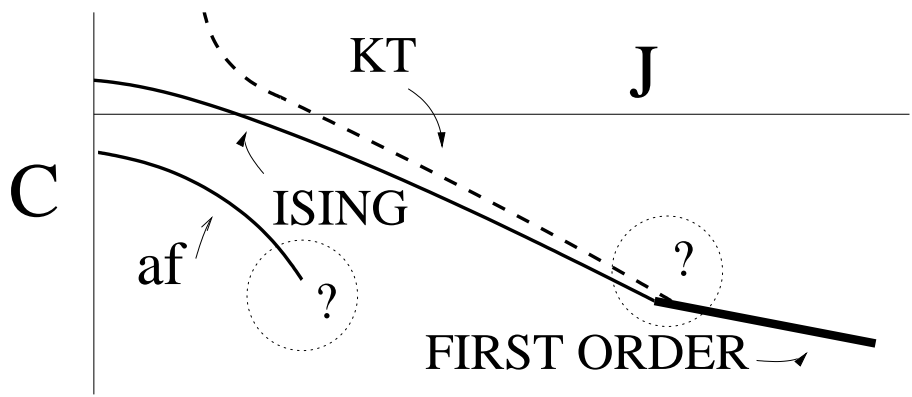}}
\vspace{2mm}
\caption{ 
Sketch of the phase diagram of the Ising-XY model.  The
continuous, dashed, and thick continuous lines represent Ising, KT,
and first-order transition lines.  The distance between the Ising and
KT lines is amplified to distinguish them.  The phase diagram within
the two circled regions is unknown.  
}
\label{phasedisxy}
\end{figure}

In this section we discuss the critical behavior of the IsXY model
(\ref{IsXYp}).  Some features of its phase diagram can be easily
determined, see Fig.~\ref{phasedisxy}.  At $J=0$ we have an Ising
transition at $C= C_{\rm Is}={1\over2} \log(1 + \sqrt{2}) =
0.4406868\ldots$.  An Ising transition line in the $J$-$C$ plane
starts from this point. For $C=\infty$ there is a KT transition at
\cite{HP-97,Has-05} $J\approx1.1199$, from which a KT line starts.  As we shall
see, with decreasing $C<C_{\rm Is}$ the KT and Ising transition lines
get closer and closer.  Another Ising transition occurs at $J=0$ and
$C=-C_{\rm Is}$, from which an Ising transition line should start and
lie in the region $C<-C_{\rm Is}$, bounding an antiferromagnetic
phase.  We have not investigated the behavior of the IsXY model along
this antiferromagnetic transition line, although it is likely (but we
have not numerically checked) that the second-order transition line
turns into a first-order one as $J$ increases, as it happens for the
ferromagnetic case that we are going to discuss below.

The phase diagram of the IsXY model was already studied in
Refs.~\cite{GKLN-91,LGK-91,NGK-95}. They found an Ising and a KT
transition for $C_{\rm Is}>C\gtrsim 0.1$ and a single first-order
transition for $C\lesssim -4$. These works also claimed the existence
of unique continuous transitions belonging to new universality
classes for $-4\lesssim C \lesssim 0.1$. Along this line critical
exponents were found to depend on $C$.  As we shall see, our results
do not support the existence of a unique continuous transition in
which spin and chiral modes become critical at the same $J$. We find
two close KT and Ising continuous transitions for $C\lesssim -5$, and
a single first-order transition beyond, in agreement with the argument
reported in Sec.~\ref{models}, which forbids a continuous transition
where chiral and spin modes become critical.

\begin{table}
\caption{ \label{resisxy}
Estimates of several quantities computed at fixed $R_c=R_{\rm Is}$
in the IsXY model for $C=0$.}
\hspace{-3mm}
\footnotesize
\begin{tabular}{@{}rllllllll}
\hline
\multicolumn{1}{c}{$L$}&
\multicolumn{1}{c}{$J_{R_c}$}&
\multicolumn{1}{c}{$\chi_c$}&
\multicolumn{1}{c}{$B_c$}&
\multicolumn{1}{c}{$d R_c/dJ$}&
\multicolumn{1}{c}{$-d B_c/dJ$}&
\multicolumn{1}{c}{$\xi_s$}&
\multicolumn{1}{c}{$dR_s/dJ$}&
\multicolumn{1}{c}{$\Upsilon$}\\
\hline 
6 & 1.4339(7)   & 23.6(2) & 1.1788(6) & 4.74(2) & 1.956(13) & 4.143(4) 
& 1.973(10) & 0.7886(11) \\
  
8 & 1.4415(4)   & 38.83(3) & 1.1842(5) & 6.51(2) & 2.822(16) & 5.371(5) 
& 2.638(10) & 0.7264(9) \\
  
12 & 1.4498(3)  & 78.02(5) & 1.1904(5) & 10.37(4) & 4.72(3) & 7.680(6) 
& 3.98(2) & 0.6394(9) \\
  
16 & 1.4538(2)  & 127.68(12) & 1.1958(7) & 14.73(7) & 6.94(6) & 9.816(14) 
& 5.42(3) & 0.5780(13) \\
  
24 & 1.4589(2)  & 257.9(3) & 1.2023(7) & 24.49(12) & 11.93(10) & 13.85(2) 
& 8.49(5) & 0.4978(16) \\
  
32 & 1.46165(7) & 426.7(4) & 1.2043(8) & 35.1(2) & 17.13(14) & 17.58(2) 
& 11.59(7) & 0.4427(16) \\
  
48 & 1.46445(4) & 868.8(1.0) & 1.2061(8) & 58.0(3) & 28.1(2) & 24.25(4) 
& 17.71(8) & 0.360(2) \\
  
64 & 1.46586(4) & 1441.7(1.6) & 1.2062(7) & 82.1(4) & 39.2(3) & 29.98(5) 
& 23.38(12) & 0.303(2) \\
  
90 & 1.46707(3) & 2627(3) & 1.2029(8) & 121.1(6) & 55.6(5) & 37.55(7) 
& 31.3(2) & 0.223(2) \\
  
128 & 1.46782(3) & 4865(7) & 1.1957(7) & 175.6(1.0) & 76.2(8) & 45.11(10) 
& 38.9(3) & 0.137(2) \\
  
180 & 1.46823(3) & 8855(19) & 1.1843(8) & 240(2) & 97.4(1.2) & 50.29(17) 
& 41.1(4) & 0.069(2) \\
  
256 & 1.46834(3) & 16247(50) & 1.1773(13) & 328(4) & 131(3) & 52.5(3) 
& 38.7(7) & 0.018(3) \\
  
360 & 1.46838(3) & 29466(133) & 1.1724(15) & 463(8) & 177(5) & 52.7(4) 
& 31.8(9) & 0.005(4)\\
\hline
\end{tabular}
\end{table}

\begin{figure}[tb]
\centerline{\psfig{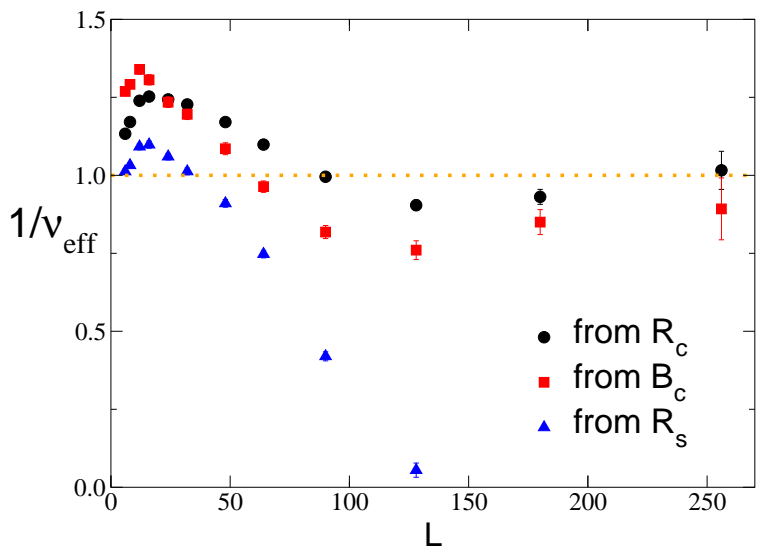}}
\caption{Effective exponents $1/\nu_{\rm eff}$ as obtained
from the derivatives of $R_c$, $B_c$, and $R_s$, for
the IsXY model at $C=0$.  The dotted line corresponds to 
the Ising value $\nu=1$.
}
\label{nueffisxyc0}
\end{figure}

We perform the same FSS analysis as described in Sec.~\ref{fssffxy}.
We first discuss in some detail the results for $C=0$.  In
Table~\ref{resisxy} we report data for $C=0$ up to $L=360$, which turn
out to be sufficient to infer the asymptotic critical behavior.  As
also shown in Fig.~\ref{rs}, the spin correlation length $\xi_s^{(c)}$
at the chiral transition clearly converges to a finite value: we find
$\xi_s^{(c)}= 52.7(4)$.  Correspondingly, the helicity modulus
vanishes in the large-$L$ limit.  Again, the best evidence of an
asymptotic Ising critical behavior is obtained from the analysis of
the chiral Binder cumulant $B_c$.  Fig.~\ref{bcffxy} shows the
difference $\Delta B_c\equiv B_c-B_{\rm Is}$ that clearly converges to
the Ising value as $L$ increases.  A linear fit to $a + b L^{-7/4}$ of
the data for the three largest lattices gives $B_c=1.1679(18)$,
to be compared with \cite{SS-00} $B_{\rm Is}=1.167923(5)$.
A linear fit to $b L^{-7/4}$ of $\Delta B_c$ for
the three largest lattices gives $b=146(6)$ with $\chi^2/{\rm
d.o.f}\approx 0.1$.  The critical coupling $J_{R_c}$ shows the
expected rate of convergence. A linear fit to $a+b L^{-11/4}$ of the
results for the largest lattices gives $J_{\rm ch}=1.4684(1)$.  The
effective exponents $\nu_{\rm eff}$ obtained from the derivative of
$R_c$, $B_c$, $R_s$, see Fig.~\ref{nueffisxyc0}, and the exponent
$\eta_{\rm eff}$ obtained from $\chi_c$ behave as observed in the FFXY
model and in the $\phi^4$ model in the region $0< D\lesssim 2$,
see also Sec.~\ref{crossover}.  The values of $1/\nu_{\rm eff}$
obtained from $R_c$ and $B_c$ are first larger than the Ising value,
then become significantly smaller, and eventually converge to 1 from
below.

Estimates of the critical value of $J$ at the KT transition have been
obtained following the procedure described in Sec.~\ref{spintrffxy},
using the estimate of $R_s$ and $\Upsilon$ with $L\le 512$.  We obtain
$J_{\rm sp}=1.493(1)$, see also Table~\ref{betaiskt}.  The chiral
correlation length $\xi_c^{(s)}$ at the KT transition is
$\xi_c^{(s)}= 4.3(5)$.  Again, the relative difference $\delta\equiv
J_{\rm sp}/J_{\rm ch}-1$ of the critical couplings is very small,
i.e. $\delta = 0.0167(7)$.

\begin{figure}[tb]
\centerline{\psfig{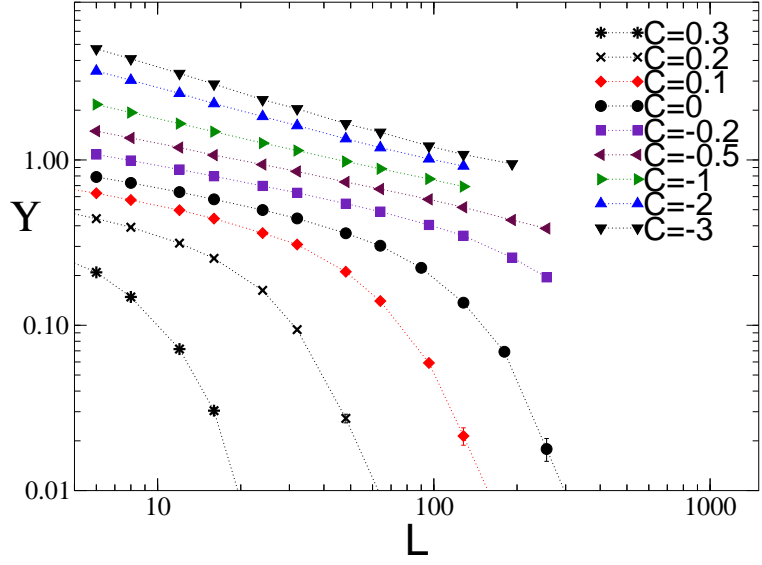}}
\caption{ 
Helicity modulus at fixed $R_c$ for several values of $C$ in the IsXY
model. 
}
\label{yupsilonisxyplain}
\end{figure}

\begin{figure}[tb]
\centerline{\psfig{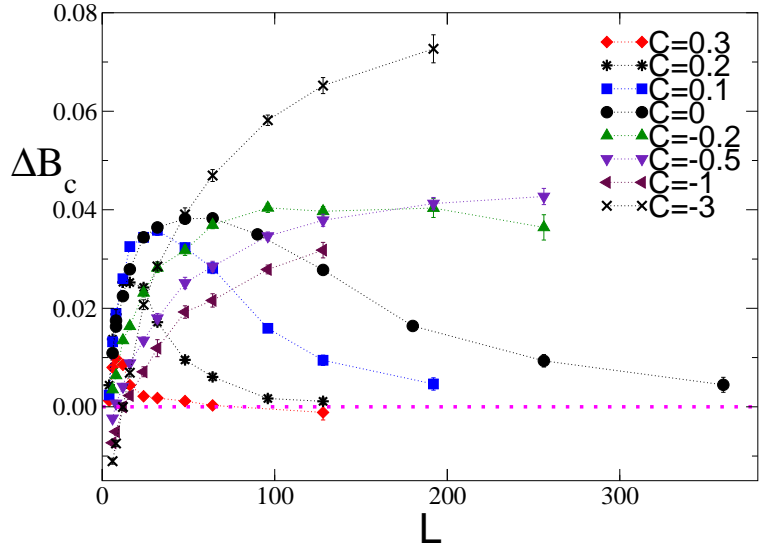}}
\caption{ 
$\Delta B_c\equiv B_c - B_{\rm Is}$ at fixed $R_c=R_{\rm Is}$ for
the IsXY at  several values of $C$.
$B_{\rm Is} = 1.167923(5)$ is the value of the
Binder parameter at the critical point in the Ising model
\cite{SS-00}.
}
\label{bicisxy}
\end{figure}

\begin{figure}[tb]
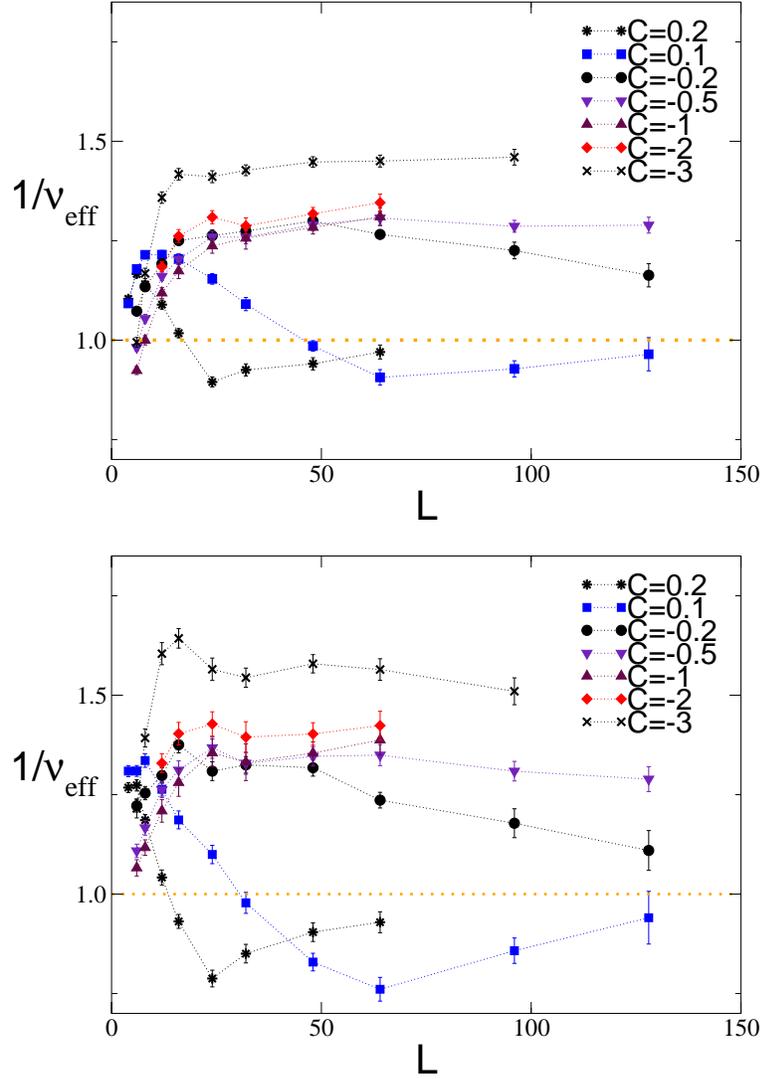

\centerline{\psfig{width=10truecm,angle=0,file=nueffisxyplain.eps}}
\vspace{4mm}
\centerline{\psfig{width=10truecm,angle=0,file=nueffisxyplainbi.eps}}
\caption{Effective exponent 
$1/\nu_{\rm eff}$ for several values of $C$ in the IsXY model,
from $R_c$ (above) and $B_c$ (below). 
}
\label{nueffisxyplain}
\end{figure}

The results for the other values of $C$ are similar.  For example, in
Figs.~\ref{yupsilonisxyplain}, \ref{bicisxy}, and \ref{nueffisxyplain}
we show the helicity modulus $\Upsilon$, the chiral Binder parameter
$B_c$, and the effective exponent $1/\nu_{\rm eff}$ for several values
of $C$.  Precise estimates of the spin correlation length at the
chiral transition are obtained for $C=0.1,0.2,0.3$:
$\xi_s^{(c)}=26.6(2),10.8(1),3.70(2)$, respectively.  As expected
$\xi_s^{(c)}$ rapidly decreases with increasing $C$, indicating that
the difference between the Ising and KT critical temperature becomes
larger and larger up to $C=C_{\rm Is}$.  We also performed simulations
for negative values of $C$, $C=-0.2,-0.5$ for lattice sizes up to
$L=256$, $C=-1,-2$ to $L=128$, and $C=-3$ to $L=192$.  The estimates
of $R_s$ and $\Upsilon$ appear to decrease as in the $C=0$ case,
suggesting that the spin modes are not critical at the chiral
transition, thus favoring the two-transition picture.  With decreasing
$C$, $\xi_s^{(c)}$ apparently increases, but we have not been able to
directly observe that $\xi_s^{(c)}$ is finite on the lattice sizes
that we considered.  However, our results are not consistent with a
unique transition. As we shall see in the next section, the results for $C\ge
-1$ can be explained in terms of an approximately universal crossover
within the standard two-transition picture.  For $C=-2$ and $C=-3$ 
the estimates of RG quantities
(see, e.g., Fig.~\ref{nueffisxyplain} for $\nu_{\rm eff}$) 
are apparently stable with $L$.
For instance, the analysis of the derivatives of $R_c$, $B_c$, $R_s$
with respect to $J$ for $C = -2$ up to $L=128$ gives $1/\nu\approx
1.31,1.40,1.10$. Each estimate appears quite stable with respect to
$L$ and thus it could be taken as evidence that the asymptotic limit
has been reached and that the critical chiral behavior 
is not Ising but belongs to a new universality class. 
This interpretation is however not possible, since, if 
the transition is unique, the estimate of $\nu$ should be independent
of the observable. Analogously, for $C = -3$ up to $L=192$ we obtain
$1/\nu\approx 1.42,1.57,1.23$ from $R_c$, $B_c$, $R_s$, which are
again not consistent with the existence of a new universality class.
Taking also into account the argument reported in
Sec.~\ref{models}, which forbids continuous transitions where chiral
and spin modes become critical, we conclude that these results are
better explained in terms of crossover effects.

The FSS behavior of the helicity modulus $\Upsilon$ at the chiral
transition and for lattice sizes $L\ll \xi_s^{(c)}$ can be related to
the fractal dimension of the geometrical Ising clusters.  As already
discussed in Sec.~\ref{models}, at the chiral transition spins
$\vec{s}$ effectively live on geometrical Ising clusters. 
If $L\ll \xi_s^{(c)}$, XY spins are strongly correlated and thus
we can use the spin-wave approximation, which leads to
the prediction
\begin{equation}
\Upsilon \propto L^{-\epsilon}
\label{yfss}
\end{equation}
where $\epsilon = 2-d_{\rm gc}>0$ and $d_{\rm gc}$ is the effective
fractal dimension of the geometrical clusters in this range of lattice
sizes.  In the case chiral and spin modes are effectively decoupled,
and therefore the properties of the geometrical clusters are not
affected by the spin modes, one expects $d_{\rm gc}$ to be equal to
the fractal dimension of the geometrical Ising clusters which is predicted
to be $d_{\rm Igc}=187/96=1.9479...$ \cite{SV-89,JS-05}, corresponding
to $\epsilon=5/96=0.0521...$. For $L \lesssim \xi_s^{(c)}$
the data shown in Fig.~\ref{yupsilonisxyplain} show a power-law behavior as in
Eq.~(\ref{yfss}), but with a
rather different exponent. For example, we find $\epsilon \approx 0.35$
for $0.2\gtrsim C\gtrsim -1$.  This is due to the interaction between
chiral and spin modes, which changes the effective
fractal dimension of the geometrical clusters when $L\lesssim
\xi_s^{(c)}$. 
Note that the power-law decay of the helicity modulus
is another piece of evidence in favor of the argument presented in 
Sec.~\ref{models}, which excludes a unique second-order transitions 
for any $C$, and therefore also for $C=-2,-3$. 

For larger values of $-C$, first-order transitions separate the LT
from the disordered phase. Our simulations indicate the presence of a
unique first-order transition for $C\lesssim -5$.  In particular, we
obtain the estimates $\sigma\approx 0.012,\, 0.006$ for the interface
tension at $C=-8,\,-7$, respectively. Moreover, for $C\le -7$, the absence of a
continuous KT transition is rather clear from the analyses of the spin
variables.  On the LT side of the transition, which is at
$J_c=15.4255(5)$ for $C=-7$, we definitely find $\eta < 1/4$:
$\eta\approx 0.035$ at $J=15.426$, and $\eta\approx 0.043$ at
$J=15.425$ in the metastable LT phase.  The presence of first-order
transitions for $C\lesssim -5$ may also explain the results for the
effective exponent $1/\nu_{\rm eff}$ at $C=-2,-3$. The apparent
increase of $1/\nu_{\rm eff}$ as $|C|$ increases on small lattices can
be interpreted as due to the presence of a nearby first-order
transition.  

In conclusion, the phase diagram of the IsXY model is characterized by
an Ising and a KT transition for $C_{\rm Is}> C \gtrsim -5$,
and by a unique first-order transition
for $C< C^*$ with $-5\gtrsim C^*\gtrsim -7$. A sketch of the phase
diagram is shown in Fig.~\ref{phasedisxy}.  Again we do not know how
the two continuous transition lines connect to the first-order
transition line. The discussion at the beginning of Sec.~\ref{phasetr}
applies also in this case, with the only difference that a unique
continuous transition is excluded by the argument reported in
Sec.~\ref{models}.  A possible scenario, consistent with our numerical
results, is shown in Fig.~\ref{phasedisxy}.  The crossover curves of
the effective exponents are quite similar to those of the $\phi^4$
model, at least for $C$ not too negative, as we will further discuss
in Sec.~\ref{crossover}.

\section{Crossover behavior} 
\label{crossover}

\begin{figure}[tb]
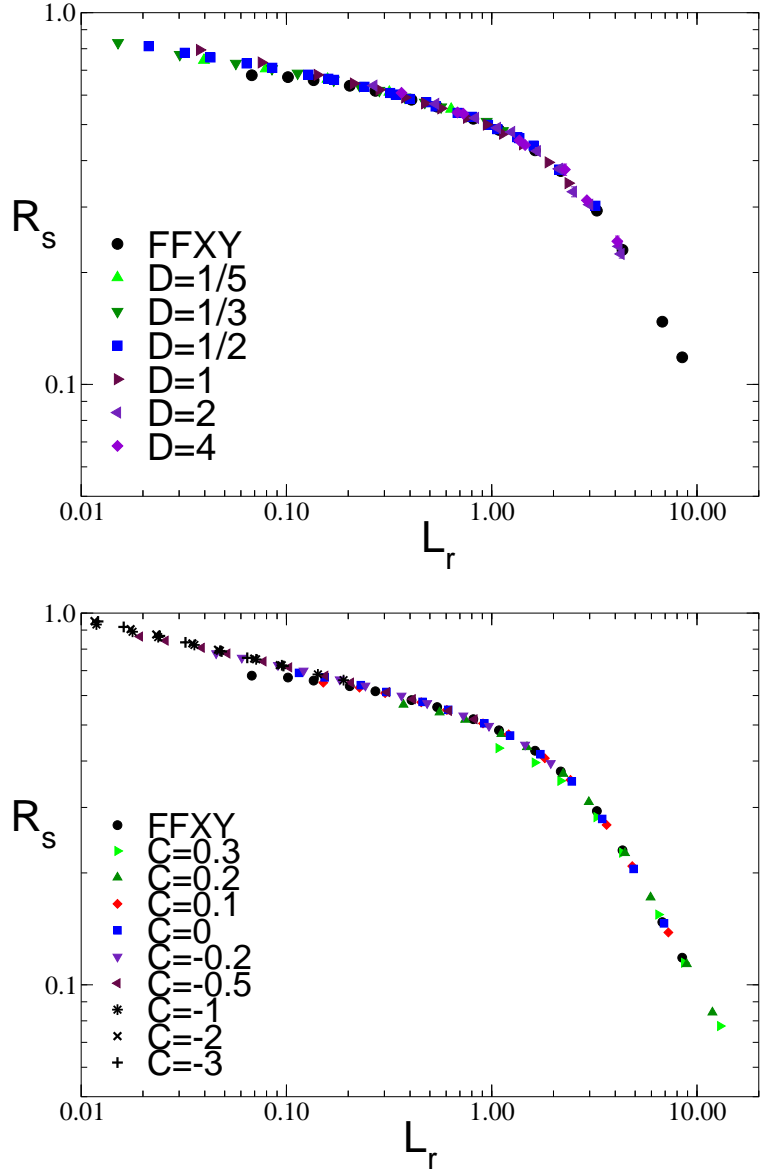

\centerline{\psfig{width=10truecm,angle=0,file=xillog.eps}}
\vspace{4mm}
\centerline{\psfig{width=10truecm,angle=0,file=xilisxylog.eps}}
\caption{ 
Correlation-length ratio $R_s$ at fixed $R_c$ vs $L_r = L/l$
for the FFXY model and $\phi^4$ (above) and IsXY (below) models. 
Note the logarithmic scale on both axes.
We set $l_{\rm FFXY}=118 \approx \xi_s^{(c)}$.
}
\label{xil}
\end{figure}

In the preceding sections we established that the FFXY
model and the $\phi^4$ and IsXY models in a large parameter region
undergo two transitions that belong to the Ising and KT universality
classes respectively.  However, the eventual Ising critical regime at
the chiral transition is reached after a crossover regime which is
surprisingly similar in all models considered. For example, in most
cases we find a rather extended preasymptotic region in which the
effective correlation-length exponent $\nu_{\rm eff}$ is approximately
equal to 0.8.  Similar values have also been observed in other models
related to the FFXY model, see Table~\ref{summaryliterature}. This
suggests that this crossover is somewhat universal.  

\begin{figure}[tb]
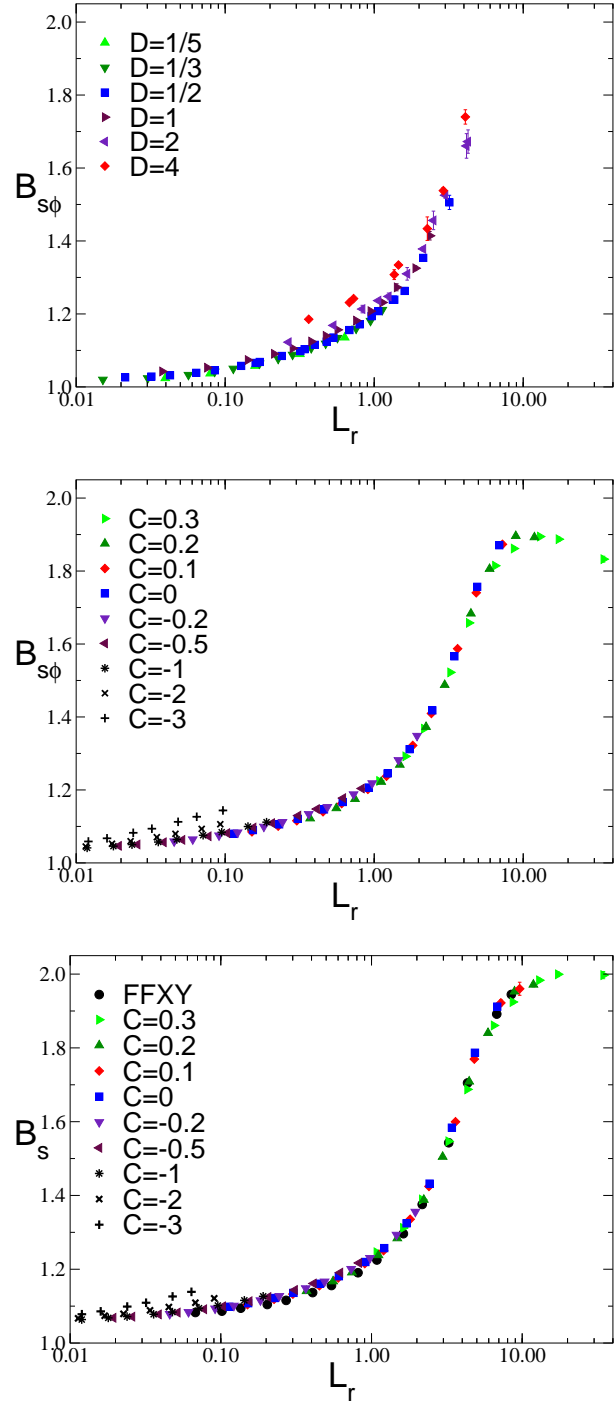

\centerline{\psfig{width=8truecm,angle=0,file=bi1.eps}}
\vspace{4mm}
\centerline{\psfig{width=8truecm,angle=0,file=bi2.eps}}
\vspace{4mm}
\centerline{\psfig{width=8truecm,angle=0,file=bi3.eps}}
\caption{ 
Binder parameters $B_{s\phi}$ (top and middle) and $B_s$ (bottom)
at fixed $R_c$, versus $L_r = L/l$,
for the FFXY, $\phi^4$, and IsXY models. 
The rescalings $l$ are the same as in Fig.~\protect\ref{xil}.  
For $L_r\to \infty$, $B_s \to 2$ and $B_{s\phi}\to 3/2$.
}
\label{bi}
\end{figure}

First, we consider the behavior at the chiral transition and show
that any RG quantity scales approximately as 
\begin{equation}
{\cal R} = f_{\cal R} (L/l),
\label{scaling-crossover}
\end{equation}
 where $l$ is a model-dependent, but observable-independent
length scale.\footnote{
Note that our results have been obtained at fixed $R_c$ and not at the
chiral critical point. However, since we are dealing with an Ising
transition and $R_c$ has been fixed to the critical-point Ising
value, this is irrelevant in the scaling limit. Had we fixed $R_c$
to a different value, we would have still observed 
a universal crossover behavior, though the 
scaling curves would have been quantitatively different. }
To verify Eq.~(\ref{scaling-crossover}), we consider $R_s$,
determining the factors $l$ so as to obtain the best collapse
of the data. The results are reported in Fig.~\ref{xil}.
The data fall on a single curve with remarkable precision.
The optimal
data collapse is obtained for $l/l_{\rm FFXY} \approx 7.0, \, 4.5,\,
3.2,\,1.8,\, 1.0,\, 0.75$ respectively for
$D=1/5,\,1/3,\,1/2,\,1,\,2,\,4$, and $l/l_{\rm
FFXY}=0.031,\,0.089,\,0.22,\,0.45,\,1.11,\,2.6,\,5.7,\, 12,\, 17$ for
$C=0.3,\,0.2,\,0.1,\,0,\,-0.2,\,-0.5,\,-1, \,-2,\,-3$ in the case of
the IsXY model. 

In order to verify the universality of the scaling Ansatz 
(\ref{scaling-crossover})
we have considered the spin Binder parameters $B_s$ and $B_{s\phi}$.
In Fig.~\ref{bi} we plot these quantities in terms of $L_r = L/l$,
using the rescaling factors determined in the analysis of $R_s$.
Again we observe a good universal behavior.\footnote{The Binder 
parameter
$B_{s\phi}$ is expected to converge to 3/2 for $L_r\to \infty$
(a simple argument predicts: $B_{s\phi} = 3/2 + a L_r^{-1/4}$, $a > 0$).
Nonetheless, for $L_r \approx 10$, such a quantity is still 
quite different from its asymptotic value.}
The quality of the data collapse worsens as $D$
or $-C$ increases.  Indeed, while for $R_s$ all data collapse on a
single curve, except for a few FFXY points that correspond to small
values of $L$ ($L\lesssim 10$), the data of $B_{s\phi}$ for $D=4$ ($\phi^4$
model) and $C = -2,-3$ (IsXY model) show some systematic
deviations. They apparently require a rescaling $l$ that differs 
by approximately a factor of 2 for $D=4$ and by a
factor of 5 for $C=-3$ from that determined by using $R_s$.
Note that these deviations are not due to small values of 
$L$ and/or of $\xi_{s}^{(c)}$, since 
$\xi_{s}^{(c)}\approx 88$ for the $\phi^4$ model at $D=4$, and
$\xi_s^{(c)}\gtrsim 200$ for the IsXY model at $C=-2,-3$.  
Most probably, the behavior is
influenced by the first-order transition line that is present nearby in the
phase diagram.

It is easy to realize that, given two different models, $l_1/l_2$
corresponds to the ratio of the corresponding spin correlation lengths
at the chiral transition.  Indeed, since $R_s = f_{R_s}(L/l)$ and
$\xi_s\to \xi_s^{(c)}$ for large $L$, we must have $f_{R_s}(x) = a/x$
for $x\to\infty$, where $a$ is model independent, and $\xi_s^{(c)} =
l/a$. Thus, the scaling we observe implies that $R_s$ is an
approximately universal function of $L/\xi_s^{(c)}$.  Assuming
universality, this allows us to estimate the spin correlation length
$\xi_s^{(c)}$ at the chiral transition even in those cases in which
our simulations did not probe the regime $L\gg \xi_s^{(c)}$ that
allows us to obtain a direct estimate of $\xi_s^{(c)}$.  For example,
using the above-reported estimates of the optimal ratio $l/l_{\rm
FFXY}$ and $\xi_s^{(c)} = 118(1)$ for the FFXY model, we infer that
$\xi_s^{(c)}\approx 830, 530, 380, 212, 118$ respectively for
$D=1/5,1/3,1/2,1,2$ ($\phi^4$ model) and $\xi_s^{(c)}\approx 131, 310,
670$ respectively for $C=-0.2,-0.5,-1$ (IsXY model).  The result for
the $\phi^4$ model at $D=2$ is consistent with the estimate
$\xi_s^{(c)}=116(3)$ obtained for $L=500$.
In Fig.~\ref{xil} and in those we shall present below we have fixed
$l_{\rm FFXY} = 118 \approx \xi_s^{(c)}$, so that we are reporting the
data in terms of $L_r \equiv L/\xi_s^{(c)}$.

\begin{figure}[tb]
\centerline{\psfig{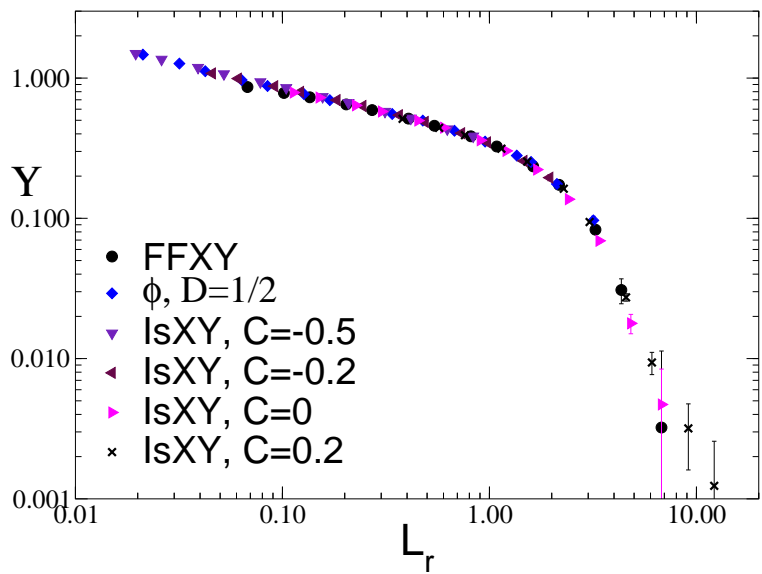}}
\caption{ 
Helicity modulus $\Upsilon$ at fixed $R_c$,
versus $L_r=L/l$.
The rescalings $l$ are the same as in Fig.~\protect\ref{xil}.  
Note the logarithmic scale on both axes.
}
\label{yupsilon}
\end{figure}

In Fig.~\ref{yupsilon} we do the same analysis for the helicity
modulus $\Upsilon$. Again, the collapse of the data is remarkable.  In
a quite large region, i.e. for $L_r\lesssim 0.5$, $\Upsilon$ decreases
as $L_r^{-\epsilon}$ with $\epsilon\approx 0.33$,
with differences of at most 10\%.\footnote{
More precisely, fitting the available data for $L\gtrsim
10$ and $L_r\lesssim 0.5$ to $\Upsilon=c L_r^{-\epsilon}$, we find
$\epsilon=0.335(8)$ ($\chi^2/{\rm d.o.f.}\approx 1.1$) for the FFXY model,
$\epsilon=0.332(2)$ ($\chi^2/{\rm d.o.f.}\approx 0.5$) for the $\phi^4$
model with $D=1/2$, $\epsilon=0.297(4)$ ($\chi^2/{\rm d.o.f.}\approx 0.5$)
for the $\phi^4$ model with $D=1/5$, $\epsilon=0.360(5)$
($\chi^2/{\rm d.o.f.}\approx 1.1$) for the IsXY model with $C=0$,
$\epsilon=0.352(3)$ ($\chi^2/{\rm d.o.f.}\approx 0.7$) for the IsXY model
with $C=-0.5$ (d.o.f. is the numer of degrees of freedom of the fit). 
The small differences in the estimates of $\epsilon$ can be easily
accounted for by scaling corrections.} 
As already discussed in Sec.~\ref{isingxy}, a power-law decay is expected in
the IsXY model for $L\ll \xi_s^{(c)}$ and gives  the 
effective fractal dimension of the geometrical clusters,
$d_{\rm gc}=2-\epsilon$.  The universal behavior
observed in Fig.~\ref{yupsilon} shows that a power-law decay also
characterizes the FFXY and $\phi^4$ models.  It is interesting to note
that the value $\epsilon\approx 0.33$ is much larger than the value
$\epsilon=5/96$ corresponding to the fractal dimension $d_{\rm
Igc}=187/96$ of Ising geometrical clusters: in the IsXY model, this is
the dimension of the geometrical clusters only for $L\gg \xi_s^{(c)}$.
This shows, once again, that chiral and spin modes are strongly
coupled for $L\lesssim \xi_s^{(c)}$.  A power-law behavior is also
observed in $R_s$. As it can be seen from Fig.~\ref{xil}, for $L_r
\lesssim 0.5$ $R_s$ scales as $L_r^{-\epsilon}$, although in this case
$\epsilon$ is much smaller, $\epsilon \approx 0.1$.  Note that, if the
behavior we observe for $0.02 \lesssim L_r \lesssim 0.5$ holds up to
$L_r = 0$ we would obtain $R_s,\Upsilon\to\infty$ for $L_r\to 0$.

\begin{figure}[tb]
\centerline{\epsfig{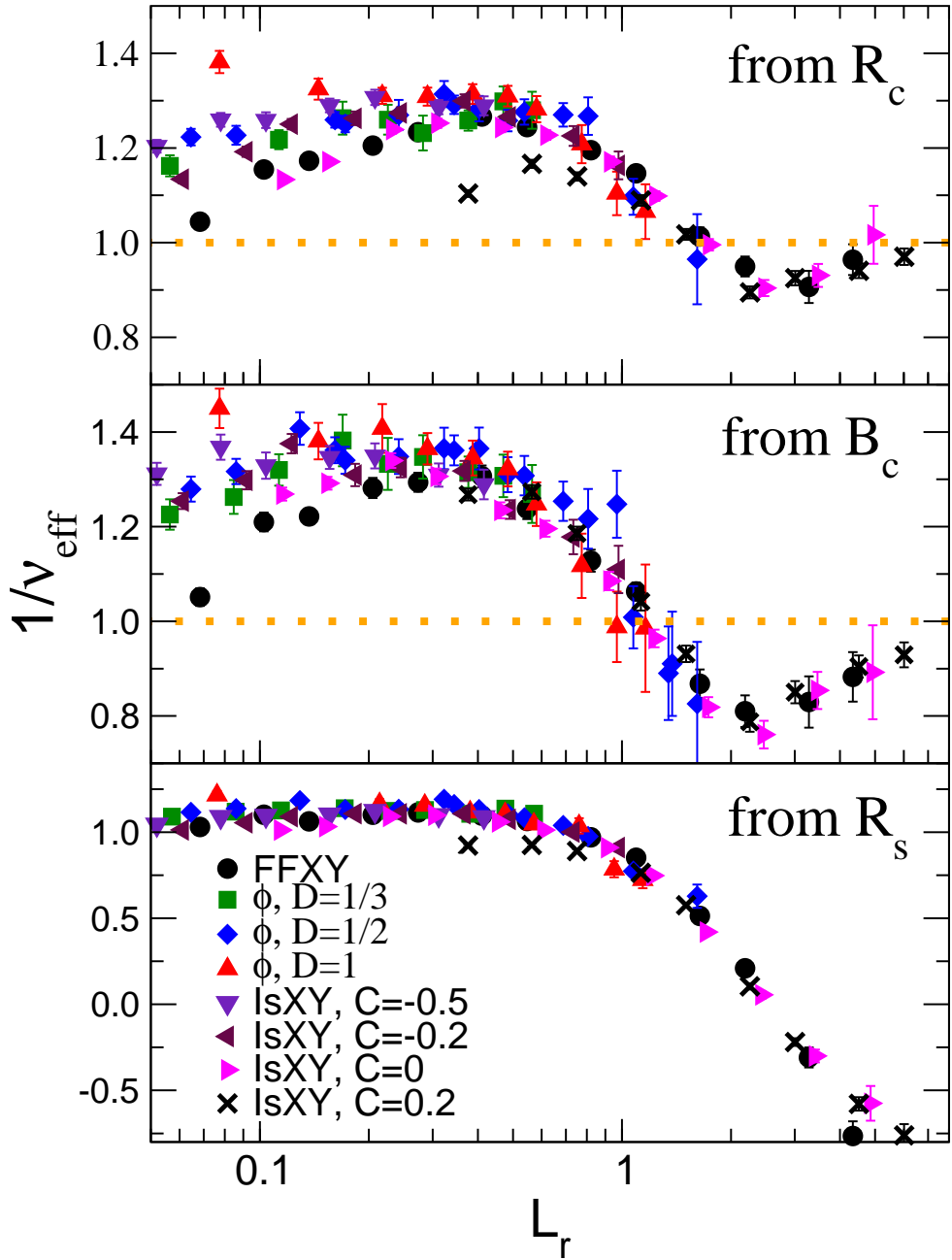}}
\caption{Effective exponent 
$1/\nu_{\rm eff}$ computed using $R_c$, $B_c$, and $R_s$, 
versus $L_r=L/l$.
The rescalings $l$ are the same as in Fig.~\protect\ref{xil}.  
For $L_r\to\infty$, $1/\nu_{\rm eff}$ should
converge to $1/\nu_{\rm Is} = 1$ for $R_c$ and $B_c$, and $-1$ for
$R_s$.  }
\label{nueff}
\end{figure}

In Fig.~\ref{nueff} we plot the effective exponents $1/\nu_{\rm eff}$,
defined as in Eq.~(\ref{nueffdef}), as obtained from the derivative of
$R_c$, $B_c$, and $R_s$ with respect to $J$, versus the rescaled
lattice size $L_{r} = L/l = L/\xi_s^{(c)}$.  We use the same
rescaling factors $l$ as above.  The agreement is quite good, although
the quality of the collapse for the chiral exponent $\nu_{\rm eff}$ is
not as impressive as what is observed for the spin-related quantities
$R_s$, $B_s$, and $B_{s\phi}$. These crossover curves show a quite
interesting pattern.  The effective exponents $1/\nu_{\rm eff}$
derived from $R_c$ and $B_c$ show first a rather flat region in which
$1/\nu_{\rm eff}\approx 1.3$, corresponding to $\nu_{\rm eff}\approx
0.8$.  Therefore, a FSS analysis limited to values $L\lesssim
\xi_s^{(c)}$ would apparently provide the estimate $\nu\approx 0.8$
for the critical exponent associated with the chiral correlation
length, as found in the FSS analysis of Sec.~\ref{sfssa}.  Only by
using larger lattices, i.e.~$L\gg \xi_s^{(c)}$, can one observe the
asymptotic Ising critical behavior.  Note that, to make things worse,
the asymptotic approach is nonmonotonic: for $L\approx \xi_s^{(c)}$
the exponent $\nu_{\rm eff}$ starts to increase, becomes larger than
the Ising value (the curves of $1/\nu_{\rm eff}$ have a minimum for
$L/\xi_s^{(c)}\approx 2$ corresponding to $\nu_{\rm eff}\approx 1.1$
for $R_c$ and $\nu_{\rm eff} \approx 1.2$ for $B_c$) and then
eventually converges to the Ising value.  The effective exponents
$1/\nu_{\rm eff}$ associated with the spin variables show a similar
crossover behavior.  The exponent derived from $R_s$ must converge to
$-1$ for $L\gg \xi_s^{(c)}$, since the spin correlation length is
finite.  However, for $L\lesssim \xi_s^{(c)}$ the data of $1/\nu_{\rm
eff}$ show a plateau at $1/\nu_{\rm eff}\approx 1.1$,
see Fig.~\ref{nueff}.  In the case of $B_s$ and $B_{s\phi}$ (not
shown) we observe an analogous plateau, but at $1/\nu_{\rm eff}\approx
1.7$.  Substantial deviations from the above behaviors are observed
for $D=4$ in the $\phi^4$ model and for $C=-2,-3$ in the IsXY model,
as one can already see from Figs.~\ref{nueffphi4plain} and
\ref{nueffisxyplain} where $1/\nu_{\rm eff}$ is plotted versus $L$.

\begin{figure}[tb]
\centerline{\psfig{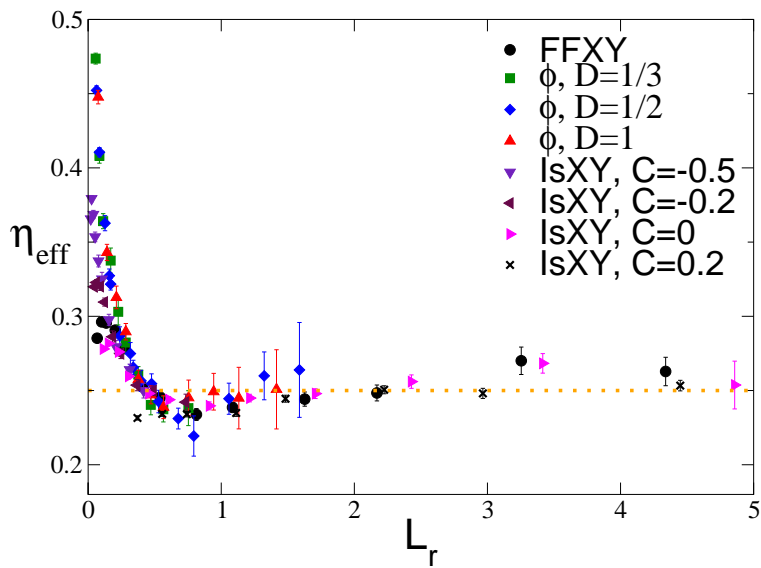}}
\caption{ 
Effective exponent $\eta_{\rm eff}$ at fixed $R_c$, as obtained
from $\chi_c$, 
versus $L_r=L/l$.
The rescalings $l$ are the same as in Fig.~\protect\ref{xil}.  
}
\label{etaeff}
\end{figure}

It is impossible to estimate from the data the behavior of the
effective exponent $\nu_{\rm eff}$ for $L_r\rightarrow 0$, since
points are quite scattered.  For small values of $D$ in the $\phi^4$
model $1/\nu_{\rm eff}$ is smaller than one for small lattice sizes,
as expected from the influence of the close O(4) multicritical point,
for which $1/\nu=0$, see Sec.~\ref{o4mc}.  On the other hand, for
$D\gtrsim 1$ and $C\lesssim -1$, the effective exponent approaches the
plateau from above, a behavior that may be due to the presence of a
first-order transition line for large $D$ and $-C$.

\begin{figure}[tb]
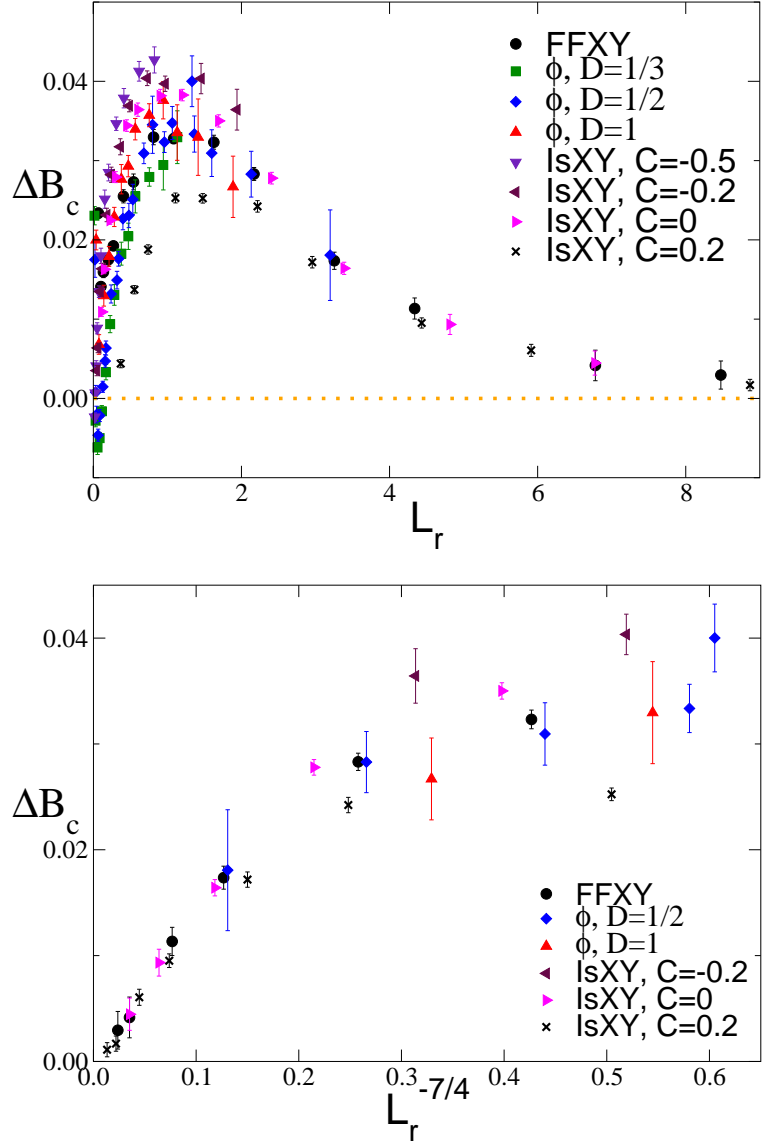

\centerline{\psfig{width=10truecm,angle=0,file=bicr.eps}}
\vspace{4mm}
\centerline{\psfig{width=10truecm,angle=0,file=bicras.eps}}
\caption{ 
Binder parameter $B_c$ at fixed $R_c$,
versus $L_r=L/l$.
The rescalings $l$ are the same as in Fig.~\protect\ref{xil}.  
We plot $\Delta B_c = B_c - B_{\rm Is}$,
where $B_{\rm Is} = 1.167923(5)$ is the value of the Binder parameter
at the critical point in the Ising model \protect\cite{SS-00}.
}
\label{bicr}
\end{figure}

In Figs.~\ref{etaeff} and \ref{bicr} we plot the effective exponent
$\eta_{\rm eff}$ obtained from $\chi_c$ and $\Delta B_c\equiv
B_c-B_{\rm Is}$. Again, we observe an approximate collapse of the
data.  A more quantitative check can be done by observing that the
approximate universality implies for $L\to\infty$
\begin{equation}
\Delta B_c = a (L/\xi_s^{(c)})^{-7/4},
\end{equation}
where $a$ is model independent. If we define $b \equiv a
(\xi_s^{(c)})^{7/4}$, the previous equation gives a relation between
the ratio of the constants $b$ in different models and the ratio of
the spin correlation lengths.  Using the estimates of $b$ reported in
Secs. \ref{fssffxy} and \ref{isingxy} for the FFXY model and IsXY
model with $C=0$, we find $b_{\rm FFXY}/b_{\rm IsXY,
C=0}=4.0(3)$. This is in perfect agreement with the prediction $b_{\rm
FFXY}/b_{\rm IsXY, C=0} =(\xi_{s,{\rm FFXY}}^{(c)}/\xi_{s,{\rm
IsXY,C=0}}^{(c)})^{7/4}=4.1(1)$.

The results reported above show that, in the 
FFXY model and in the $\phi^4$ and IsXY models for a rather extended region of 
parameters (respectively for $0<D\lesssim 2$
and $-1\lesssim C \lesssim 0.2$), the finite-size behavior at the
chiral transition is somewhat universal. A natural 
conjecture is that this universality is controlled by 
a {\it multicritical} point (or by a line of multicritical points of the 
same type) where chiral and spin degrees of freedom become both critical.
If this conjecture is correct, one must observe a universal crossover behavior 
also at the spin transition, with the same rescaling factors. 

\begin{figure}[tb]
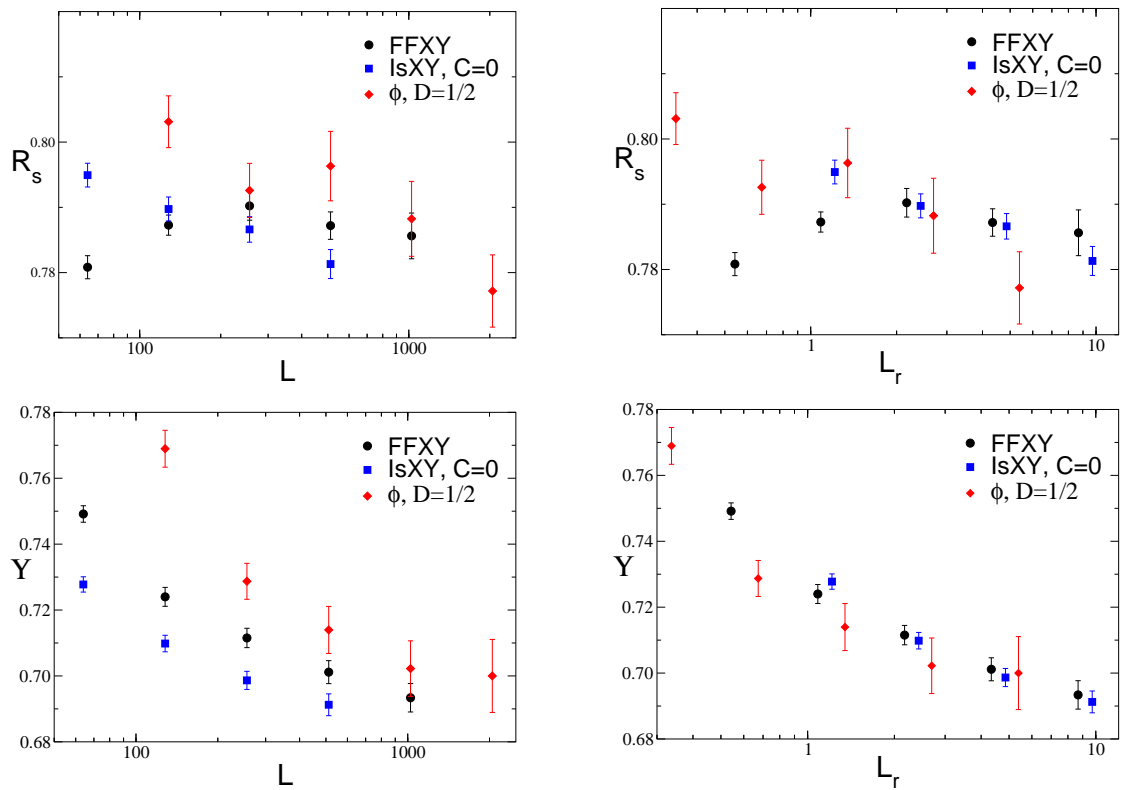

\begin{minipage}{17pc}
\includegraphics[width=16pc]{xisp2.eps}
\end{minipage}\hspace{2pc}%
\begin{minipage}{17pc}
\includegraphics[width=16pc]{xisprs2.eps}
\end{minipage} 
\begin{minipage}{17pc}
\vspace{2mm}
\includegraphics[width=16pc]{ysp2.eps}
\end{minipage}\hspace{2pc}%
\begin{minipage}{17pc}
\vspace{2mm}
\includegraphics[width=16pc]{ysprs2.eps}
\end{minipage} 
\caption{
Ratio $R_s\equiv \xi_s/L$ (above) and helicity modulus (below) at the
spin transition.  On the left we plot the data versus $L$, on the
right versus $L_r=L/l$. The rescalings $l$ are the same
as in Fig.~\protect\ref{xil}.  }
\label{resc-spin}
\end{figure}

In Fig.~\ref{resc-spin} we report $R_s \equiv \xi_s/L$ and $\Upsilon$
at the spin transitions for the three models for which we have 
a reliable estimate of $J_{\rm sp}$. 
The rescaling factors $l$ are those determined at the chiral
transition. The agreement is again quite good. The existence of scaling
at the two transitions with the same rescaling factors is another
piece of evidence in favor of the multicritical origin of
the universality we observe.
This hypothesis is also supported by the fact that the
ratio $\xi_s^{(c)}/\xi_c^{(s)}$ is approximately constant. 
Indeed, we have 
$\xi_c^{(s)}=8.0(5)$, $\xi_s^{(c)}=118(1)$, thus
$\xi_s^{(c)}/\xi_c^{(s)}=15(1)$, for the FFXY model;
$\xi_c^{(s)}=22(3)$, $\xi_s^{(c)}\approx 380$, thus
$\xi_s^{(c)}/\xi_c^{(s)}=17(3)$
(assuming an uncertainty of approximately 10\% 
on $\xi_s^{(c)}$), for the $\phi^4$ model for $D=1/2$;
$\xi_c^{(s)}=4.3(5)$, $\xi_s^{(c)}=52.7(4)$, thus
$\xi_s^{(c)}/\xi_c^{(s)}=12(2)$, for the IsXY model at $C=0$.
These results are consistent with a 
constant ratio $\xi_s^{(c)}/\xi_c^{(s)} \approx 15$. 

While for $L_r\to \infty$ one observes Ising and KT behavior,
in the opposite limit $L_r\to 0$, the scaling functions converge to 
the value of the corresponding observable at the multicritical point 
(assuming it exists, of course).
The relevant scales are
$\xi_s^{(c)}$ and $\xi_c^{(s)}$ at the two transitions and thus
information on the nature of the multicritical point can be
obtained by studying the crossover curves in the regimes 
$L\ll \xi_s^{(c)}$ (chiral transition) and
$L\ll \xi_c^{(s)}$ (spin transition). Unfortunately, in the models in which 
we have a precise knowledge of $J_{\rm sp}$, $\xi_c^{(s)}$ is quite small
and thus we are not able to probe the multicritical regime at the 
spin transition. 
On the other hand, since $\xi_s^{(c)}$ is quite large, the behavior at
the chiral transition gives us some indications on the multicritical point.

\begin{figure}[tb]
\centerline{\psfig{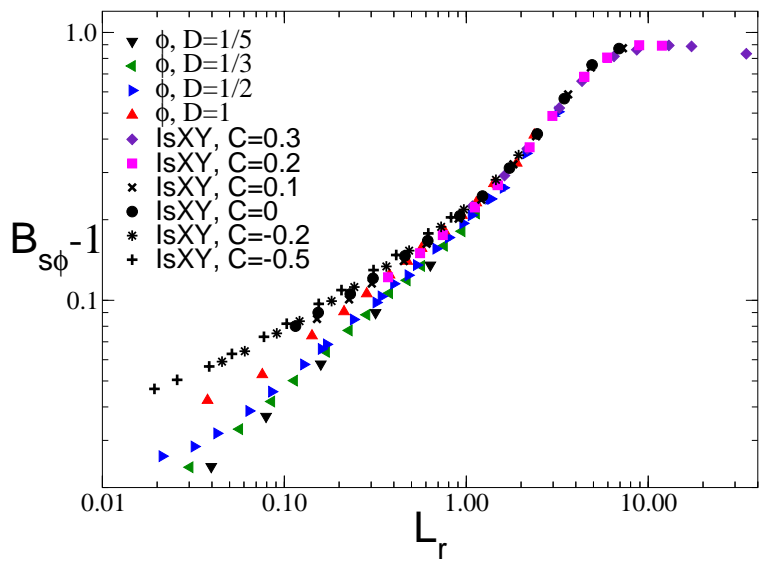}}
\caption{ 
Binder cumulant $B_{s\phi}$ at fixed $R_c$,
versus $L_r=L/l$.
The rescalings $l$ are the same as in Fig.~\protect\ref{xil}.  
Note the logarithmic scale on both axes.
}
\label{Bsphilog}
\end{figure}

From the data we can certainly exclude an Ising-XY decoupled multicritical 
point. Indeed, see Fig.~\ref{yupsilon}, the helicity modulus is clearly larger 
than the XY value $\Upsilon_{XY} \approx 0.6365$. A more likely
possibility is that the behavior for $L_r$ small is controlled 
by a zero-temperature transition point. Indeed, as we already discussed,
our data are compatible with $R_s,\Upsilon\to \infty$ for 
$L_r\to0$. In Fig.~\ref{Bsphilog} we report the results for 
$B_{s\phi}$ using a double logarithmic scale in order to make the 
region of small $L_r$ more visible. Even though scaling corrections
are present, it is quite plausible that $B_{s\phi}\to 1$ for $L_r\to 0$, 
as expected in the case of a zero-temperature transition.

\section{Conclusions}
\label{conclusions}

In this paper we investigated the phase diagram of the
square-lattice FFXY model and of two related models, a lattice
discretization of the LGW Hamiltonian for the
critical modes of the FFXY model, cf.~Eq.~(\ref{HLi}), and the
IsXY model (\ref{IsXYp}).  For this purpose we presented a
FSS analysis of the numerical results obtained by
high-statistics MC simulations on square lattices $L^2$, for quite
large values of $L$, up to $L=O(10^3)$.  Our main results can be
summarized as follows.

(i) In all models the LT phase is characterized by
the breaking of the ${\mathbb Z}_2$ symmetry with a chiral
magnetization, and by quasi-long-range order associated with the
breaking of the O(2) rotational symmetry.  We find that the critical
behavior of the spin modes is controlled by the same line of Gaussian fixed
points as in the standard XY model.  Indeed, we verify that the spin
correlation length, the helicity modulus, and the exponent $\eta$
satisfy the same universal finite-size relations that hold in the LT phase
of the XY model.

(ii) We conclusively show that the square-lattice 
FFXY model undergoes two very close
transitions, respectively at $J_{\rm sp}$ and $J_{\rm ch}$, with
$J_{\rm sp}> J_{\rm ch}$.  The transition at $J_{\rm sp}$ is a KT
transition associated with the spin modes, while for $J = J_{\rm ch}$
there is an Ising transition where the chiral ${\mathbb Z}_2$ symmetry
is restored.  For $J_{\rm ch} < J < J_{\rm sp}$, spins are
paramagnetic but chiral symmetry is broken.  The Ising and
KT transitions are very close: $\delta\equiv (J_{\rm sp}-J_{\rm
ch})/J_{\rm ch} = 0.0159(2)$.

(iii) The phase diagram of the $\phi^4$ model for $U=1$ is shown in
Fig.~\ref{phased}.  It presents two qualitatively different regions.
There is a region in which the behavior is analogous
to that of the FFXY model, with an Ising and a KT transition at
$J_{\rm ch} < J_{\rm sp}$.  These transitions are very close: we find
$\delta\equiv (J_{\rm sp}-J_{\rm ch})/J_{\rm ch} \lesssim 0.003$ 
for all values of $D$ in the two-transition region
($\delta=0.0025(2)$ for $D=1/2$). For
$D>D^*$ with $34 < D^*\approx 49$, we find
instead a single first-order transition.  We do not know how the two
continuous transition lines connect to the unique first-order one (see
the discussion at the beginning of Sec.~\ref{phasetr}). A
possibility is shown in Fig.~\ref{phased}: the Ising line ends at a
tricritical Ising point, which is the endpoint of the first-order
transition line; then the KT line meets the first-order one.  
Since for $D = 0$ there is no finite-$J$ transition,
it is natural to assume that the two transition lines get closer and
closer as $D\to 0$ and meet at $1/J=0, D=0$, which thus acts as an
O(4)-symmetric multicritical point.  
The critical lines approach the multicritical point according to
$J_{\rm ch} \sim \ln D^{-1}$ and $\delta\sim 1/\ln D^{-1}$ for
$D\rightarrow 0$.

(iv) The phase diagram of the IsXY model (\ref{IsXYp}) is shown in
Fig.~\ref{phasedisxy}. In Sec.~\ref{models} we argue that the IsXY
model cannot undergo a continuous transition in which chiral and spin
modes become both critical.  Our MC results support this fact.  As in
the $\phi^4$ model, there is a rather extented region in which there
is an Ising and a KT transition with $J_{\rm ch} > J_{\rm sp}$.  The
Ising and KT transitions are close in most of the cases.  For example,
we find $\delta\equiv (J_{\rm sp}-J_{\rm ch})/J_{\rm ch} =0.0167(7)$
for $C=0$.  For $C < C^*$, $-7\lesssim C^* \lesssim -5$, we find a
single first-order transition.  Again, we do not know how the two
continuous transition lines connect to the unique first-order one.  A
possibility is shown in Fig.~\ref{phasedisxy}.

(v) We do not find evidence of continuous transitions---but we cannot
completely exclude their presence---where chiral and
spin modes become both critical, giving rise to a new universality
class.  Thus, our results do not support the field-theoretical results of
Refs.~\cite{COPS-03,CP-01}, which provided some evidence for the
existence of a stable fixed point in the RG
flow of the LGW continuous
theory (\ref{LGWH}).  However, they are not necessarily in
contradiction, since the lattice models considered here may be outside
the attraction domain of the stable fixed point found in
Refs.~\cite{COPS-03,CP-01}. Note also that the existence of two
continuous transitions at fixed bare parameters (i.e., at fixed 
$D$ and $U$, or $C$) means that the critical modes are not
appropriately described by the LGW theory, at least in the cases
considered in this paper. Actually, it suggests that a field theory
with an Ising and an XY field, for instance the multicritical theory of 
Refs.~\cite{KNF-76,CPV-03}, might be more appropriate.

(vi) In most cases the spin correlation length $\xi_{s}^{(c)}$ at the
chiral transition is quite large. 
For example, $\xi_{s}^{(c)} \approx 118 $ in the FFXY model,
$\xi_{s}^{(c)} \approx 53$ in the IsXY model for $C=0$, and
$\xi_{s}^{(c)} \approx 380$ in the $\phi^4$ model for $D=1/2$.
The presence of this additional
length scale makes the observation of the expected Ising behavior
quite problematic. We find that Ising behavior is observed only for $L
\gg \xi_{s}^{(c)}$, which in turn requires very large lattices. 

(vii) At the chiral and spin transitions we observe a preasymptotic 
regime that is universal to some extent.  Indeed, in the FFXY model,
in the $\phi^4$ model with $U=1$ and $0< D \lesssim 2$, and in
the IsXY model with $-1\lesssim C \lesssim 0.2$, we find that RG
quantities at the transitions scale approximately as
\begin{equation}
{\cal R}(L) \approx f_{\cal R}(L/l),
\label{crossoverscaling}
\end{equation}
where $f_{\cal R}(x)$ is a model-independent function that is of course 
different at the two transitions. The rescalings $l$ are model dependent
and set the length scale. Given two different models, $l_1/l_2$ is
equal to the ratios $\xi_{s,1}^{(c)}/\xi_{s,2}^{(c)}$ or 
$\xi_{c,1}^{(s)}/\xi_{c,2}^{(s)}$.
One might expect scaling (\ref{crossoverscaling}) 
to be valid for $L\to\infty$, $l\to \infty$ (or, equivalently, 
$\xi_{s}^{(c)}, \xi_{c}^{(s)} \to \infty$) at
fixed ratio. However, these conditions are not sufficient to observe this
universal behavior. Substantial deviations are observed in other parameter 
regions, e.g., for $D=4$ in the $\phi^4$ model and for $C=-2,-3$ in
the IsXY, even though $\xi_s^{(c)}$ is large. This may be due to the 
influence of the first-order transition present for large values of $D$ and 
$-C$. Note that, at the chiral transition, the effective chiral exponent $\nu$ 
shows an intermediate region in which  it is approximately equal to 0.8. This 
explains earlier results obtained by using smaller lattices. 
By using only data with $L\lesssim \xi_s^{(c)}$
one would obtain $\nu = 0.8$ instead of the Ising value.
Scaling (\ref{crossoverscaling}) may be due to a multicritical
point where chiral and spin modes are both critical. 
The analysis of the crossover curves for $L/l\to 0$ allows us to 
exclude a decoupled Ising-XY multicritical transition. A zero-temperature
point is instead compatible with the numerical data.
Note that the universal behavior
we have observed here concerns the finite-size behavior at
criticality, but of course similar phenomena are expected in the
thermodynamic limit close to the critical points.
If $t_{\rm ch}$ and $t_{\rm sp}$ are the reduced
temperatures and $\delta$ the temperature difference of the two
transitions, Ising and KT behavior can only be observed if $t_{\rm
ch}, t_{\rm sp} \ll \delta$. In the opposite limit, $1\gg t_{\rm ch},
t_{\rm sp} \gtrsim \delta$ we expect instead a universal crossover
behavior.  

The identification of the multicritical point is a problem that is
still open and requires further investigation.  If the relevant
multicritical point is a zero-temperature transition, as the MC data
suggest, then the O(4) point discussed in
Sec.~\ref{o4mc} is a possible candidate. Otherwise, one should also 
the consider the finite-temperature multicritical point expected
in the phase diagram of the generalized IsXY model \cite{LGK-91} with
Hamiltonian
\begin{equation}
{\cal H}_{\rm IsXY} = - J \sum_{\langle xy\rangle} 
\left[ (A+B\sigma_x \sigma_y) \vec{s}_x \cdot \vec{s}_y + C \sigma_x\sigma_y \right],
\label{IsXY}
\end{equation}
in frustrated XY models with modulated couplings
\cite{BDGL-86,EHKT-89,GKS-98} (they are obtained by generalizing the
FFXY model (\ref{ffxyim}), setting $j_{xy}=-\eta$ with $\eta>0$ on the
antiferromagnetic links), and in $\phi^4$ theories in which the
field-interchange symmetry is broken or in which one of the two O(2)
symmetries is broken by adding, for instance, an interaction term of
the form $(\vec{\phi}_{1,x}\cdot \vec{\phi}_{2,x})^2$.

\section*{Acknowledgments}

We thank Pasquale Calabrese for useful discussions, and Alberto Ciampa
and Maurizio Davini for their technical assistance to
manage the computer clusters where the Monte Carlo simulations have been done.
The numerical simulations were performed at the Computational Centers 
of the Physics Department and of INFN in Pisa.

\appendix

\section{Some details on the Monte Carlo algorithms}
\label{mcalg}

\subsection{The FFXY model}

The algorithm we used for the simulation of the FFXY model is based on
the results of Ref.~\cite{PP-98}.  These authors found that the 
addition of a certain number of overrelaxation (OR) sweeps 
to a local Metropolis sweep makes the algorithm more efficient.
They also suggested an efficient method to generate the proposed 
new spin in the Metropolis update. Every sweep over the lattice
\footnote{ We have generated a new $\vec{r}$ more frequently, namely
for each row of the lattice. We have not tested whether or not this
gives an advantage over choosing a new $\vec{r}$ just once every
sweep.} one chooses a single unit vector $\vec{r}=(\cos\alpha,\sin\alpha)$,
where $\alpha$ is uniformly distributed in $[0,2\pi)$. 
Then, for each site $x$ the proposal is 
\begin{equation}
 {\vec{s}_x}\! ' =2 (\vec{r} \cdot \vec{s}_x) \vec{r} - \vec{s}_x .
\end{equation}
This proposal is accepted with the Metropolis probability
\begin{equation}
 A=\mbox{Min}\left[ e^{-({\cal H}'-{\cal H})},1\right] .
\end{equation}
The advantage of this proposal compared with more standard ones is
that no random numbers are needed and also no expensive  
trigonometric function is evaluated.
Furthermore, the authors of Ref.~\cite{PP-98} showed that the
autocorrelation times in terms of sweeps of the Metropolis
update with this type of proposal (in the following
we will refer to it as MR update) are essentially the same as those of 
a Metropolis update in which the new spin is obtained by 
a random rotation of the old one.
At the chiral transition the acceptance rate of the MR update is 
approximately $25 \%$, in agreement with Ref.~\cite{PP-98}.

In most of our simulations we used a demon version \cite{Creutz}
of the Metropolis update. We indicate the demon version of any update 
by a adding a letter D. The demon version of the MR update is 
therefore denoted by DMR. In this framework, one introduces 
a demon field $d_x \in [0,\infty)$ with distribution $\exp(-\sum_x d_x)$
and replaces the Metropolis accept/reject step in the following way.
One defines 
\begin{equation}
 d_x' = d_x - {\cal H}'+{\cal H}.
\end{equation}
If $d_x' \ge 0$, the proposal is accepted and 
$\vec{s}_x$, $d_x$ are replaced by  ${\vec{s}_x}\! '$, $d_x'$. Otherwise,
the demon and the spin keep their old value. This way the 
generation of a random number and,
more importantly, the evaluation of the exponential is avoided. 
However, these updates do not change the sum
\begin{equation}
\sum_x d_x +  {\cal H} .
\end{equation}
Therefore, one must also perform heat-bath updates of all demons,
by taking $d_x = -\ln z_x$,
where $z_x$ is a random number uniformly distributed in (0,1]. 
It is easy to see that by 
updating $d_x$ before the update of $\vec{s}_x$, one recovers
the standard Metropolis algorithm. Also note that the acceptance 
rate of the demon version of the Metropolis update is exactly the same 
as that of the standard one, although the autocorrelation times might increase.

The local overrelaxation (OR) update is given by
\begin{equation}
{\vec{s}_x}\! ' =  \frac{2 \vec{s}_x \cdot \vec{S}}{S^2} \vec{S} - 
\vec{s}_x \;\;,
\end{equation}
where $\vec{S}=\sum_{y.nn.x} j_{xy} \vec{s}_y$ 
is the sum  of $\vec{s}_y$ over the nearest neighbors $y$ of site $x$. 
One advantage of the OR update is that no random numbers and 
no evaluations of the exponential  
function are needed. In the case of the standard XY model one finds 
that for the optimal mixture of Metropolis and OR updates the dynamical
critical exponent is reduced to $z \approx 1.2$ \cite{GuDeBaFoBaAp88}. 
In contrast, for the FFXY model at the chiral
transition the authors of Ref. \cite{PP-98} find 
only a speed-up by a constant factor.

A complete cycle of the algorithm we used is:
\begin{itemize}
\item Heat-bath of the demons;
\item
$5\times$[(DMR$+$OR) sweep $+$ $N_{\rm or}$ OR sweeps].
\end{itemize}
Here (DMR $+$ OR) means that a single-site DMR update is followed by an 
elementary  OR update at the same site.
At the chiral transition we used $N_{\rm or}=3$, while at the KT transition
larger values of $N_{\rm or}$ were used.

We have implemented a parallel program using the MPI library.  To this
end, we have divided the lattice in one direction into sublattices of
size $l \times L$, where $l \times n_P =L$ and $n_P$ is
the number of processes that is used. 
In order to speed up the
simulation, we also did some kind of parallelization in the second direction,
running the OR sweep in the second direction in the following way:
\begin{verbatim}
    for (x1=0;x1<L/2;x1++) {OR site (x0,x1); OR site (x0,x1+L/2)},
\end{verbatim}
where ${\tt (x0,x1)}$ is a lattice point.
This small change allows a better use of the floating-point pipelines
and gives a speedup of approximately 30\%. On a 
server with 4 $1.8$GHz Opteron CPUs, one complete update cycle
takes $0.26$s for $L=1000$.
%(on mercury 2 x 2 dual core $2.2$ GHz Opteron machine  $0.24$ s.)
Roughly speaking, one cycle takes
$1$ s on one Opteron processor with one core.

In table \ref{tauffXY} we report the integrated
autocorrelation times of the chiral susceptibility for 
$L \ge 256$. The couplings $J$ are close to the chiral critical value
$J_{\rm ch}=2.20632(5)$.  Fitting all data given in table \ref{tauffXY} we get
$N_{\rm local} \tau_{\rm ch} = 0.022(8) \times L^{1.91(6)}$ with
$\chi^2/$d.o.f.$=0.99$.

\begin{table}
\caption{\label{tauffXY}
 Integrated autocorrelation times $\tau_{\rm ch}$ for the chiral susceptibility 
 of the FFXY model close to the chiral transition.
 $\tau_{\rm ch}$ is reported in number of measurements.
 $\#$ meas. is the total number of measurements and
 $N_{\rm local}$ is the number of update cycles between two measurements.
}
\begin{indented}
\item[]
\begin{tabular}{@{}rlccl}
\hline
\multicolumn{1}{c}{$L$}&
\multicolumn{1}{c}{$J$}&
\multicolumn{1}{c}{$\#$ meas.}&
\multicolumn{1}{c}{$N_{\rm local}$}&
\multicolumn{1}{c}{$\tau_{\rm ch}$} \\
\hline
 256 & 2.2053  & 22000   & 500 & 1.76(10) \\
 384 & 2.2062  & 56600   & 700 & 2.67(13) \\
 512 & 2.2063  & 39500   &1000 & 3.3(3) \\
 800 & 2.2066  & 20500   &2000 & 3.6(3) \\
1000 & 2.2063  & 35700   &2000 & 6.2(5) \\
\hline 
\end{tabular}
\end{indented}
\end{table}

At the KT transition, critical slowing down can be efficiently reduced
by using overrelaxation updates. In the case of the standard XY model, at
the transition one should increase $N_{\rm or}$ as $N_{\rm or} \propto
L$.  We found that also in the case of the FFXY model at the KT
transition this procedure leads to very small autocorrelation
times. For $J=2.242$ [$J_{\rm sp} = 2.2415(5)$] 
and $L=1024$, by using $N_{\rm or}=12$ we find an
integrated autocorrelation time $\tau_{\rm sp} \approx 5.4(5)$ in
units of cycles (heat bath of the demons followed by
5$\times$[(DMR$+$OR)$+$12 OR] sweeps).  For $L\approx 1000$ 
the autocorrelation time is smaller by a factor of 500 than 
that at the chiral transition.

\subsection{The Ising-XY model}

For the IsXY model we used three types of updates. We performed
single-cluster \cite{Wolff-89} and overrelaxation updates 
of the spins $\vec{s}$ keeping the chiral field $\sigma$ fixed, 
and Metropolis updates in which $\sigma$ and $\vec{s}$ are both changed.
The proposal for the Metropolis update is
\begin{equation}
 \sigma_x' = - \sigma_x,\qquad   
{{\vec{s}}_x}\! ' = (\cos\alpha,\sin\alpha),
\end{equation}
where $\alpha$ is a random number uniformly distributed in $[0,2 \pi)$. 
Close to the chiral transition, the acceptance rate of this update 
is approximately $11 \%$ for $C=0$ and $3 \%$ for $C=-7$.
We mostly used the demon version of this update
(in the following we refer to it as DM update).

Between two measurements we perform  
$N_{\rm local}$ local cycles [a local cycle
consists in a heat bath of the demons and in 5$\times$(DM sweep $+$ OR sweep)]
followed by $N_{\rm single}$ single-cluster updates. 
On one 2.2 GHz Opteron CPU, one local cycle 
%% (i.e. 5 $\times$ (DM+OR)) 
takes approximately $0.20$s for $L=512$.

In Table \ref{tauIsingXY} we report the integrated
autocorrelation times of the chiral susceptibility for lattices with
$64 \le L \le 360$.  The couplings $J$ are close to the chiral critical
value $J_{\rm ch} = 1.4684(1)$.  The number of single-cluster updates was 
chosen {\it ad hoc}, since 
the cluster update has little influence on the autocorrelation times.
Fitting $N_{\rm local}\tau_{\rm ch}$ for $L \ge 128$
(we are neglecting here the role of the single-cluster updates)
we obtain $N_{\rm local} \tau_{\rm ch} = 0.12(4) \times
L^{1.95(6)}$ with $\chi^2/$d.o.f.$=0.53$.
At the chiral transition,
the cluster algorithm leaves the dynamic critical exponent
substantially unchanged, i.e.~$z \approx 2$.  A similar observation
holds for other values of $C$, whenever the transition is of second
order.

\begin{table}
\caption{\label{tauIsingXY} Integrated autocorrelation times $\tau_{\rm ch}$ 
for the chiral susceptibility in the IsXY model at $C=0$, close to the
chiral transition. $\tau_{\rm ch}$ is reported in units of
measurements.  $\#$ meas. is the total number of measurements,
$N_{\rm local}$ is the number of local update cycles, 
and $N_{\rm single}$ is the number of single-cluster updates per
measurement.}
\begin{indented}
\item[]
\begin{tabular}{rrcccl}
\hline
\multicolumn{1}{c}{$L$}&
\multicolumn{1}{c}{$J$}&
\multicolumn{1}{c}{$\#$ meas.}&
\multicolumn{1}{c}{$N_{\rm local}$}&
\multicolumn{1}{c}{$N_{\rm single}$}&
\multicolumn{1}{c}{$\tau_{\rm ch}$}\\
\hline
%  64 &1.4660 & 100000  & 500 & 2 $\times$ 10  &  0.92(2)\phantom{0} \\
%  90 &1.4672 & 100000  & 700 & 2 $\times$ 10  &  1.22(3)\phantom{0}  \\
% 128 &1.4678 & 100000  &1000 & 2 $\times$ 20  &  1.62(5)\phantom{0}  \\
% 180 &1.4681 & 100000  &1000 & 2 $\times$ 100  &  3.05(12) \\
% 256 &1.4681 &  75000  &1000 & 2 $\times$ 50  &  6.25(27) \\
% 360 &1.4585 &  50000  &2000 & 2 $\times$ 40  &  6.14(42) \\
  64 &1.4660 & 100000  & 500 & 20  &  0.92(2)\phantom{0} \\
  90 &1.4672 & 100000  & 700 & 20  &  1.22(3)\phantom{0}  \\
 128 &1.4678 & 100000  &1000 & 40  &  1.62(5)\phantom{0}  \\
 180 &1.4681 & 100000  &1000 & 200  &  3.05(12) \\
 256 &1.4681 &  75000  &1000 & 100  &  6.2(3) \\
 360 &1.4585 &  50000  &2000 & 80  &  6.1(4) \\
\hline
\end{tabular}
\end{indented}
\end{table}

Critical slowing down at the KT transition and in the LT
phase is essentially eliminated by the single-cluster updates. 
For instance, 
for $L = 512$ and $J=1.492$ we get $\tau_{\rm sp}= 0.94(1)$ in
units of measurements, where a measurement was performed after 4
single-cluster updates and 20 local cycles.

\subsection{The $\phi^4$  model}

Also here, we used a combination of local Metropolis updates and
single-cluster updates. 
The single-cluster algorithm was used to update 
the angle of either $\vec{\phi}_1$ or $\vec{\phi}_2$. 

We used different versions of the local Metropolis update, depending 
on the value of $D$. 
For $D \le 4$ we used a Metropolis update with the  following proposal
(update M4):
\begin{equation} 
\label{standardupdate}
\phi_{ij,x}'= \phi_{ij,x} + s (r_{ij,x}-0.5) \;\;,
\end{equation}
where $\vec{\phi}_{j,x}\equiv (\phi_{1j,x},\phi_{2j,x})$
and $r_{ij,x}$ are random numbers with uniform distribution
in $[0,1)$.  The step size $s$ was chosen so that the acceptance
rate is roughly $50 \%$. 

This update becomes inefficient for large values of $D$. Indeed,
for $D$ large,
$|\vec{\phi}_{1,x}| \gg |\vec{\phi}_{2,x}|$ or vice versa, so
that the Metropolis proposal (\ref{standardupdate}) leads 
to changes of the fields that are of the typical size of the smaller one.  
Hence, the algorithm becomes inefficient.
To deal with this problem, we constructed proposals that are specifically
adapted to this situation: 
\begin{itemize}
\item[(M2)] We change only one field at each time.
With probability 1/2 we change either $\phi_1$  or $\phi_2$ using the 
proposal
\begin{eqnarray}
\phi_{11,x}' = \phi_{11,x} + s (r_{1,x} -0.5), \qquad
\phi_{12,x}' = \phi_{12,x} + s (r_{2,x} -0.5)
\end{eqnarray}
or the analogous one for $\phi_{2,x}$. 
Here $r_{1,x}$ and $r_{2,x}$ are random numbers with uniform 
distribution in $[0,1)$. The step size $s$ is chosen so that 
the acceptance rate is approximately $25 \%$. This way, we achieve an efficient 
update of the large component of the field.
\item[(MLE)]
We exchange the lengths of the two fields.
The proposal is given by
\begin{eqnarray}
\vec{\phi}_{1,x}' =
  \frac{|\vec{\phi}_{2,x}|}{|\vec{\phi}_{1,x}|} \vec{\phi}_{1,x},\qquad
\vec{\phi}_{2,x}' = 
  \frac{|\vec{\phi}_{1,x}|}{|\vec{\phi}_{2,x}|} \vec{\phi}_{2,x} \;. 
\label{MLE}
\end{eqnarray}
This type of update should speed up the 
chiral degrees of freedom. The acceptance rate is $14 \%$ at the chiral
transition at $D=49$. 
%%% ($37 \%$ at $D=0.5$). 
One could have also 
exchanged the components of the two fields, using 
\begin{eqnarray}
 \vec{\phi}_{1,x}' = \vec{\phi}_{2,x}, \qquad
 \vec{\phi}_{2,x}' = \vec{\phi}_{1,x}.
\end{eqnarray}
However, this proposal has a smaller acceptance rate than (\ref{MLE}).
\end{itemize}

Beside these updates we also used local updates that leave invariant
part of the Hamiltonian and are close in spirit to 
overrelaxation updates. A first possibility (OR4)  consists in 
performing a field reflection that keeps 
the O(4)-invariant part of the Hamiltonian constant:
\begin{equation}
{\vec{\phi}_x}\! ' =
2 \frac{\vec{\phi}_x \cdot \vec{S}}{\vec{S}^2} \vec{S}  - \vec{\phi}_x
\end{equation}
where $\vec{S}$ is the sum over the fields of the four neighbors of $x$. 
For convenience, we define the four-component vector 
$\vec{\phi}_x=(\phi_{11,x},\phi_{12,x},\phi_{21,x},\phi_{22,x})$.
This type of update has no free parameters. The 
remaining part of the Hamiltonian is taken into account 
in an accept/reject step. We expect 
the acceptance rate to decrease with increasing $D$. 
At the chiral transition we find 
$87 \%$, $70 \%$, $61 \%$, $43 \%$, $40 \%$   for $D=1/2, 2, 4, 34, 99$.
Note that it seems to have a finite limit as $D \rightarrow \infty$.   
In the simulations we used a demon version (DOR4) with a single demon 
(not one for each site).

A second possibility is the overrelaxation of one component of the field (OR1).
The Hamiltonian for a single component $\phi_{ij,x}$ 
of the field, keeping all other fields fixed, has the form 
\begin{equation}
 \bar{H}(\phi_{ij,x}) = a \phi_{ij,x} + b \phi_{ij,x}^2 +  c \phi_{ij,x}^4.
\end{equation}
The proposal
\begin{equation}
  \phi_{ij,x}' =- \phi_{ij,x} -a/b 
\end{equation}
keeps the value of $a \phi_{ij,x} + b \phi_{ij,x}^2$ constant. 
The quartic part of the 
Hamitonian is taken into account in an accept/reject step. 
The acceptance rate for this type of move is  $67 \%$ at the chiral
transition for $D=49$  ($56 \%$ for $D=0.5$). 
Such a step, in a demon version, was performed subsequently
for each of the four components of the field at a given site.
We used a single demon for the whole lattice, refreshing it before 
the update of each site (this is equivalent to using a demon for each
lattice site and updating it before each sweep).

Because of the different features of the model for large and small
values of $D$ we used different local update cycles in the two regimes. 
For $D\le 4$ we used a local cycle with
an M4 sweep followed by $N_{\rm or}$ DOR4 sweeps.
In most of the cases we used $N_{\rm or} = 8$, but there is 
no sharp dependence of the efficiency on $N_{\rm or}$.
For large values of $D$ a typical local cycle is instead:
(M2+DOR4) sweep, 2 DOR4 sweeps, DOR1 sweep, 2 DOR4 sweeps, (MLE+DOR4) sweep, 
and finally 2 DOR4 sweeps. 
Here (M2+OR4) means a local M2 update followed by a local 
DOR4 update on the same site; (MLE+DOR4) has the same meaning.
Beside the local cycle we also performed 
single-cluster updates: between two measurements we perform
$N_{\rm local}$ local update cycles followed 
by $N_{\rm single}$ single-cluster updates on each field.

Let us look in more detail at the performance at $D=1/2$.  At the
chiral transition, by using the local cycle (M4 + 8 DOR4),
we obtain the autocorrelation times reported in Table
\ref{tauphi4}.  They are consistent with a critical exponent $z = 2$.

At the KT transition the cluster algorithm allows us to eliminate
the slowing down of the spin degrees of freedom. For $J=1.47$ and 
$L = 2048$ we find $\tau_{\rm sp}=2.2(3)$ in units of
measurements. Between two measurements we performed 10 local cycles and $10$ 
single-cluster updates on each field.

\begin{table}
\caption{\sl \label{tauphi4}
 Integrated autocorrelation times $\tau_{\rm ch}$ for the chiral susceptibility
 in the $\phi^4$ model at $D=1/2$, close to the chiral transition.
 $\tau_{\rm ch}$ is reported in units of measurents. 
$\#$ meas. is the total number of measurements, $N_{\rm local}$
is the number of local update cycles (one cycle corresponds to 
M4 + 8 DOR4) and 
$N_{\rm single}$ the number of single-cluster updates for each field 
per measurement. 
}
\begin{indented}
\item[]
\begin{tabular}{@{}rlcccl}
\hline
\multicolumn{1}{c}{$L$} &
\multicolumn{1}{c}{$J$} &
\multicolumn{1}{c}{$\#$ meas.}&
\multicolumn{1}{c}{$N_{\rm local}$}&
\multicolumn{1}{c}{$N_{\rm single}$}&
\multicolumn{1}{c}{$\tau_{\rm ch}$}\\
\hline
 360 & 1.4663 & 100000  & 200 & 10 &  11.0(6) \\
 512 & 1.4665 &  43000  & 320 & 10 &  14.5(1.1)\\
 600 & 1.4666 &  40000  & 200 & 10 &  40(8) \\
 800 & 1.4665 & 127000  & 100 & 10 &  95(11)\\ 
1200 & 1.4666 &  37600  & 200 & 10 & $>$100   \\
\hline
\end{tabular}
\end{indented}
\end{table}

\section{Some results for the
low-temperature phase of 2-$d$ XY models}
\label{elmodeta}

In the spin-wave limit the 
partition function of the 2-$d$ XY model is 
\begin{eqnarray}
 Z &=&  \sum_{n_1,n_2} \int \mbox{D} [\phi] 
 \exp\left[ -\frac{\beta_{SW}}{2} \sum_{x,\mu} 
 (\phi_x - \phi_{x+\hat \mu} - 2 \pi n_\mu  \delta_{x_\mu,L_\mu})^2\right] 
 \nonumber \\
  &=& \sum_{n_1,n_2} W(n_1,n_2) \int \mbox{D} [\phi]
 \exp\left[ -\frac{\beta_{SW}}{2} \sum_{x,\mu} (\phi_x - \phi_{x+\hat \mu})^2
  \right],
\end{eqnarray}
where $n_1$ and $n_2$ are integer.
The shift $2 \pi n_\mu  \delta_{x_\mu,L_\mu}$ at the boundary 
takes into account winding XY configurations on lattices with periodic 
boundary conditions.   The weights are given by
\begin{equation}
 W(n_1,n_2) = \exp\left[ - 2 \pi^2 \beta_{SW} 
\left( \frac{L_2}{L_1} n_1^2 + \frac{L_1}{L_2} n_2^2 \right)  \right] .
\end{equation}
Note that the contributions from $(n_1,n_2) \ne (0,0)$ are tiny: for instance,
we have $W(1,0)=3.487... \times 10^{-6}$ for $L_1=L_2$ at $\beta_{SW}=2/\pi$. 
The LT phase of the XY model is effectively described by the 
spin-wave model for  $\beta_{SW} \ge 2/\pi$. For smaller values of 
$\beta_{SW}$, vortices become relevant and disorder the system. 
Setting $s_x=\cos \phi_{x}$,
the XY correlation functions are given by
\begin{eqnarray}
\label{correlation1}
\langle s_x s_y \rangle = 
\phantom{0000000000000009999999999999999999000000000000000000000} \\
\frac{\sum_{n_1,n_2} W(n_1,n_2)  
\cos[p_1 n_1 (x_1-y_1)  + p_2 n_2 (x_2-y_2)] 
\langle  \exp[i 2 \pi (\phi_x-\phi_y)] \rangle_{G}
}
{ \sum_{n_1,n_2} W(n_1,n_2) } ,
\nonumber 
\end{eqnarray}
where $p_{\mu}= 2 \pi /L_{\mu}$. $\langle . \rangle_{G}$ denotes the 
expectation value in a Gaussian system without boundary shift. Using the 
Wick rule 
\begin{equation}
  \langle \phantom{0} (\phi_x-\phi_y)^{2 n} \rangle_{G}
= (2n-1)!! \; \langle  (\phi_x-\phi_y)^{2} \rangle_{G}^n,
\end{equation}
we can express the exponential in terms of the propagator of the 
Gaussian field:
\begin{equation} 
\label{correlation2}
 \langle \exp[ i 2 \pi (\phi_x-\phi_y)] \rangle_{G}
 = \exp\left[- 2 \pi^2  \langle (\phi_x-\phi_y)^2 \rangle_{G}\right] .
\end{equation} 
The exponent $\eta$ of the magnetic susceptibility is given by
\begin{equation}
\eta = \frac{1}{2 \pi \beta_{SW}}\, .
\label{etaxy}
\end{equation}
The ratio $R\equiv \xi/L$ is computed numerically using 
Eqs.~(\ref{correlation1}) and (\ref{correlation2}). 
One can also derive the helicity modulus. 
For a lattice with $L_1=L_2$ we have \cite{Has-05}
\begin{equation}
\Upsilon=\beta_{SW} - 
{4 \pi^2 \beta_{SW}^2 \sum_{n=-\infty}^\infty n^2 \exp (- 2\pi^2 \beta_{SW} n^2)\over 
\sum_{n=-\infty}^\infty \exp (-2 \pi^2 \beta_{SW} n^2)}.
\label{xyupsilon}
\end{equation}
One may use Eq.~(\ref{etaxy}) to replace $\beta_{SW}$ with $\eta$ in
Eq.~(\ref{xyupsilon}), and obtain the universal relation between the
helicity modulus and critical exponent $\eta$ in the large-$L$ limit
of a square $L\times L$ lattice. An analogous relation can be obtained
for the ratio $\xi/L$.  

According to the KT theory, the leading corrections to $\chi$, $\xi/L$
and $\Upsilon$ in the LT phase are proportional to $L^{-2x}$ with
$x=\pi \beta -2$.  Using Eq.~(\ref{etaxy}), we have $2x=1/\eta-4$.  In
the case of $\chi$ and $\xi$ we also have corrections from the
analytic background---they are of order $L^{\eta-2}$--- which dominate
for $2-\eta< 2 x$.

\end{document}